\documentclass[aps,prb,twocolumn,10pt,longbibliography,floatfix,nofootinbib,superscriptaddress]{revtex4-2}

\usepackage{physics}
\usepackage{bbold}
\usepackage[english]{babel}
\usepackage[utf8]{inputenc}
\usepackage{amssymb}
\usepackage{amsmath}
\usepackage{dsfont}
\usepackage{multirow,tabularx,makecell}
\usepackage[pdftex]{graphicx}
\usepackage{xspace}
\usepackage{dsfont}
\usepackage{bm}
\usepackage[export]{adjustbox}
\usepackage[pdfencoding=auto, psdextra]{hyperref}
\hypersetup{
    colorlinks,%
    citecolor=blue,%
    filecolor=blue,%
    linkcolor=blue,%
    urlcolor=blue
}

\usepackage[usenames, dvipsnames]{xcolor}
\providecommand{\red}[1]{\textcolor{red}{#1}}

\providecommand{\mean}[1]{\expval{#1}}

\begin{document}

\title{Spin drift-diffusion for two-subband quantum wells}
\author{I. R. de Assis}
\affiliation{Instituto de Física, Universidade Federal de Uberlândia, Uberlândia, MG 38400-902, Brazil}

\author{R. Raimondi}
\affiliation{Dipartimento di Matematica e Fisica, Università degli Studi Roma Tre, Via della Vasca Navale 84, 00146 Rome, Italy}

\author{G. J. Ferreira}
\affiliation{Instituto de Física, Universidade Federal de Uberlândia, Uberlândia, MG 38400-902, Brazil}
\date{\today}

\begin{abstract}
Controlling the spin dynamics and spin lifetimes is one of the main challenges in spintronics. To this end, the study of the spin diffusion in two-dimensional electron gases (2DEGs) shows that when the Rashba and Dresselhaus spin-orbit couplings (SOC) are balanced, a persistent spin helix regime arises. There, a striped spin pattern shows a long lifetime, limited only by the cubic Dresselhaus SOC, and its dynamics can be controlled by in-plane drift fields. Here, we derive a spin diffusion equation for non-degenerate two-subbands 2DEGs. We show that the intersubband scattering rate, which is defined by the overlap of the subband densities, enters as a new nob to control the spin dynamics, and can be controlled by electric fields, being maximum for symmetric quantum wells. We find that for large intersubband couplings the dynamics follow an effective diffusion matrix given by approximately half of the subband-averaged matrices. This extra $\frac{1}{2}$ factor arises from Matthiessen's rule summing over the intrasubband and intersubband scattering rates, and leads to a reduced diffusion constant and larger spin lifetimes. We illustrate our findings with numerical solutions of the diffusion equation with parameters extracted from realistic Schrödinger-Poisson calculations.
\end{abstract}

\maketitle

\section{Introduction}

The Datta-Das spin transistor \cite{DattaDas, Chuang2014SFET} has established the full electric control of the spin dynamics as one of the main goals of spintronics \cite{Kikkawa1999SpinCoherence, Wolf2001Spintronics, Zutic2004Spintronics, Fabian2007Review, Awschalom2007Spintronics, Wu2010ReviewSpin, Manchon2015PerspectRashba}. To this end, one explores the spin-orbit couplings (SOCs) \cite{Winkler2003book}, which can be seen as an effective momentum-dependent magnetic field. For two-dimensional electron gases (2DEGs) hosted in quantum wells (QWs) of zinc-blende semiconductors, the SOCs arise from the bulk and structural inversion asymmetries, yielding the Dresselhaus \cite{Dresselhaus55} and Rashba \cite{Rashba84} terms, respectively. While these allow for the spin control via external fields, they also introduce spin relaxation in diffusive systems, \textit{e.g.}, via the Dyakonov-Perel (DP) \cite{DP71a} and Elliott-Yafet (EY) \cite{Elliott1954EY, Yafet1963EY} mechanisms. Nevertheless, when the linear-in-momentum Rashba and Dresselhaus SOCs are balanced, the effective SOC field becomes uniaxial and it emerges a conserved SU(2) symmetry \cite{Schliemann2003, Bernevig2006PSH}, yielding an helical spin mode with long lifetime, named \textit{persistent spin helix} (PSH). The PSH has been observed experimentally by means of spin-grating spectroscopy \cite{koralek2009emergence, weber2007nondiffusive}, and via time- and spatially-resolved magneto-optical Kerr rotation \cite{walser2012direct, Ishihara2013DirectImage}. The first observes a spin-lifetime enhancement when the Rashba SOC is tuned to match the linear Dresselhaus SOC, while the latter maps the actual formation of the PSH. The PSH dynamics was first described theoretically by the spin drift-diffusion equation in Refs.~\cite{Mishchenko2003, Mishchenko2004Kinetic, Saikin2004, Bernevig2006PSH}, derived from the Keldysh formalism for the kinetic equation \cite{Rammer1986RMP, rammer2007quantum, HaugJauhoBook}, and an intuitive picture can be drawn from a random walk process \cite{Yang2010RandomWalk, Ferreira2017SpinDiff2}. For single subband 2DEGs, these have been extensively studied \cite{Froltsov2001,Pershin2005,Schwab2006,Stanescu2007, Tokatly2008, Tokatly2010, Sinova2012UTheorySpin, Salis2014PSH, shen2014theory}. A detailed derivation of the spin-diffusion equation can be found in Refs.~\cite{Sinova2012UTheorySpin, shen2014theory, mastersIsmael}.  Experimentally, the PSH has also been recently exploited in many different forms. Applying in-plane fields induces a controllable precession frequency linear in the cubic Dresselhaus SOC and drift velocity \cite{Altmann2016CubicSOCPSH, Kunihashi2016Drift}. The pitch of the spin helix can be controlled by tuning the QW asymmetry and electron density \cite{dettwiler2017stretchable}. The high symmetry of the PSH uniaxial spin regime allows for a perturbative approach to measure weak (anti)-localization \cite{Weigele2020WeakLocal}. The spin relaxation anisotropy can be enhanced by tilting the precession angle with magnetic fields \cite{Iizasa2021Anisotropy}. Under intense pump fields with short delay, the anisotropy might also be induced by scattering with holes \cite{Anghel2021EHoles}. Additionally, detrimental effects on the spin lifetime due to random local fluctuation on the Rashba coupling was discussed in \cite{Glazov2010PRB, Glazov2010PhysE, Bindel2016NatPhys}.
For a comprehensive review of the PSH, see Ref.~\cite{schliemann2017colloquium}.

For two-subbands QWs \cite{bernardes2006spin, EsmerindoPRL2007, calsaverini2008intersubband, Fu2015TwoSubbands}, the spin drift-diffusion has only been studied recently \cite{Fu2015Skyrmion, Hernandez2016CISP, Ferreira2017SpinDiff2, Luengo2017Gate, Hernandez2020Anisotropy}. Particularly, it has been observed strong spin relaxation anisotropy with long lifetimes \cite{Hernandez2020Anisotropy}. Theoretically,
it has been predicted that the Rashba and Dresselhaus SOCs from distinct subbands can be set to match along perpendicular axes \cite{Fu2015Skyrmion}, indicating the formation of a \textit{persistent Skyrmion lattice} (PSL). Even though, the PSH is a robust effect in a single-subband system, the role of intersubband scattering in a two-subband 2DEG has been shown to be relevant for the evaluation of the transport properties \cite{Zaremba1992Boltzmann2subbands}. It is then an open problem whether the PSH and PSL are robust in the presence of intersubband scattering \cite{Ferreira2017SpinDiff2}. This paper aims to shed light on this problem, by showing that the intersubband scattering reveals itself as a new knob to actually control the spin lifetime in quantum well devices.

In this paper we derive and analyze the spin drift-diffusion equation for two-subbands 2DEGs accounting for both intrasubband and intersubband scattering. The derivation follows from the the Keldysh formalism for the kinetic equation \cite{HaugJauhoBook, rammer2007quantum, shen2014theory, mastersIsmael}. We show that the intersubband scattering introduces a timescale $t_c$ (to be properly defined later) for the subband relaxation, and its comparison with the momentum relaxation time $\tau_0$ and spin lifetime $\tau_S$ allows for a classification of the spin dynamics between two extreme limits that match the intuitive pictures taken from a random walk process \cite{Ferreira2017SpinDiff2}. First, if $t_c$ is comparable to $\tau_S$, the intersubband dynamics is slow, defining a weak-coupling regime where electrons from each subband diffuse and precesse in time independently. On the other hand, for $t_c \ll \tau_S$ the fast intersubband dynamics leads to a subband-averaged strong-coupling regime. In both limits, and for strong enough disorder, the spin relaxation time $\tau_S$ follows the Dyakonov-Perel (DP) behavior $\tau_S \propto 1/\tau_0$. For intermediate regimes, the intersubband coupling breaks this proportionality, since we find $t_c \propto \tau_0$. Consequently, the competition between the DP mechanism and the EY-type subband relaxation $t_c$ leads to a non-monotonic transition between strong- and weak-coupling regimes as $\tau_0$ increases. Therefore, the control over the diffusion regime requires a careful choice of the disorder strength and of the degree of asymmetry of the quantum well.

Applying these results to the PSL system from Ref.~\cite{Fu2015Skyrmion}, we find that the skyrmion checkerboard pattern may exist only in the weak-coupling regime. An intuitive physical picture can be understood by recalling that, as also remarked in Ref.~\cite{Fu2015Skyrmion}, the PSL inherits its robustness from the PSHs in the separate subbands. From this point of view, it is crucial to achieve a weak-coupling or intermediate regime where the spin dynamics is {\it faster} than the intersubband dynamics. As shown in Sec.~\ref{sec:PSL}, we are able to demonstrate how the checkerboard spin pattern, initially killed by the switching on of the intersubband scattering, rises again upon increasing the momentum relaxation time.

The rest of the paper is organized as follows. In Sec.~\ref{sec:model} we introduce the two-subband model with both Rashba and Dresselhaus SOC, and as well as intrasubband and intersubband disorder scattering.  In Sec.~\ref{sec:spindiff2} we derive the  spin drift-diffusion equation for two non-degenerate subbands. In Sec.~\ref{sec:discussion} we present our results with an extended discussion of the emergence of the PSL in a two-subband quantum well device. We provide our conclusions in Sec.~\ref{sec:conclusions}. Finally, a number of technical appendices give further details about our derivation and solution of the diffusive equation.

\section{2DEG model}
\label{sec:model}

\begin{figure}[tb]
    \centering
    \includegraphics[width=\columnwidth]{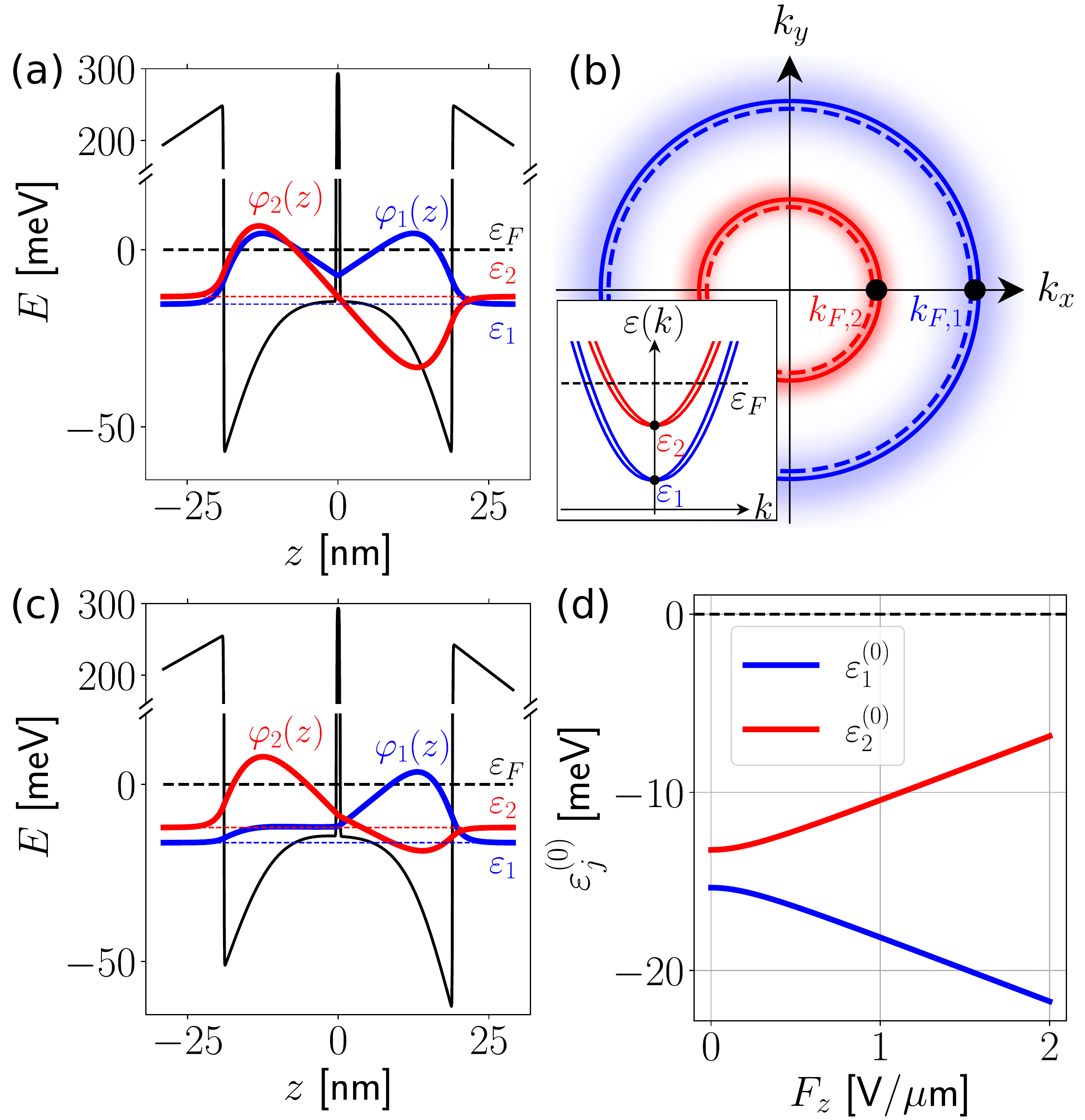}
    \caption{
    (a) Symmetric GaAs double quantum well with two occupied subbands. The dashed line marks the Fermi level at $\varepsilon_F = 0$ for a total density $n_T = 8 \times 10^{11}$~cm$^{-2}$, and the $\varphi_j(z)$ envelope functions are shown for the first ($j=1$, blue) and second ($j=2$, red) subbands.
    (b) Fermi circles of both subbands labeled by their color with respect to (a). The blurring illustrates a small impurity broadening. The inset illustrates the parabolic dispersion with a small SOC. 
    (c) For a quantum well tilted by an electric field $F_z = 0.5$~V/$\mu$m the $\varphi_1(z)$ [$\varphi_2(z)$] shifts to the right (left) edges of the well, which increases the energy splitting, as shown in (d).}
    \label{fig:system}
\end{figure}

Let us start with a clean two-dimensional electron gas (2DEG) with two-subbands given by the 2D Hamiltonian \cite{bernardes2006spin, EsmerindoPRL2007, calsaverini2008intersubband, Fu2015TwoSubbands, Fu2015Skyrmion},
\begin{align}
	H_{\rm 2D} &=
	\begin{pmatrix} 
	H_1 & H_{12} 
	\\ 
	H_{12}^\dagger & H_2
	\end{pmatrix}
	\approx
	\begin{pmatrix} 
	H_1 & 0
	\\ 
	0 & H_2
	\end{pmatrix},
	\label{eq:Hclean}
\end{align}
where $H_j$ are $2\times2$ spinfull Hamiltonians for each subband $j=\{1,2\}$, which includes the Rashba and Dresselhaus spin-orbit couplings (SOC). This 2D model is defined in the basis of the quantum well eigenstates for each subband, i.e., $\braket{\bm{r}}{j\bm{k}\sigma} = \varphi_j(z) [e^{i\bm{k}\cdot\bm{r}}/\sqrt{A}]\xi_\sigma$, where $\bm{k}=(k_x,k_y)$ is the in-plane wave-vector, $\xi_\sigma$ are the spin eigenvectors with $\sigma=\{\uparrow, \downarrow\}$, $A$ is the normalization area, and $\varphi_j(z)$ are the confined eigenstates of the quantum well, which are illustrated in Fig.~\ref{fig:system}(a). We are interested in a nondegenerate regime, i.e., the subband energies $\varepsilon^{(0)}_1 \neq \varepsilon^{(0)}_2$ and the Fermi circles do not overlap, as shown in Fig.~\ref{fig:system}(b). Therefore we can neglect off diagonal intersubband SOC in $H_{12}$~\cite{Fu2015Skyrmion}. The 2D Hamiltonian for each subband reads as
\begin{align}
    \label{eq:Hj}
	H_j &= \Big[\varepsilon_j(k) - e\bm{F}_\parallel\cdot\hat{\bm{r}}_\parallel\Big]\sigma_0 + H_j^{\rm soc}(\bm{k}),
	\\
	H_j^{\rm soc} &= \lambda_{j,+} \sigma_x k_y + \lambda_{j,-} \sigma_y k_x 
	\nonumber
	\\
	& \hspace*{1cm} + 2\beta_{3,j} \dfrac{k_x^2-k_y^2}{k^2}(k_y \sigma_x - k_x \sigma_y),
\end{align}
where $\varepsilon_j(k) = \varepsilon_j^{(0)} + \hbar^2 k^2/2m$, $\bm{F}_\parallel = (F_x, F_y)$, and $\hat{\bm{r}}_\parallel = (\hat{x},\hat{y})$ are the in-plane and external electric field coordinates in $k$-representation (i.e., $\hat{x} = i\partial_{k_x}$, $\hat{y} = i\partial_{k_y}$), $\sigma_0 = \mathds{1}_{2\times2}$ and $\sigma_{x|y|z}$ are the Pauli matrices in spin space.
$H_j^{\rm soc}$ contains the Rashba ($\alpha_j$), linear ($\beta_{1,j}$), and cubic ($\beta_{3,j} = \gamma_D k^2/4$) Dresselhaus intrasubband couplings, with $\lambda_{j,\pm} = \beta_{1,j} \pm \alpha_j$. As presented later in Sec.~\ref{sec:discussion}, the SOC and other parameters can be controlled by the $z$-component of the electric field $F_z$ \cite{bernardes2006spin, EsmerindoPRL2007, calsaverini2008intersubband, Fu2015TwoSubbands}.

For the spin diffusion, in the next section, we consider scattering by random short-range impurities. Therefore, we must add it to the clean $H_{\rm 2D}$ of Eq.~\eqref{eq:Hclean} as a perturbation. The three-dimensional (3D) form of the impurity potential is $V(\bm{r}) = \sum_i v_0 \delta(\bm{r}-\bm{R}_i)$, where $v_0$ is the intensity and $\bm{R}_i=(X_i,Y_i,Z_i)$ is the position of each of the $N$ random impurities, labeled by $i=\{1, N\}$. The 2D projection of $V(\bm{r})$ into the basis $\ket{j\bm{k}\sigma}$ reads as
\begin{align}
    \label{eq:Vij}
    \bra{j'\bm{k}'\sigma'}V(\bm{r})\ket{j\bm{k}\sigma} &= \delta_{\sigma,\sigma'} \sum_{i=1}^N \tilde{v}_{j', j}(Z_i)e^{i(\bm{k}-\bm{k}')\cdot\bm{R}_i},
	\\
	\tilde{v}_{j',j}(Z_i) &= \dfrac{v_0}{A}\varphi_{j'}^\dagger(Z_i)\varphi_{j}(Z_i).
\end{align}
Notice that the impurities allow for subband scattering, since $V_{12} = \bra{1\bm{k}_1\sigma}V(\bm{r})\ket{2\bm{k}_2\sigma} \neq 0$. In practice, we are interested in the impurity ensemble self-average~\cite{Rammer1986RMP, bruus2004many, rammer2007quantum}, which reads as
\begin{align}
    \mean{V_{ba}} &= \delta_{b,a} n_{\rm imp} v_0 \equiv 0,
\\
	\mean{V_{b,c}V_{c,a}} &=
	\delta_{k_b, k_a}
	\dfrac{n_{\rm imp} v_0^2}{A}
	\Lambda_{j_a, j_c}
	\approx \delta_{b, a}
	\dfrac{n_{\rm imp} v_0^2}{A}
	\Lambda_{j_a, j_c},
	\label{eq:VVaverage}
\\
	\Lambda_{j_a, j_c} &=
	\int 
	|\varphi_{j_a}(z)|^2
	|\varphi_{j_c}(z)|^2
	dz,
	\label{eq:overlap}
\end{align}
where we have introduced a compact notation for the indices as $a \rightarrow (j_a, \bm{k}_a, \sigma_a)$, and $n_{\rm imp}$ is the impurity density. The first-order average, $\mean{V_{ba}} \equiv 0$, is set as the energy reference. For more details on the derivation of these quantities, see Appendix~\ref{app:selfenergy}. The second order $\mean{V_{b,c}V_{c,a}}$ conserves energy and momentum, but it is not block diagonal in subband space \textit{a priori}. However, in the non-degenerate scenario the Fermi circles from different subbands do not overlap [see Fig.~\ref{fig:system}(b)], and $\delta_{k_b, k_a}$ implies subband conservation, yielding the full $\delta_{b, a}$ in the approximate expression in Eq.~\eqref{eq:VVaverage}. Nevertheless, it still allows for a virtual jump to a different subband $j_c$, which will lead to the intersubband coupling on the collision integral of the kinetic equation, as shown later in Eqs. \eqref{eq:I0coll} and \eqref{eq:I1coll}. The density overlap integral $\Lambda_{j,j'}$ defines the intensity of this coupling.

\section{Spin diffusion with two subbands}
\label{sec:spindiff2}

To derive the spin drift-diffusion equation for two subband quantum wells, we follow the kinetic equation approach~\cite{Rammer1986RMP, rammer2007quantum, Sinova2012UTheorySpin, shen2014theory} and extend the formalism to the two-subband model. Our starting point is the well-known quantum Boltzmann transport equation for a Wigner distribution function $\rho(\bm{k},\bm{R},T)$, which is a $4\times4$ matrix in spin and subband spaces, and $(\bm{R},T)$ are the Wigner coordinates~\cite{rammer2007quantum, HaugJauhoBook}. Here we take a block-diagonal approximation with $\rho = \rho_1 \oplus \rho_2$, where $\rho_j(\bm{k},\bm{R},T)$ are $2\times2$ matrices in spin space for each subband $j$, and obeys the kinetic equation
\begin{align}
    \nonumber
	\partial_T \rho_j 
	- \dfrac{1}{i\hbar}[H_j, \rho_j]
	&- \dfrac{1}{2\hbar} \{ \nabla_{\bm{R}} H_j, \nabla_{\bm{k}} \rho_j \}
	\\
	&+ \dfrac{1}{2\hbar} \{ \nabla_{\bm{k}} H_j, \nabla_{\bm{R}} \rho_j \}
	=
	I_j^{\rm coll},
	\label{eq:boltzmann}
\\
    I^{\rm coll} 
    = \dfrac{i}{4\pi\hbar} &\int dE
    \Big(\{A, \Sigma^K\} - \{\Gamma, G^K\}\Big).
    \label{eq:Icoll}
\end{align}
The block-diagonal approximation is valid in the non-degenerate regime presented in the previous section, where the Fermi circles of each subband are distinguishable, see Fig.~\ref{fig:system}(b), and it is consistent with the self-consistent Born approximation (SCBA), as shown next as we present the collision integral $I^{\rm coll}$, which will couple $\rho_1$ and $\rho_2$ through intersubband impurity scattering. 

The $I^{\rm coll}$ above is defined in terms of the spectral function $A(E,\bm{k}) = i[G^R(E,\bm{k})-G^A(E,\bm{k})]$, broadening $\Gamma(E,\bm{k})=i[\Sigma^R(E,\bm{k})-\Sigma^A(E,\bm{k})]$, and the Keldish component of the self-energy $\Sigma^K(E,\bm{k},\bm{R},T)$ and Green's function $G^K(E,\bm{k},\bm{R},T)$. The retarded and advanced Green's functions are block diagonal $G^{R(A)} = G_1^{R(A)}\oplus G_2^{R(A)}$, with $G_j^{R(A)} \approx [E-\varepsilon_j(k) \mp i\eta]^{-1}$ for each subband $j$. For simplicity, here we neglect small SOC contributions, while we show these perturbative corrections in Appendix~\ref{app:collision}. Thus, the spectral function becomes $A = A_1 \oplus A_2$, with $A_j \approx 2\pi \delta(E-\varepsilon_j(k))$. The $G^K$ plays a central role in deriving Eq.~\eqref{eq:boltzmann}, as shown in Refs.~\cite{rammer2007quantum, HaugJauhoBook}, with $\rho(\bm{k},\bm{R},T) = -\int \frac{dE}{2\pi i} G^K(E,\bm{k},\bm{R},T)$. 

The last missing pieces to characterize our system are the self-energies $\Sigma^R$, $\Sigma^A$, and $\Sigma^K$. To derive these, we follow the standard impurity self-average within the self-consistent Born approximation (SCBA)~\cite{bruus2004many} and extend it to the two-subbands scenario using the impurity-averaged potentials from Eq. \eqref{eq:VVaverage}, yielding for each subband $j$
\begin{align}
	\Sigma_j^\nu = \dfrac{n_{\rm imp} v_0^2}{A} \sum_c  \Lambda_{j,j_c} \mean{G_c^\nu},
	\label{eq:sigma}
\end{align}
which is written generically for all components $\nu={R,A,K}$. The block-diagonal form here, $\Sigma^\nu = \Sigma^\nu_1 \oplus \Sigma^\nu_2$, is a consequence of the non-degenerate scenario introduced in Eq. \eqref{eq:VVaverage}, and it justifies the block-diagonal form of $G^K$ and $\rho$, since it allows for the decoupling of the block-diagonal and non-diagonal components. This approximation will fail if the Fermi circles from different subbands overlap with the range defined by the broadening of the spectral function. In this case, the intersubband SOC~\cite{bernardes2006spin, EsmerindoPRL2007, calsaverini2008intersubband, Fu2015Skyrmion, Ferreira2017SpinDiff2} in $H_{12}$ could also play a significant role. 

Collecting the approximations above, we calculate the leading-order contribution to the collision integral $I^{(0)} = I^{(0)}_1 \oplus I^{(0)}_2$, where
\begin{align}
	I^{(0)}_j(\bm{k})
	= 
	\sum_{j_c} \dfrac{
	\mean{\rho_{j_c}(\varepsilon_j(k), \bm{R},T)}
	- 
	\rho_j(\varepsilon_j(k), \theta,\bm{R},T)
	}{\tau_{j,j_c}},
	\label{eq:I0coll}
\end{align}
$\bm{k} = (k \cos\theta, k \sin\theta)$ is now in polar coordinates, and $\mean{\rho} = (2\pi)^{-1} \int \rho d\theta$ is the $\theta$-average over the Fermi circle. 

The intra- and inter-subband momentum relaxation rates are
\begin{align}
    \label{eq:tauj}
	\dfrac{1}{\tau_{j,j_c}} 
	&= 
	 \dfrac{\phi_{j,j_c}}{\tau_0},
\end{align}
where $\tau_0^{-1} = \Lambda_{1,1} n_{\rm imp} m v_0^2/\hbar^3$ is the first subband momentum relaxation rate, and the overlap integral ratio $\phi_{j,j_c} = \Lambda_{j,j_c}/\Lambda_{1,1}$ defines the relative intensity between the intra- and intersubband scattering times ($\tau_{1,1}$, $\tau_{2,2}$, and $\tau_{1,2}$). Essentially, $\phi_{j,j_c}$ measures the extension of the wave-functions of the quantum well $\varphi_j(z)$, Fig.~\ref{fig:system}, which defines how many impurities couple the initial and final states. In the next section, the spin diffusion will be written in terms of
\begin{align}
    \dfrac{1}{\tau_j} = \sum_{j_c} \dfrac{1}{\tau_{j,j_c}} = \dfrac{1}{\tau_0} \sum_{j_c} \phi_{j,j_c},
    \label{eq:taueff}
\end{align}
where $\tau_j$ is the total relaxation rate for each subband $j$, thus obeying Matthiessen's rule.

Considering $I_j^{\rm coll} \approx I^{(0)}_j$ in Eq. \eqref{eq:boltzmann} already captures the intersubband coupling between $\rho_1$ and $\rho_2$ and leads to the Dyakonov-Perel spin relaxation. However, for consistency it is important to carry the perturbation expansion to the same order in SOC in both sides of Eq. \eqref{eq:boltzmann}. Therefore, in the Appendix~\ref{app:collision} we show that considering the SOC corrections in $G^{R(A)}$ and in the spectral function, we obtain
\begin{align}
    I_j^{(1)}(\bm{k}) &\approx \sum_{j_c} \dfrac{1}{2\tau_{j,j_c}}
    \Big\{
    H_j^{\rm soc}(\bm{k}), \dfrac{\partial}{\partial \varepsilon} \mean{\rho_{j_c}(\varepsilon,\theta,\bm{R},T)}
    \Big\}_{\varepsilon=\varepsilon_j(k)}.
    \label{eq:I1coll}
\end{align}
Now we can proceed with $I_j^{\rm coll} \approx I^{(0)}_j + I^{(1)}_j$ in Eq. \eqref{eq:boltzmann} to derive the spin diffusion equation.

\subsection{Spin diffusion equation}

The Wigner distribution in Eqs. \eqref{eq:boltzmann}, \eqref{eq:I0coll}, and \eqref{eq:I1coll} can be expanded~\cite{shen2014theory} into charge and spin components as $\rho_j(\bm{k},\bm{R},T) = g_j^i(\bm{k},\bm{R},T)\sigma_i$ and $\mean{\rho_j(k,\bm{R},T)} = \mean{g_j^i(k,\bm{R},T)}\sigma_i$ (Einstein's notation for the sums over repeated indices is implied). Thus, $g_j^i$ and $\mean{g_j^i}$ are the $\sigma_i$ component of $\rho_j$ for each subband $j$. Using these, straightforward calculation of the (anti-)commutators allow us to cast the Boltzmann equation, Eq. \eqref{eq:boltzmann}, into the matrix form
\begin{align}
    \Big[
      \mathcal{K}_j - \dfrac{e \tau_0}{\hbar}\bm{F}_\parallel \cdot \bm{\nabla}_k + \dfrac{\tau_0}{\tau_j}
    \Big]  
    \bm{g}_j = \sum_{j'} \phi_{j,j'}
      (1+\mathcal{T}_j)\mean{\bm{g}_{j'}},
    \label{eq:matrixform}
\end{align}
where the matrices $\mathcal{K}_j$ and $\mathcal{T}_j$ are presented in Appendix~\ref{app:matrices}, and the vector $\bm{g}_j = (g_j^0, g_j^x, g_j^y, g_j^z)$. Since these are components of the distribution function, integrating over $\bm{k}$ gives $\sum_{\bm{k}} \bm{g}_j = \{N_j, S^x_j,  S^y_j, S^z_j\}$, \textit{i.e.,} the charge and spin densities of each subband $j$.

To obtain a closed set of equations for $\mean{\bm{g}_j}$, we first rearrange Eq. \eqref{eq:matrixform} as~\cite{Sinova2012UTheorySpin, shen2014theory}
\begin{multline}
    \bm{g}_j = \sum_{j'} \phi_{j,j'}
    \Big[
      \mathcal{K}_j + \dfrac{\tau_0}{\tau_j}
    \Big]^{-1}
    \Big[
      (1+\mathcal{T}_j)\mean{\bm{g}_{j'}} 
    \Big]
    \\
    + \dfrac{e\tau_0}{\hbar}
    \Big[
      \mathcal{K}_j + \dfrac{\tau_0}{\tau_j}
    \Big]^{-1}
    \bm{F}_\parallel \cdot \bm{\nabla}_k \bm{g}_j.
\end{multline}
The inverses can be expressed as the geometric series expansion for $||\mathcal{K}_j|| \ll 1$ and $\tau_0/\tau_j$ of order one. Recursively replacing $\bm{g}_j$ on the right-hand side up to first order in the electric field $\bm{F}_\parallel$ and integrating over $\bm{k} = (k, \theta)$, we obtain
\begin{multline}
   \begin{pmatrix}
        \mathbb{1} & 
        \frac{\phi_{1,2}}{\phi_{1,1}}
        \\ 
        \frac{\phi_{2,1}}{\phi_{2,2}} 
        & \mathbb{1} 
    \end{pmatrix} \dfrac{\partial \vec{\mathcal{S}}}{\partial t}
    = 
    \begin{pmatrix}
    -\mathcal{D}_{1} - \gamma_1  &  \gamma_1 - \frac{\phi_{1,2}}{\phi_{1,1}}\mathcal{D}_{1}
    \\
     \gamma_2 - \frac{\phi_{2,1}}{\phi_{2,2}}\mathcal{D}_{2} & -\mathcal{D}_{2} - \gamma_2
    \end{pmatrix}
    \vec{\mathcal{S}},
    \label{eq:spindiff}
\end{multline}
where $\mathbb{1}$ is the $4\times4$ identity matrix, the 8-vector $\vec{\mathcal{S}} = [N_1, \vec{S}_1, N_2, \vec{S}_2]^T$ is written in a compact notation, with $N_j \equiv N_j(\bm{q},t)$, $\vec{S}_j \equiv \vec{S}_j(\bm{q},t)$, and $\bm{q} = (q_x, q_y)$ are the coordinates of the reciprocal space on the Fourier transform $\mathcal{F}_{\bm{r} \rightarrow \bm{q}}[N_j(\bm{r},t)] = N_j(\bm{q},t)$, and equivalently for $\vec{S}_j(\bm{q},t)$. For simplicity, starting in Eq. \eqref{eq:spindiff} and hereafter, we refer to the Wigner coordinates $(\bm{R},T)$ simply as $(\bm{r}, t)$.

To express the diffusion matrices for each subband $j$, we split it as
\begin{equation}
    \mathcal{D}_j = 
    [D_j \bm{q}^2 - i\bm{q}\cdot\bm{v}_j]
    +
    \begin{pmatrix}
        0 & \Theta_0^T
        \\
        \Theta_0+\Theta_1 & \mathcal{D}_j^{\rm S}
    \end{pmatrix},
    \label{eq:Djmat}
\end{equation}
where the spin block $\mathcal{D}_j^{\rm S}$ contains the spin precession and relaxation terms,
\begin{multline}
    \mathcal{D}_j^{\rm S} 
    = 
    \begin{pmatrix}
      \dfrac{1}{\tau_{j,-}} &
      0 & 
      +2iD_jq_x Q_{j,-}^*
    \\
      0 &
      + \dfrac{1}{\tau_{j,+}} &
      -2iD_jq_y Q_{j,+}^*
    \\
      -2iD_jq_x Q_{j,-}^* &
      +2iD_jq_y Q_{j,+}^* &
      + \dfrac{1}{\tau_{j,-}} + \dfrac{1}{\tau_{j,+}}
    \end{pmatrix}
\\
    +
    \begin{pmatrix}
      0 &
      0 & 
      +v_{j,x}Q_{j,-}^{**}
    \\
      0 &
      0 &
      -v_{j,y}Q_{j,+}^{**}
    \\
      -v_{j,x}Q_{j,-}^{**} &
      +v_{j,y}Q_{j,+}^{**} &
    0
    \end{pmatrix},
\end{multline}
while the spin-charge couplings are given by the rectangular blocks $\Theta_0$ and $\Theta_1$, which are presented in Appendix~\ref{app:spincharge} together with the spin Hall angles. Above, $Q_{j,\pm}^* = Q_{j,\pm} - Q_{j,3}$, $Q_{j,\pm}^{**} = Q_{j,\pm} - 2Q_{j,3}$ and the other elements are
\allowdisplaybreaks
\begin{align}
    D_{j} &= \dfrac{\tau_j}{2}\left(\dfrac{\hbar k_{Fj}}{m}\right)^2,
    \label{eq:diff}
\\
    \label{eq:vdrift}
    \bm{v}_j &= \dfrac{e \tau_j}{m} \bm{F}_\parallel,
\\
    Q_{j,\pm} &= \dfrac{2m}{\hbar^2}\lambda_{j,\pm} = \dfrac{2m}{\hbar^2}(\beta_{1,j}\pm\alpha_j),
\\
    Q_{j,3} &= \dfrac{2m}{\hbar^2}\beta_{3,j},
\\
    \label{eq:taujpm}
    \dfrac{1}{\tau_{j,\mp}} &= D_j \Big(Q_{j,\mp}^2  - 2 Q_{j,\mp} Q_{j,3} + 2 Q_{j,3}^2\Big),
\\
    \label{eq:gammaj}
    \gamma_j &= \dfrac{1}{\tau_j}\dfrac{\phi_{1,2}}{\phi_{j,j}},
\end{align}
which refer, for each subband $j$, to the diffusion constant $D_j$, the drift velocity $\bm{v}_j = (v_{x,j}, v_{y,j})$, the linear $Q_{j,\pm}$ and cubic $Q_{j,3}$ spin-orbit momenta, the Dyakonov-Perel relaxation rates $1/\tau_{j,\mp}$, and $\gamma_j$ is the intersubband relaxation rate.

\subsection{Strong and weak intersubband coupling regimes}

The intersubband coupling is dictated by the $\phi_{1,2}$ overlap ratio. A particularly interesting scenario arises when $\phi_{j,j'} \approx 1$ for all $(j,j')$, which allow us to combine the equations for $[N_1, \vec{S}_1]$ and $[N_2, \vec{S}_2]$ from Eq. \eqref{eq:spindiff} as
\begin{equation}
    \dfrac{\partial}{\partial t}
    \begin{pmatrix}
        N_T \\ \vec{S}_T
    \end{pmatrix}
    =
    - \mathcal{D}_{\rm avg}
    \begin{pmatrix}
        N_T \\ \vec{S}_T
    \end{pmatrix},
    \label{eq:spindiffNST}
\end{equation}
where $N_T = N_1 + N_2$ and $\vec{S}_T = \vec{S}_1 + \vec{S}_2$ are the total charge and spin densities. In this regime the charge-spin dynamics is led by the subband averaged diffusion matrix $\mathcal{D}_{\rm avg} = (\mathcal{D}_1+\mathcal{D}_2)/2$. A qualitative picture of this scenario is clearly seen as a random walk of the particle quickly jumping between subbands~\cite{Ferreira2017SpinDiff2}, thus feeling only the subband averaged drift and diffusion forces.

In the opposite limit of intersubband coupling $\phi_{1,2} \approx 0$, the dynamics of the first and second subband charges and spin densities, $[N_1, \vec{S}_1]$ and $[N_2, \vec{S}_2]$, decouples as Eq. \eqref{eq:spindiff} becomes block diagonal. In this case each subband evolves in time independently~\cite{Ferreira2017SpinDiff2} as
\begin{equation}
    \dfrac{\partial}{\partial t}
    \begin{pmatrix}
        N_j \\ \vec{S}_j
    \end{pmatrix}
    =
    - \mathcal{D}_j
    \begin{pmatrix}
        N_j \\ \vec{S}_j
    \end{pmatrix}.
    \label{eq:spindiff0}
\end{equation}

\begin{figure}[tb]
    \centering
    \includegraphics[width=\columnwidth]{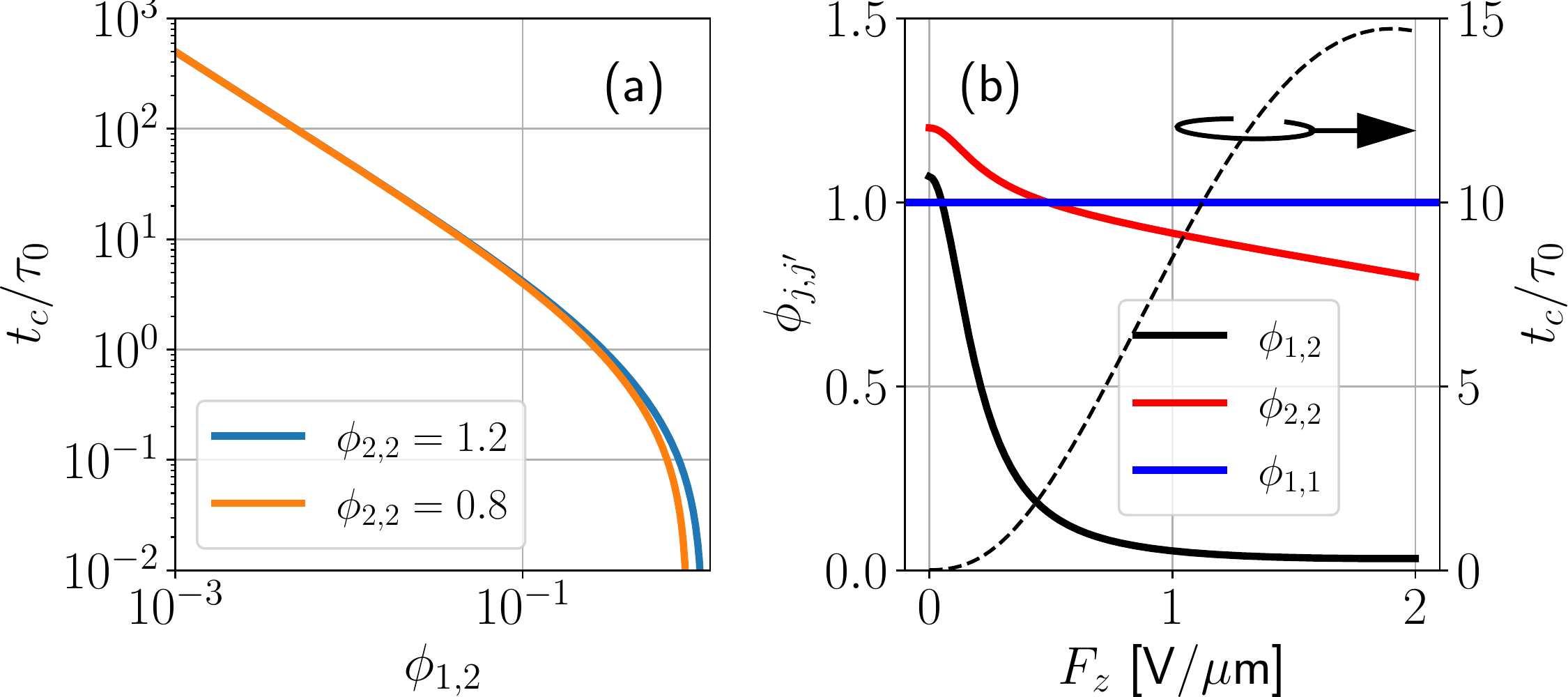}
    \caption{(a) Intersubband relaxation time $t_c/\tau_0$ from Eq.~\eqref{eq:tc} as a function of $\phi_{1,2}$. For small $\phi_{1,2}$, $t_c/\tau_0 \approx 1/(2\phi_{1,2})$ shows as a straight line in the log-log scale, and quickly drops to zero as $\phi_{1,2}^2$ approaches $\phi_{2,2}$.
    (b) Density overlap ratios $\phi_{j,j'}$ as a function of $F_z$ calculated from the Schrödinger-Poisson model with parameters from Table \ref{tab:parameters}. The dashed line refers to the right axis and shows $t_c/\tau_0$ calculated from the $\phi_{j,j'}$ shown in this panel.}
    \label{fig:tc}
\end{figure}

While the $\phi_{1,2}$ coupling defines the intensity of the intersubband coupling, a more insightful quantity is the intersubband relaxation time
\begin{align}
    t_c &\approx
    \dfrac{
        \phi_{2,2} - \phi_{1,2}^2
    }{
        2\phi_{1,2}(1+\phi_{1,2})(\phi_{1,2}+\phi_{2,2})
    } \tau_0.
    \label{eq:tc}
\end{align}
This expression is derived in Appendix \ref{app:weakstrong} and it characterizes the time scale for the initial condition to relax towards the subband averaged dynamics given by the strong-coupling regime. Therefore, the weak-coupling regime is only dominant for $t \ll t_c$ and the strong-coupling dominates for $t \gg t_c$. As shown in Fig. \ref{fig:tc}(a), $t_c/\tau_0$ vanishes as $\phi_{1,2}^2 \rightarrow \phi_{2,2}$. More importantly, for small $\phi_{1,2}$ the $t_c/\tau_0 \approx 1/(2\phi_{1,2})$, which shows that $t_c/\tau_0 \gg 1$  only if $\phi_{1,2} \ll 1$. Consequently, for typical $\tau_0 = 1$~ps, one would get $t_c \approx 100$~ps only for $\phi_{1,2} \approx 10^{-3}$. Typical values of $\phi_{1,2}$ are much larger than that, as shown in Fig.~\ref{fig:tc}, which favors the strong-coupling regime. Therefore, we expect the weak-coupling regime to only occur for extreme cases of nearly isolated quantum wells.

Equations \eqref{eq:spindiffNST} and \eqref{eq:spindiff0} above are nearly identical. However, two important differences between the strong- and weak-coupling regimes are hidden in the details. First, let us assume that the diffusion constants $D_1 \approx D_2$ for simplicity. Then, for $\alpha_2 \approx -\alpha_1$ the $(Q_{1,\pm} + Q_{2,\pm})/2$ subband average will lead to an effective $\alpha \approx 0$ dynamics for $\phi_{12} = 1$ in Eq. \eqref{eq:spindiffNST}, which typically shows circular patterns on the spin maps $S_z(x,y)$~\cite{Stanescu2007, Ferreira2017SpinDiff2}, while for $\phi_{12} = 0$ in Eq. \eqref{eq:spindiff0} the subbands evolve independently, each with its $\alpha_j$, which could lead to the persistent skyrmion lattice \cite{Fu2015Skyrmion} if the limit $t_c/\tau_0 \gg 1$ can be achieved. Second, and more importantly, the diffusion constant $D_j$ is proportional to $\tau_j$, thus $D_j \propto 1/(\phi_{j,1} + \phi_{j,2})$ [see Eq.~\eqref{eq:diff}]. Therefore, for $\phi_{1,1} \approx \phi_{2,2} \approx 1$,  the $D_j$ drops by a factor of $\frac{1}{2}$ as $\phi_{1,2}$ is increased from 0 to 1. Consequently, the smaller diffusion constant enhances the spin lifetime by a factor $\sim 2$ in the strong-coupling regime, assuming other parameters remain the same.

\section{Discussion}
\label{sec:discussion}

Hereafter, we are interested in the spin-diffusion dynamics and the spin lifetime. Therefore, we may neglect the small spin-charge coupling in Eq.~\eqref{eq:Djmat} for simplicity, which reduces the diffusion matrices from $4\times4$ to $3\times3$ in Eqs.~\eqref{eq:spindiffNST} and \eqref{eq:spindiff0} for the strong- and weak-coupling regimes. The advantage here is that in the $3\times3$ form it is possible to find analytical solutions for the spin lifetime in particular scenarios, as shown in Appendix \ref{app:spinrelax}. On the other hand, for the complete diffusion equation from Eq.~\eqref{eq:spindiff} we could not find analytical solutions neither in the full $8\times8$ nor $6\times6$ form obtained by neglecting the spin-charge couplings. Therefore, here we show numerical results for the full $8\times8$ [Eq.~\eqref{eq:spindiff}], and compare it with the spin lifetimes obtained from the $3\times3$ approximate forms of the strong- and weak-coupling regimes.

The numerical parameters for the spin diffusion equation are extracted from a representative two-subband quantum well presented in the next section. Then, we consider that our initial packet is a narrow spin excitation, i.e., $S_{z,j}(\bm{r},t=0) \approx \delta(\bm{r})/2$ (for both $j=1$ and 2), which models the pump-and-probe time-resolved Kerr rotation experiments \cite{walser2012direct}. In the reciprocal $\bm{q}$ space this corresponds to a uniform initial $S_{z,j}(\bm{q},t=0)$. The numerical solutions are obtained by propagating Eq. \eqref{eq:spindiff} in time independently for each $\bm{q}$, and applying the inverse Fourier transform to return to $\bm{r}$ space. 

Additionally, in Sec.~\ref{sec:PSL} we consider the quantum well from Ref.~\cite{Fu2015Skyrmion}, where it is proposed the \textit{persistent skyrmion lattice} (PSL) regime where the subbands are set with orthogonal PSH regimes. This system clearly shows an insightful transition from strong- to weak-intersubband-coupling regimes as $\tau_0$ is increased.

\subsection{System parameters}

\begin{table}[b]
\centering
\caption{Typical set of parameters obtained for a wide 38-nm GaAs quantum well with a 0.5-nm central barrier, and total density $n_{\rm 2D} = 8  \times 10^{11} \text{ cm}^{-2}$. The subband energies $\varepsilon_j^{(0)}$, densities $n_j$, and SOCs $\alpha_j$, $\beta_{1,j}$, and $\beta_{3,j}$, and overlaps $\phi_{j,j'}$ are listed for each subband $j=\{1,2\}$ at transverse field $F_z = 0$. The energy reference is set at the Fermi level $\varepsilon_F = 0$. The subband split is $\Delta \varepsilon = \varepsilon_2^{(0)} - \varepsilon_1^{(0)} \approx 2$~meV, and the non-degeneracy condition requires $\hbar/\tau_0 \ll \Delta \varepsilon$, thus $\tau_0 \gg \hbar/\Delta \varepsilon \approx 0.3$~ps. Here we use $\tau_0 = 1$~ps. To calculate $\beta_{3,j} = \gamma_D \pi n_j/2$ we use the bulk Dresselhaus parameter $\gamma_D = 11$~meVnm$^3$ \cite{Altmann2016CubicSOCPSH}.}
\begin{tabular}{c c l} 
 \hline \hline
 Parameter & Value & Description
 \\ \hline
 $n_{1}$ & $4.3$ & \multirow{2}{*}{Subband densities [$10^{11} \text{ cm}^{-2}$]}
 \\
 $n_{2}$ & $3.7$ &
 \\ \hline
 $\alpha_1$ & 0 & \multirow{2}{*}{Rashba couplings [meV~nm]}
 \\
 $\alpha_2$ & 0 &
 \\ \hline
 $\beta_{1,1}$ & $ 0.16$ & \multirow{2}{*}{Linear Dresselhaus couplings [meV~nm]}
 \\
 $\beta_{1,2}$ & $ 0.26$ &
 \\ \hline
 $\beta_{3,1}$ & $0.074$ & \multirow{2}{*}{Cubic Dresselhaus couplings [meV~nm]}
 \\
 $\beta_{3,2}$ & $0.064$ & 
 \\ \hline
 $\phi_{1,1}$ & $1.00$ & \multirow{3}{*}{Overlap integrals [dimensionless]}
 \\
 $\phi_{2,2}$ & $1.07$ &
 \\
 $\phi_{1,2}$ & $1.20$ &
 \\ \hline
 $\eta_{1,2}$ & $0.25$ & \multirowcell{2}[0pt][l]{Intersubband Rashba \\ and Dresselhaus [meV~nm]}
 \\
 $\Gamma_{1,2}$ & 0 &  
 \\ \hline \hline
\end{tabular}
\label{tab:parameters}
\end{table}

Let us consider a two-subband 2DEG defined by a GaAs quantum well confined along $z \parallel [001]$, with $x \parallel [110]$ and $y \parallel [1\bar{1}0]$. To achieve a two-subbands regime we use a wide quantum well with a central barrier as defined in Table~\ref{tab:parameters} and shown in Fig.~\ref{fig:system}. Both the central barrier and the lateral regions are composed by Al$_{0.3}$Ga$_{0.7}$As, and symmetric doping is considered to be far away from the quantum well. Thus, the Schrödinger-Poisson equations \cite{bernardes2006spin, EsmerindoPRL2007, calsaverini2008intersubband, Fu2015TwoSubbands} are solved self-consistently for effective mass $m^* = 0.067m_0$ and dielectric constant $\kappa = 12.9$, where $m_0$ is the bare electron mass. To break the structural inversion symmetry and induce a Rasbha SOC we consider a transverse electric field $F_z \parallel z$ and vary its intensity. The SOC coefficients are calculated following Refs. \cite{bernardes2006spin, EsmerindoPRL2007, calsaverini2008intersubband, Fu2015TwoSubbands}, yielding the 2D model in Eq. \eqref{eq:Hj}. 

\begin{figure}[tb]
    \centering
    \includegraphics[width=\columnwidth]{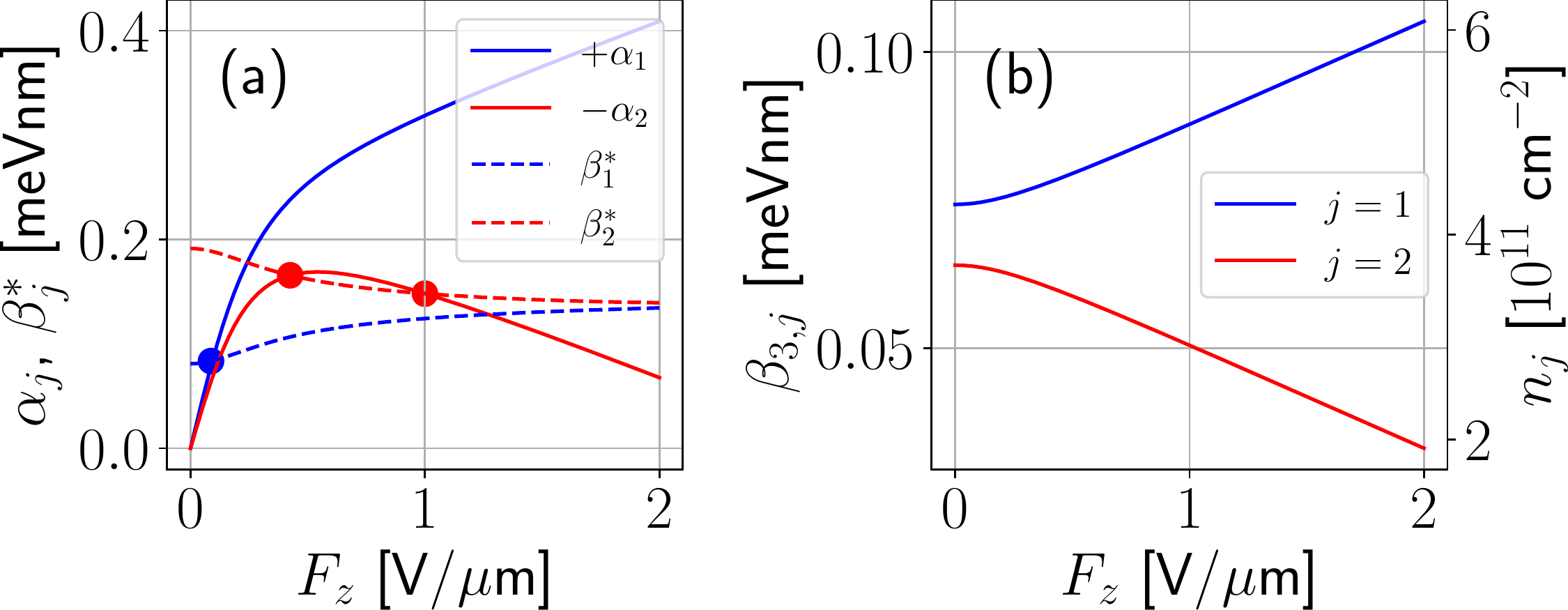}
    \caption{Spin-orbit couplings for each subband $j=\{1,2\}$ as a function of the electric field $F_z$ for the quantum well parameters in Table \ref{tab:parameters}. (a) Comparing $\pm \alpha_j$ with $\beta_j^* = \beta_{1,j} - \beta_{3,j}$ shows a ${\rm PSH}_1^+$ (blue dot) and two ${\rm PSH}_2^-$ (red dots) crossings. (b) The cubic Dresselhaus $\beta_{3,j} = \gamma_D \pi n_j/2$ and subband densities $n_j$ (right axis) are approximately constant for small $F_z$ and vary linearly for large $F_z$.}
    \label{fig:SOC}
\end{figure}

The numerically calculated overlap integrals $\phi_{j,j'}$ are shown in Fig.~\ref{fig:tc}(b). In the symmetric regime, $F_z = 0$, the envelope functions $\varphi_j(z)$ are spread along both quantum wells as symmetric and anti-symmetric solutions, leading to a maximal $\phi_{1,2}$. As $F_z$ increases, the $\varphi_j(z)$ split into left and right quantum wells, as shown in Fig.~\ref{fig:system}(a), reducing $\phi_{1,2}$ in Fig.~\ref{fig:tc}(b). However, $t_c$ only reaches reasonably large values for large $F_z$, where $\phi_{1,2} \approx 0.03$ yields $t_c \approx 15 \tau_0$. The spin-orbit couplings and subband densities are shown in Fig.~\ref{fig:SOC} as a function of $F_z$. For uncoupled subbands, one would expect PSH regimes for $\alpha_j = \pm \beta_j^*$, which we label as ${\rm PSH}_j^\pm$. In Table \ref{tab:parameters} we show the intersubband SOC parameters $\eta_{1,2}$ and $\Gamma_{1,2}$ \cite{bernardes2006spin, EsmerindoPRL2007, calsaverini2008intersubband, Fu2015TwoSubbands} as a reference, but these are neglected in our model since we consider non-overlapping Fermi circles [see Fig.~\ref{fig:system}(b)].

\subsection{Spin lifetime}

The numerical solution of Eq.~\eqref{eq:spindiff} give us $S_z(\bm{q}, t)$, and a Fourier transform yields $S_z(\bm{r}, t) = \mathcal{F}[S_z(\bm{q}, t)]$. To analyze these numerical data, it is interesting to look at well-known analytical solutions at particular cases for single-subband systems and gain insight. Here we consider drift velocities $\bm{v}_j = 0$ for simplicity. There are two scenarios that allow for simple analytical solutions: (i) $Q_{j,\pm}^* = 0$; and (ii) $Q_{j,+}^* = Q_{j,-}^*$, and $\tau_{j,+} = \tau_{j,-}$. The first occurs near PSH regimes \cite{Bernevig2006PSH}, and the second is the isotropic regime \cite{Stanescu2007}. As shown in Appendix \ref{app:spinrelax}, at $\bm{r}=0$, the asymptotic $t \gg \tau_S$ solutions for these scenarios are
\begin{align}
    S_z(\bm{r}=0,t \gg \tau_S) &\propto \dfrac{e^{-t/\tau_S}}{t^n},
    \label{eq:SzAsymptotic}
\end{align}
where (i) $n=1$ and (ii) $n=1/2$ for the PSH and isotropic scenarios, respectively. Approximate expressions for the spin lifetime $\tau_S$ are shown in Appendix \ref{app:spinrelax}.

\begin{figure}[b]
    \centering
    \includegraphics[width=\columnwidth]{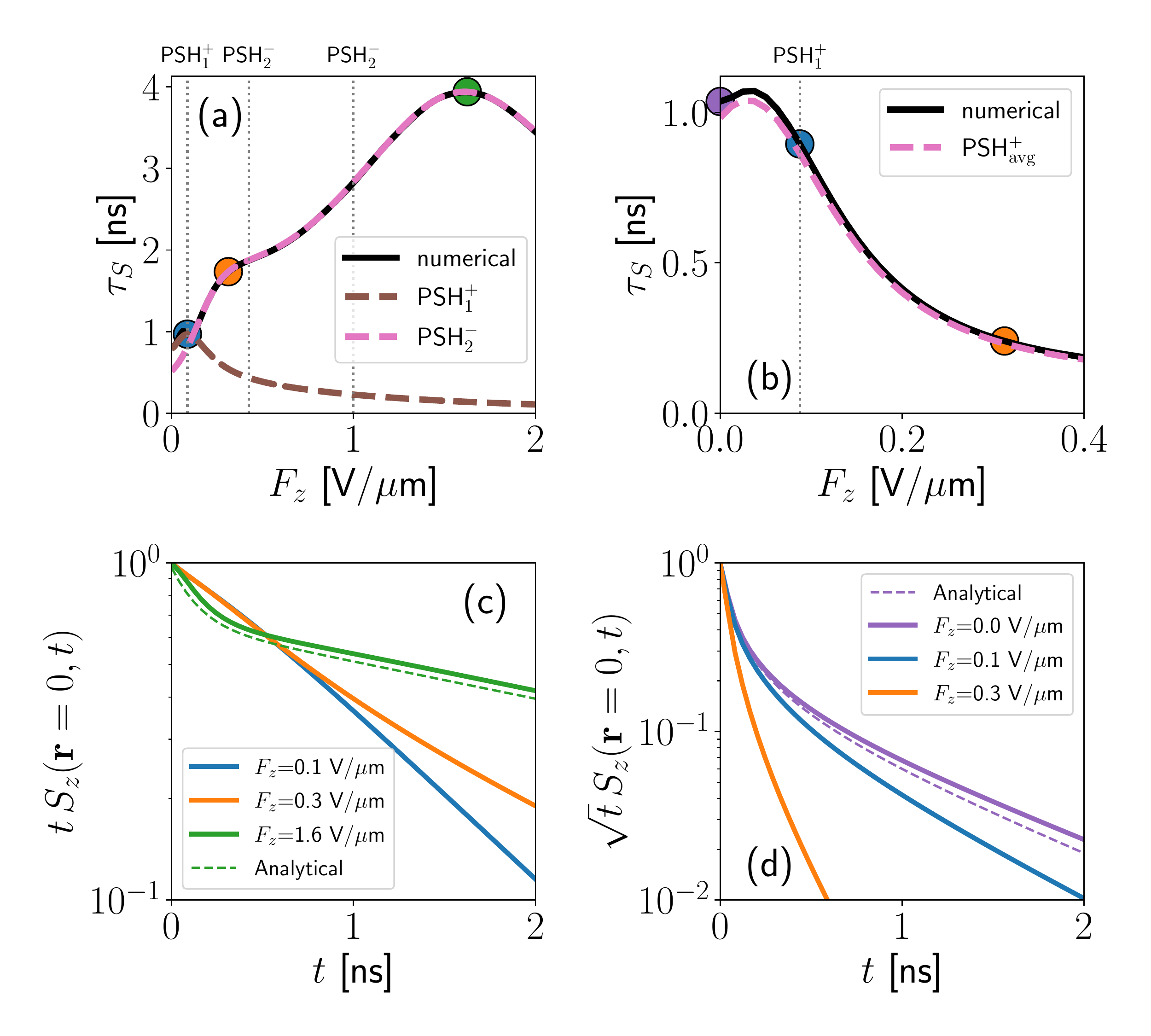}
    \caption{
    (a), (b) Spin lifetime $\tau_S$ as a function of the electric field $F_z$. In (a) we force $\phi_{12} = 0$, while in (b) we use $\phi_{12} \neq 0$ from Fig.~\ref{fig:tc}(b). Gray-dotted vertical lines mark the PSH$_j^\pm$ regimes from Fig.~\ref{fig:SOC}(a), and the colored dashed lines are taken from Eq.~\eqref{eqa:tauSi}.
    Representative $t^n S_z(\bm{r}=0,t)$ are shown for (c) $\phi_{1,2}=0$ and (d) $\phi_{1,2} \neq 0$, with a color code matching the panels above. Dashed lines are analytical solutions from (c) Eq.~\eqref{eqa:Sz12} and (d) Eq.~\eqref{eqa:SzSG}.
    The numerical $\tau_S$ in (a) and (b) are taken in the asymptotic $t \gg \tau_S$ limit, where we expect for (c) $S_z \propto e^{-t/\tau_S}/t$ and for (d) $S_z \propto e^{-t/\tau_S}/\sqrt{t}$.}
    \label{fig:lifetime}
\end{figure}

Assuming that the numerical solutions of Eq.~\eqref{eq:spindiff} approximately follow the functional form of Eq.~\eqref{eq:SzAsymptotic}, the spin lifetime $\tau_S$ can be numerically obtained from
\begin{align}
    \dfrac{\partial}{\partial t}\ln[S_z(0,t)] = -\dfrac{1}{\tau_S} - \dfrac{n}{t}.
    \label{eq:tauSnum}
\end{align}
Since we assume $t \gg \tau_S$, the second term can be neglected and the numerical derivative yields $\tau_S$ independently of $n$ in the asymptotic limit. Indeed, Fig.~\ref{fig:lifetime} shows that this is a good approximation.

To understand the role of the intersubband coupling $\phi_{1,2}$ on the spin lifetime $\tau_S$, in Fig.~\ref{fig:lifetime} we plot $\tau_S$ as a function of $F_z$. In Fig.~\ref{fig:lifetime}(a) we enforce the weak-coupling regime by setting $\phi_{1,2} \equiv 0$, while keeping other parameters with their original values from the Schrödinger-Poisson simulation. In Fig.~\ref{fig:lifetime}(b) we restore $\phi_{1,2}$ to its original values from Fig.~\ref{fig:tc}(b), which leads to the strong-coupling regime. In both Figs.~\ref{fig:lifetime}(a) and \ref{fig:lifetime}(b) the solid black line is the numerical $\tau_S$ obtained from Eq.~\eqref{eq:tauSnum}, while the dashed lines are the analytical expressions derived in Appendix \ref{app:spinrelax}, which generically reads
\begin{align}
    \dfrac{1}{\tau_S} &\approx
    \dfrac{1}{2 \tau_{\mp}}
    + \dfrac{1}{\tau_{\pm}}
    - \dfrac{1}{(4Q^*_{\pm} \tau_{\mp})^2 D}
    - (Q_{\pm}^*)^2 D.
    \label{eq:tauSi}
\end{align}
For the weak-coupling regime of Fig.~\ref{fig:lifetime}(a) we use $\tau_{\pm} \rightarrow \tau_{j,\pm}$, $Q^*_\pm \rightarrow Q^*_{j,\pm}$, and $D \rightarrow D_j$ for $j=1$ and $2$, while in the strong-coupling regime of Fig.~\ref{fig:lifetime}(b) these parameters are taken as the subband averages, which follows from the approximate Eq.~\eqref{eq:spindiffNST}. In both cases, the agreement between the numerical $\tau_S$ and the analytical approximations is patent. 

Complementarily, Figs.~\ref{fig:lifetime}(c) and \ref{fig:lifetime}(d) show $t S_z(0,t)$ and $\sqrt{t} S_z(0,t)$, respectively, for illustrative $F_z$ points. In Fig.~\ref{fig:lifetime}(c) all lines match the sum of two independent exponentials [see Eq.~\eqref{eqa:Sz12}] for the first and second subbands, but for simplicity we highlight the analytical solution only for $F_z = 1.6$~V/$\mu$m, which is the one that shows the largest deviation from a simple exponential. For the strong-coupling we only have approximate analytical solutions [Eq.~\eqref{eqa:SzSG}] for the isotropic regime of $F_z = 0$, which follows from Ref.~\cite{Stanescu2007}. In this case, the small deviation seen in Fig.~\ref{fig:lifetime}(d) is due to the small imprecision of $\tau_S$ in Fig.~\ref{fig:lifetime}(b).

A striking feature seen in Fig.~\ref{fig:lifetime}(a) is that the $\tau_S$ peaks are shifted from the PSH conditions ($\alpha_j = \pm \beta_j^*$) from Fig.~\ref{fig:SOC}(a). The shift occurs due to the strong dependence of $\beta_{3,j}$ upon $F_z$ at large fields, as shown in Fig.~\ref{fig:SOC}(b). Since the subbands are decoupled for $\phi_{1,2}=0$, the spin lifetime of each subband simplifies to $\tau_S\red{^{-1}} \propto 3\beta_{j,3}^2 + (\beta_j^* \pm \alpha_j)^2$. If $\beta_{j,3}$ were constant, the peak would occur at the expected PSH condition $\alpha_j = \pm \beta_j^*$. Indeed for small $F_z$ this is approximately true and the peaks in Fig.~\ref{fig:lifetime}(a) show only a small deviation. However, for large fields the peak occurs when $d\tau_S/dF_z = 0$, and the $F_z$ dependence of $\beta_{j,3}$ leads to the shift. This feature is not limited to the two-subbands system, but can also happen in single-subband quantum wells if the 2DEG is grounded and the total density changes with $F_z$.

For realistic $\phi_{1,2} \neq 0$, the $\tau_S$ always shows a peak near the isotropic case of $F_z = 0$, as in Fig.~\ref{fig:lifetime}(b). This is due to the Matthiessen's rule, Eq.~\eqref{eq:taueff}, and the strong dependence of $\phi_{1,2}$ with $F_z$ shown in Fig.~\ref{fig:tc}. Since $1/\tau_{j,\pm} \propto D_j$ [see Eq.~\eqref{eq:taujpm}], we find that $\tau_S \propto (1+\phi_{1,2})$, which arises from Eq.~\eqref{eq:taueff} with $\phi_{1,1} = \phi_{2,2} \approx 1$. This is a good qualitative approximation, as seen from Fig.~\ref{fig:tc}(b). Thus, as $\phi_{1,2}$ drops from 1 to 0, $\tau_S$ falls by a factor of 2 as $F_z$ increases. In practice, the peak seen in Fig.~\ref{fig:lifetime}(b) changes by a factor even larger than 2, which is due to changes in the prefactors beyond this qualitative picture. Nevertheless, this shows $\phi_{1,2}$ plays the role of an efficient knob to control the spin lifetime around the isotropic $F_z = 0$ configuration and does not require the PSH fine tuning of the SOC.

\subsection{Spin patterns}

\begin{figure}[tb]
    \centering
    \includegraphics[width=\columnwidth]{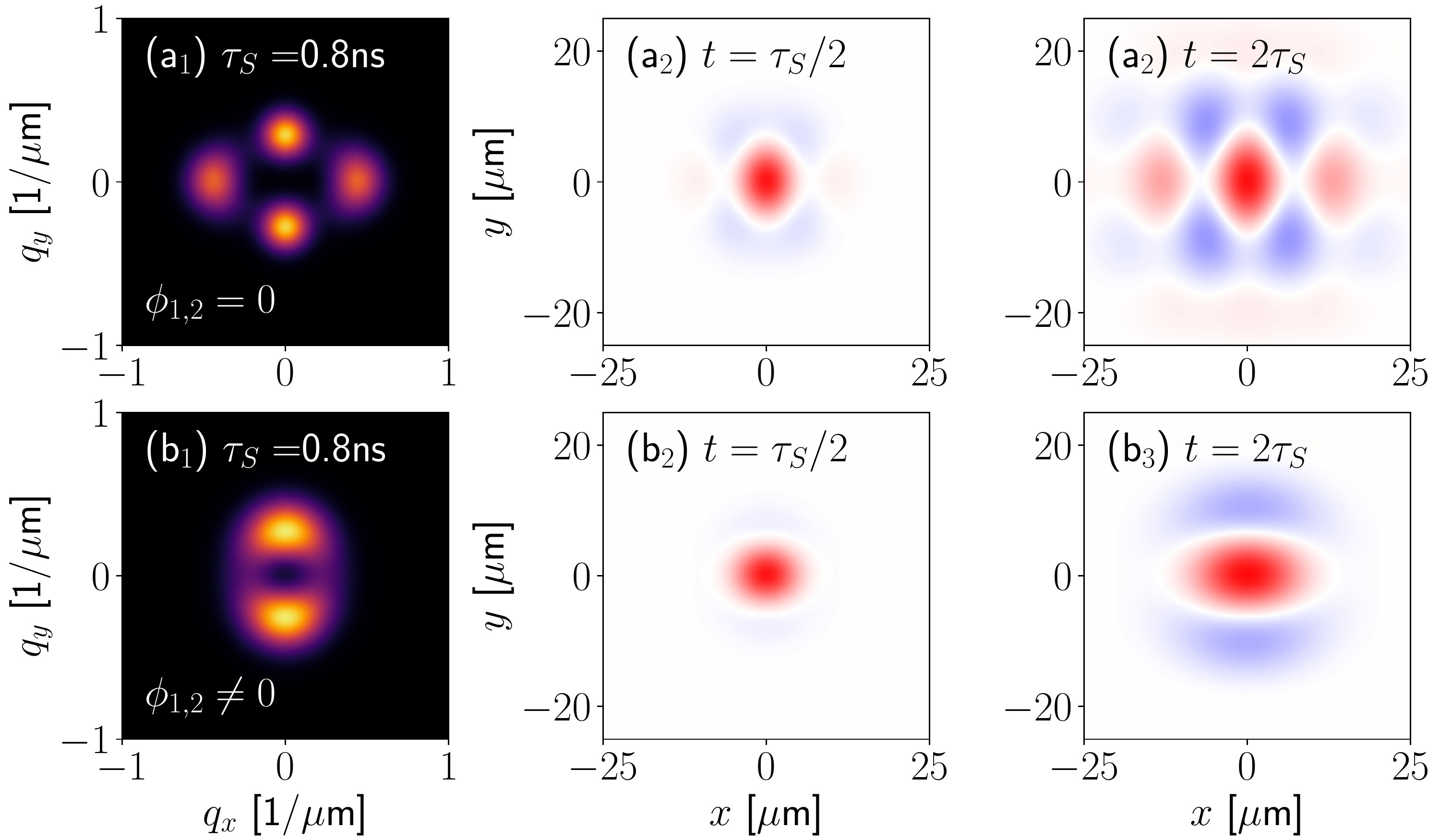}
    \caption{Comparing the spin maps evolution for (a$_n$) $\phi_{1,2} = 0$ and (b$_n$) $\phi_{1,2} \neq 0$ in $q$ ($n=1$) and $r$ spaces ($n>1$), with $F_z \approx 0.09$~V/$\mu$m (blue dots in Fig.~\ref{fig:lifetime}). (a$_n$) Enforcing the weak-coupling regime by setting $\phi_{1,2} = 0$, the $S_z(\bm{q},t)$ show poles at $\pm q_y^0$ ($\pm q_x^0$) for the first (second) subbands, which leads to checkerboard patterns in (a$_2$) and (a$_3$) $S_z(\bm{r},t)$. (b$_n$) For a finite $\phi_{1,2}$ the strong-coupling regime dominates and the subband-averaged dynamics shows broadened poles in $S_z(\bm{q},t)$ at $\pm q_y^0$, which gives the nearly striped patterns in (b$_2$) and (b$_3$) $S_z(\bm{r},t)$. 
    }
    \label{fig:SzMaps7}
\end{figure}

\begin{figure}[tb]
    \centering
    \includegraphics[width=\columnwidth]{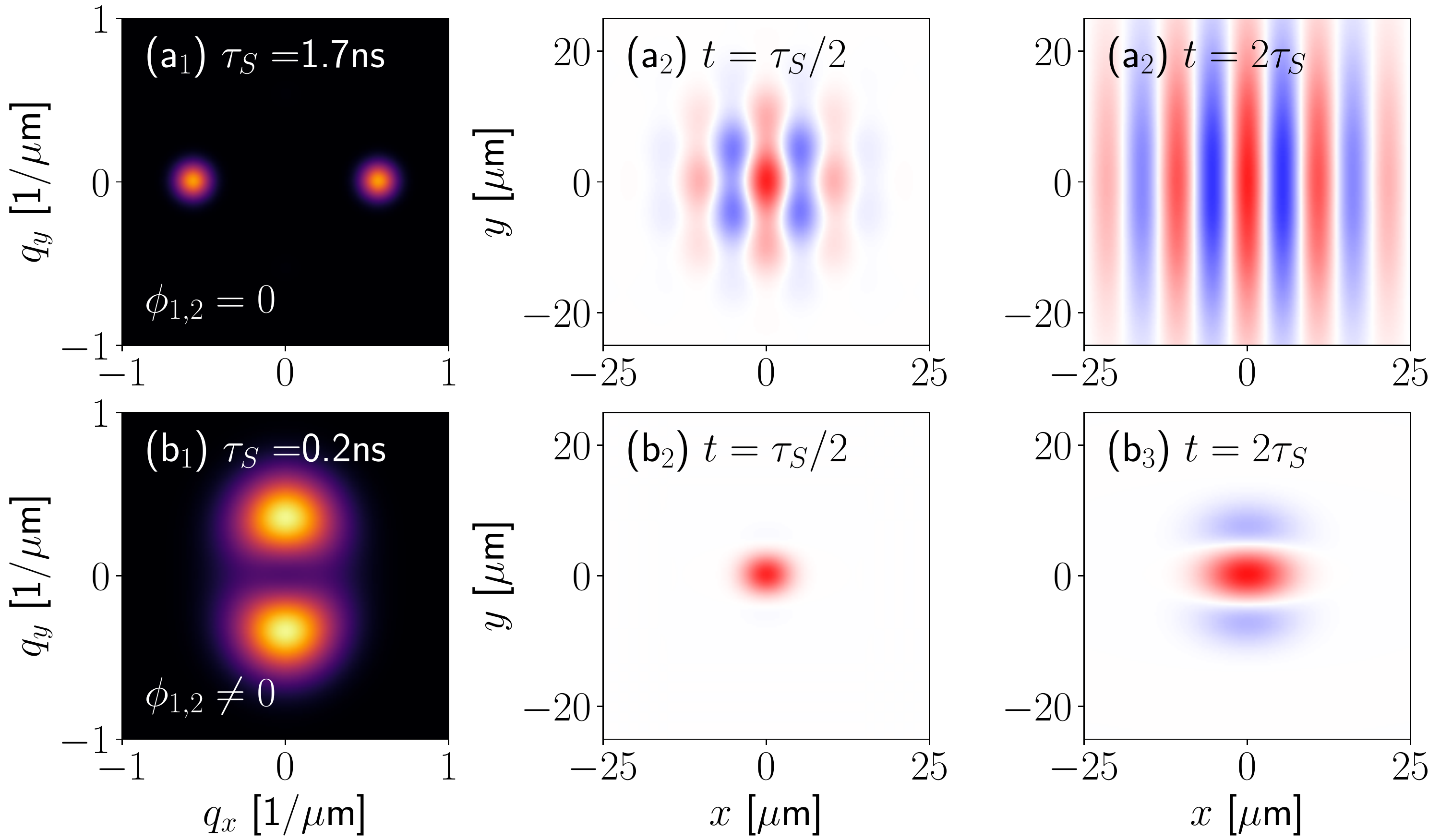}
    \caption{Equivalent to Fig.~\ref{fig:SzMaps7}, but for $F_z = 0.3$~V$\mu$m (orange dots in Fig.~\ref{fig:lifetime}). (a$_n$) For the weak-coupling regime ($\phi_{1,2}=0$) the checkerboard pattern is transient due to the large difference between the subband lifetimes in Fig.~\ref{fig:lifetime}(a). For large $t$ we see only the stripes due to the second subband. (b$_n$) In the strong-coupling regime the subband-averaged dynamics leads to striped patterns.}
    \label{fig:SzMaps25}
\end{figure}

\begin{figure}[tb]
    \centering
    \includegraphics[width=\columnwidth]{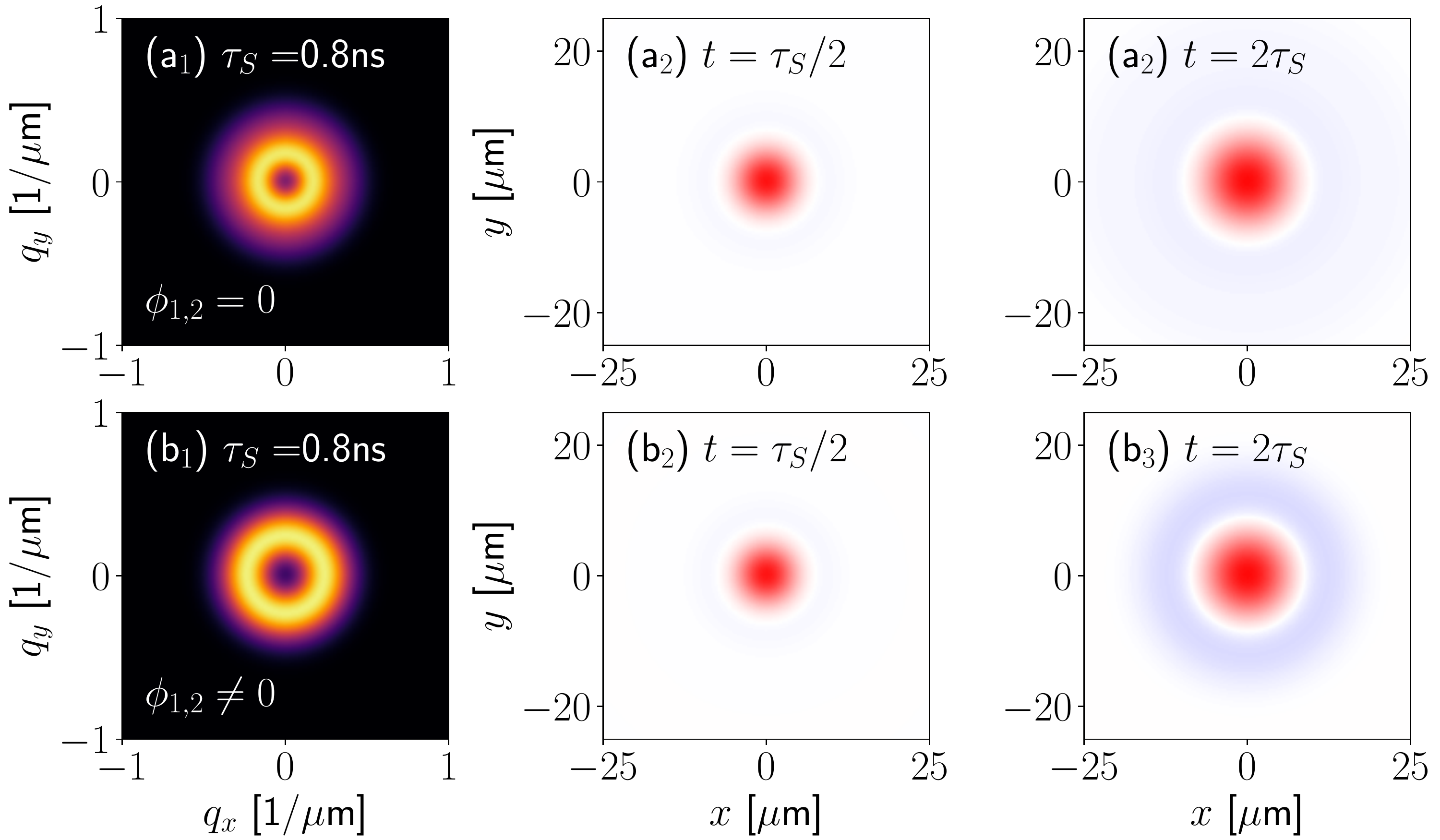}
    \caption{Equivalent to Fig.~\ref{fig:SzMaps7}, but for $F_z = 0$, which is an isotropic limit due to the symmetric quantum well. In this case the circular pattern is present in both (a$_n$) weak-coupling regime with $\phi_{1,2} = 0$, and (b$_n$) strong-coupling regime with $\phi_{1,2} \neq 0$. }
    \label{fig:SzMaps0}
\end{figure}

Illustrative examples of the spin patterns, $S_z(\bm{q}, t)$ and $S_z(\bm{r}, t)$, are shown in Figs.~\ref{fig:SzMaps7}, \ref{fig:SzMaps25}, and \ref{fig:SzMaps0} for $F_z = 0.09$, $0.3$, and $0$~V/$\mu$m, respectively. These correspond to the blue, orange, and purple dots in Figs.~\ref{fig:lifetime}(a) and \ref{fig:lifetime}(b). In these figures we compare the spin patterns for $\phi_{1,2} = 0$ (top panels) and the realistic $\phi_{1,2} \neq 0$ (bottom panels).

For $F_z = 0.09$~V/$\mu$m in Fig.~\ref{fig:SzMaps7}, the spin lifetimes for $\phi_{1,2} = 0$ are similar for both subbands, as shown in Fig.~\ref{fig:lifetime}(a). Moreover, since in this case $\alpha_1$ and $\alpha_2$ have opposite signs, the $S_z(\bm{q},t)$ in Fig.~\ref{fig:SzMaps7}(a$_1$) shows four peaks in $\pm q_x^0$ and $\pm q_y^0$, corresponding to crossed spin excitations, which leads to the checkerboard pattern seen in Figs.~\ref{fig:SzMaps7}(a$_2$) and \ref{fig:SzMaps7}(a$_3$). However, for $F_z = 0.09$~V/$\mu$m, the actual subband overlap ratio is $\phi_{1,2} \approx 0.9$, which gives $t_c \sim 10^{-5}$~ns, and the strong-coupling regime dominates. In this case the dynamics follows the subband-averaged diffusion matrix, as in Eq.~\eqref{eq:spindiffNST}. Qualitatively, this leads to a partial cancellation of $D_1\alpha_1 + D_2\alpha_2$ (since $\alpha_1$ and $\alpha_2$ have opposite signs) in the average of the non-diagonal terms of $(\mathcal{D}_1^S + \mathcal{D}_2^S)/2$ [see Eq.~\eqref{eq:Djmat}]. Consequently, the spin patterns approach the circular shape of the isotropic regime (where $\alpha_j = 0$). Since this cancellation is not exact, it gives $Q_+^* > Q_-^*$ and favors $S_z(\bm{q},t)$ peaks in $\pm q_y^0$, as seen in Fig.~\ref{fig:SzMaps7}(b$_1$). This leads to the intermediate pattern between circular and vertical stripes seen in Figs.~\ref{fig:SzMaps7}(b$_2$) and \ref{fig:SzMaps7}(b$_3$). Numerical experiments show that to recover the checkerboard pattern of the weak-coupling regime one would need to reduce $\phi_{1,2}$ by three orders of magnitude and obtain a large $t_c = 0.5$~ns.

The results for $F_z = 0.3$~V/$\mu$m shown in Fig.~\ref{fig:SzMaps25} are similar to those discussed above for Fig.~\ref{fig:SzMaps7}. However, the $\tau_S$ for $\phi_{1,2}=0$ in this case are significantly different for the first and second subbands [see Fig.~\eqref{fig:lifetime}(a)]. Consequently, the checkerboard pattern is transient, while for large $t \gg \tau_S$ the second subband spin excitation dominates and forms the vertical striped pattern. Nevertheless, for realistic $\phi_{1,2} \neq 0$, the strong coupling dominates and leads to horizontal striped pattern seen in Fig.~\ref{fig:SzMaps25}(b$_n$).

The strong-coupling regime is maximum at $F_z = 0$, where $\phi_{1,2}$ peaks. As discussed in the previous section, this $\phi_{1,2}$ peak yields an enhancement of $\tau_S$ around $F_z = 0$. However, since this is the isotropic limit, there is no quantitative difference between the circular spin patterns in both $\phi_{1,2}=0$ and finite $\phi_{1,2}\neq 0$ shown in Fig.~\ref{fig:SzMaps0}.

\subsection{Persistent Skyrmion Lattice}
\label{sec:PSL}

\begin{figure}[tb]
    \centering
    \includegraphics[width=\columnwidth]{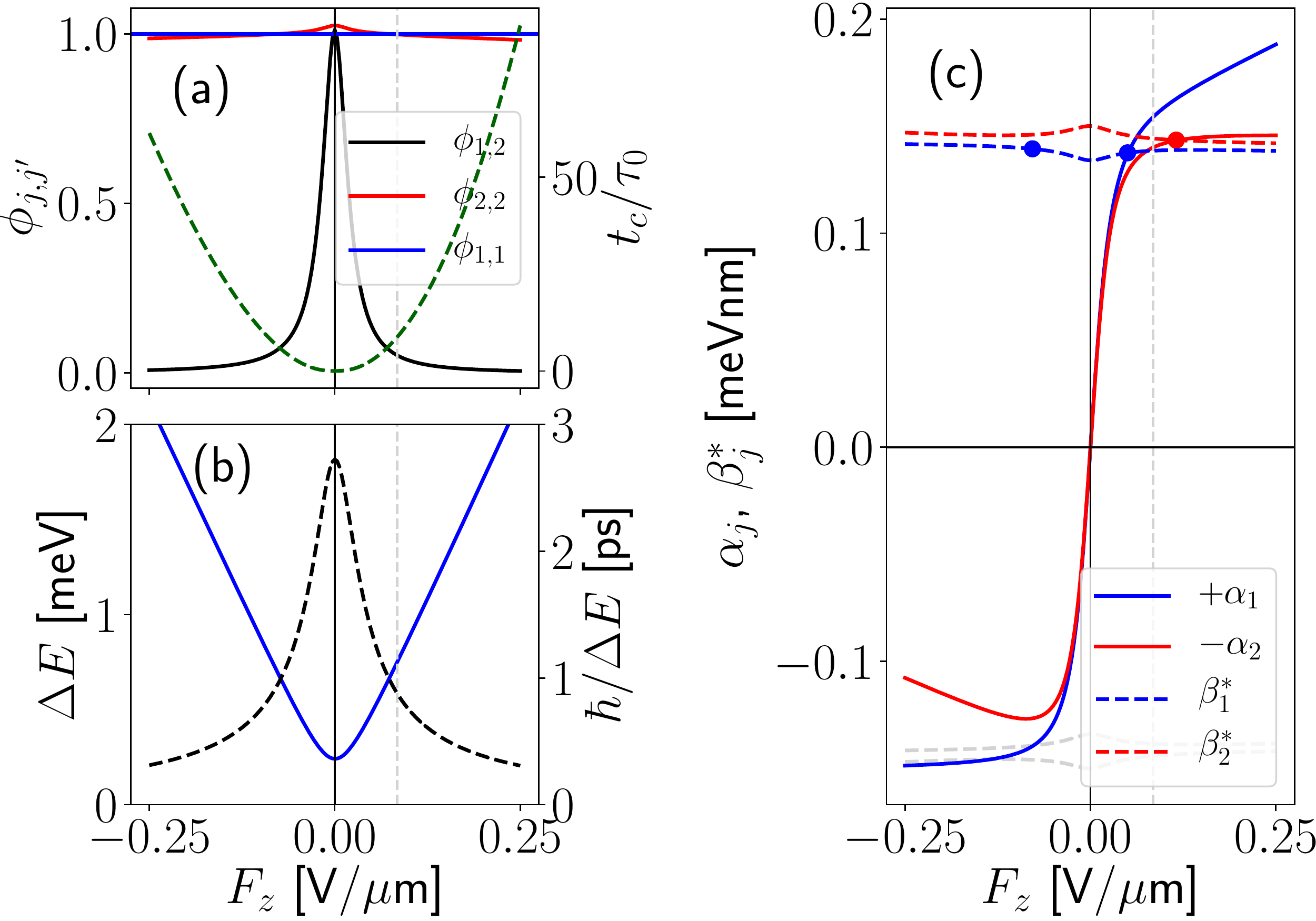}
    \caption{Data from the Schrödinger-Poisson solution for the quantum well from Ref.~\cite{Fu2015Skyrmion}.
    (a) Density overlap ratios $\phi_{j,j'}$ as a function of the electric field $F_z$ and the intersubband relaxation time $t_c/\tau_0 \approx 1/(2\phi_{1,2})$ (dashed line, right axes).
    (b) The subband energy splitting $\Delta E = \varepsilon_2-\varepsilon_1$ (blue line, left axes) defines the scale of the momentum scattering time that satisfies the non-degeneracy condition $\tau_0 \gg \hbar/\Delta E$ (dashed line, right axes).
    (c) The SOC as a function of $F_z$ with the PSH conditions $\beta_j^* = \pm \alpha_j$ marked by the circles.
    In all panels, the vertical dashed gray line mark the point $F_z = 0.084$~V/$\mu$m where we consider the PSL condition with $\alpha_1 \approx +\beta_1^*$ and $\alpha_2 \approx -\beta_2^*$.
    }
    \label{fig:tsegues}
\end{figure}

The checkerboard pattern seen previously in Figs.~\ref{fig:SzMaps7} and \ref{fig:SzMaps25} in the weak-coupling limit ($\phi_{1,2} = 0$) was predicted in Ref. \cite{Fu2015Skyrmion} as a novel topological spin texture, and dubbed \textit{persistent skyrmion lattice} (PSL). It arises when the subbands are set with orthogonal PSH regimes, \textit{e.g.,} with $\alpha_1 = +\beta_1^*$ and $\alpha_2 = -\beta_2^*$. Consequently, one subband shows vertical stripes and the other horizontal ones, such that their sum yields the checkerboard pattern. However, while the usual single-subband PSH is robust against impurity and electron-electron scattering \cite{Bernevig2006PSH}, the two-subband PSL is not robust against intersubband scattering. To see this, let us consider the total Hamiltonian composed by $H_{\rm 2D}$ from Eq.~\eqref{eq:Hclean} and the projected impurity potential $V_{j'j}$ from Eq.~\eqref{eq:Vij}, which reads as
\begin{align}
    H_T &=
	\begin{pmatrix} 
	H_1 + V_{11} & V_{12}
	\\ 
	V_{12}^\dagger & H_2 + V_{22}
	\end{pmatrix},
\end{align}
where we have neglected the intersubband SOC $H_{12}$ for simplicity. Now, let us assume that each single-subband block commutes with a spin operator as $[H_j + V_{jj}, \sigma_j] = 0$. For instance, the PSL conditions $\alpha_1 = +\beta_1^*$ and $\alpha_2 = -\beta_2^*$ imply that one subband commutes with $\sigma_1 = \sigma_x$ and the other is orthogonal with $\sigma_2 = \sigma_y$. For a generic spin operator $\sigma_T = \sigma_1 \oplus \sigma_2$, one finds
\begin{align}
    [H_T, \sigma_T] &= 
    \begin{pmatrix}
        0 & V_{12}(\sigma_2 - \sigma_1)
        \\
        V_{12}^\dagger(\sigma_1 - \sigma_2) & 0
    \end{pmatrix},
\end{align}
where we have used that the $V_{12}$ potential is scalar. Consequently, $[H_T, \sigma_T] = 0$ only if (i) the subbands are in parallel PSH regimes with $\sigma_1 = \sigma_2$; or (ii) $\sigma_1 \neq \sigma_2$, but the intersubband impurity coupling $V_{12} \approx 0$. The first case is a trivial superposition of two identical PSH regimes. The second case corresponds to the PSL regime with $\sigma_1 \perp \sigma_2$. Next, the numerical results of our spin-diffusion equation show that the PSL checkerboard pattern is killed in the strong-coupling regime and it is recovered as we increase $t_c$ to approach the weak-coupling regime, for which $V_{12}$ is indeed negligible.

\subsubsection{Transition from strong- to weak-coupling regime}

The quantum well discussed in the previous sections approximately satisfy the PSL criteria ($\alpha_1 \approx +\beta_1^*$ and $\alpha_2 \approx -\beta_2^*$). However, it shows an overall small $t_c/\tau_0$ in Fig.~\ref{fig:tc}(b), which favors the strong coupling and the spin pattern becomes nearly circular (Figs.~\ref{fig:SzMaps7} and \ref{fig:SzMaps25}). To contrast these results, in this section we now consider the system proposed in Ref.~\cite{Fu2015Skyrmion}. It consists of a $45$-nm-wide quantum well with a wider central barrier of $3$~nm, and total density $n_{\rm 2D} = 4.8\times 10^{11}$~cm$^{-2}$. The data extracted from the self-consistent Schrödinger-Poisson calculation for this system are shown in Fig.~\ref{fig:tsegues}. Here, the wider central barrier (3~nm instead of 0.5~nm) allows for an enhanced separation of the quantum wells with a reduced overlap $\phi_{1,2}$, yielding larger $t_c/\tau_0$, as seen in Fig.~\ref{fig:tsegues}(a). This helps to favor the weak-coupling limit and reach the PSL regime. However, at $F_z \approx 0$, it also leads to the small subband energy split shown in Fig.~\ref{fig:tsegues}(b). The non-degenerate regime considered in this paper requires $\Delta E \gg \hbar/\tau_0$, which implies $\tau_0 \gg \hbar/\Delta E \approx 2.5$~ps for $F_z \approx 0$. For such small $\Delta E$ the subband correlations neglected by our block-diagonal approximations of $H$, $G$, and $\Sigma$ might fail. Therefore, hereafter we shall rely only on the data for finite $F_z$, such that $\Delta E$ is large enough to justify the block-diagonal approximations, see Fig.~\ref{fig:tsegues}(b).

From the SOC data in Figs.~\ref{fig:tsegues}(c) we see that the PSL condition ($\alpha_1 \approx \beta_1^*$ and $\alpha_2 \approx -\beta_2^*$) is approximately satisfied near $F_z \sim 0.084$~V/$\mu$m. Around this electric field we find $\Delta E \approx 0.75$~meV, which satisfies the non-degenerate condition for a reasonable $\tau_0 \gg 0.8$~ps. Additionally, it shows an interesting $t_c \approx 8.8\tau_0$ that is large enough to allow us to analyze a transition from the strong- to the weak-coupling regime as we increase $\tau_0$ from 1 to 10~ps.

\begin{figure}[tb]
    \centering
    \includegraphics[width=\columnwidth]{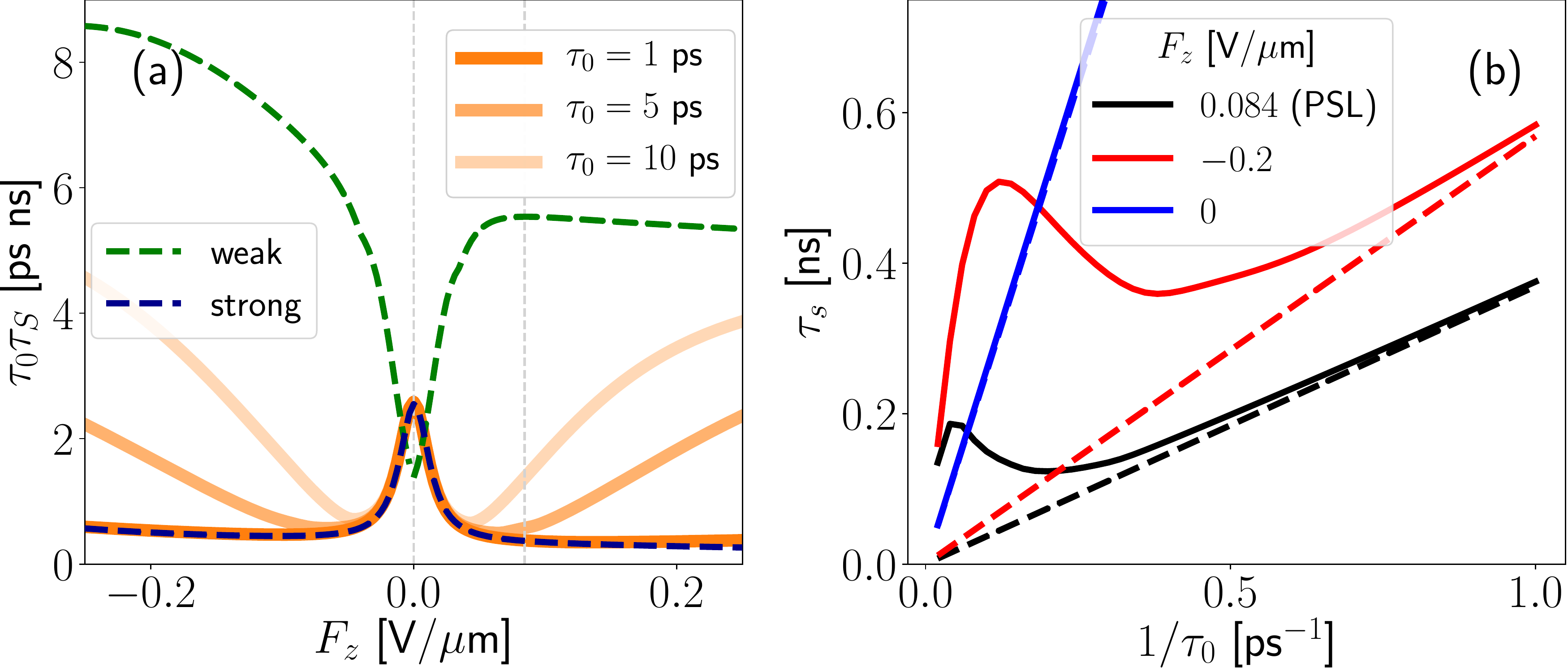}
    \caption{
    (a) The product $\tau_0 \tau_S$ shows the transition from the strong- to the weak-coupling regime as $\tau_0$ increases. At $F_z\approx 0$ the system is always at the strong-coupling regime due to the small $t_c$ (see Fig.~\ref{fig:tsegues}), while the transition emerges for finite $F_z$ due to the larger $t_c$. 
    (b) $\tau_S$ as a function of $1/\tau_0$ for selected $F_z$ from (a), showing the linear DP trend for both large and small $1/\tau_0$, with a non-monotonic transition for the intermediate regime. In (b) the dashed lines correspond to the strong-coupling limit.
    In both panels the dashed lines are taken from Eq.~\eqref{eq:tauSi}, where both strong and weak-coupling limits follow the DP relaxation with $\tau_S \propto 1/\tau_0$. Solid lines are extracted from the numerical solution of the spin diffusion using Eq.~\eqref{eq:tauSnum}.}
    \label{fig:tsegues2}
\end{figure}

In Fig.~\ref{fig:tsegues2}(a) we show the spin lifetime $\tau_S$ multiplied by $\tau_0$, \textit{i.e.,} $\tau_0 \tau_S$. This composition is interesting because both in the weak- and strong-coupling regimes we expect $\tau_S \propto 1/\tau_0$, due to the DP spin relaxation mechanism. Therefore, the product $\tau_0 \tau_S$ becomes a universal quantity to characterize the extreme weak- and strong-coupling limits. However, in-between these extreme regimes, the intersubband couplings $\gamma_j$ breaks this proportionality in Eq.~\eqref{eq:spindiff}. Consequently, in Fig.~\ref{fig:tsegues2}(a) we see a transition from the strong- to the weak-coupling regime as we increase $\tau_0$. In Fig.~\ref{fig:tsegues2}(b) we show $\tau_S$ as a function of $1/\tau_0$ for selected values of $F_z$. For $F_z \approx 0$ the system is always in the strong-coupling regime due to the small $t_c$ [see Fig.~\ref{fig:tsegues}(a)], thus it shows the linear DP trend $\tau_S \propto 1/\tau_0$ over the full range in Fig.~\ref{fig:tsegues2}(b). On the other hand, for $F_z = -0.2$ and $+0.084$~V/$\mu$m we see a non-monotonic evolution of $\tau_S$ with $1/\tau_0$ due to the larger $t_c$ for finite $F_z$. For large $1/\tau_0$ all cases fall into the DP linear trend of the strong-coupling regime (dashed lines). Similarly, for small $1/\tau_0$ they approach the weak-coupling regime, which also shows the DP linear trend, but it only converges to the weak coupling $\tau_S$ for extremely small $1/\tau_0$. Notice in Fig.~\ref{fig:tsegues2}(a) that even for $\tau_0 = 10$~ps the $\tau_0\tau_S$ lines are approaching, but still far from the weak-coupling limit. In-between these limits, for intermediate $1/\tau_0$, we see the ``N''-shaped transition, which is more pronounced for cases with larger $t_c$.

\begin{figure}[tb]
    \centering
    \includegraphics[width=\columnwidth]{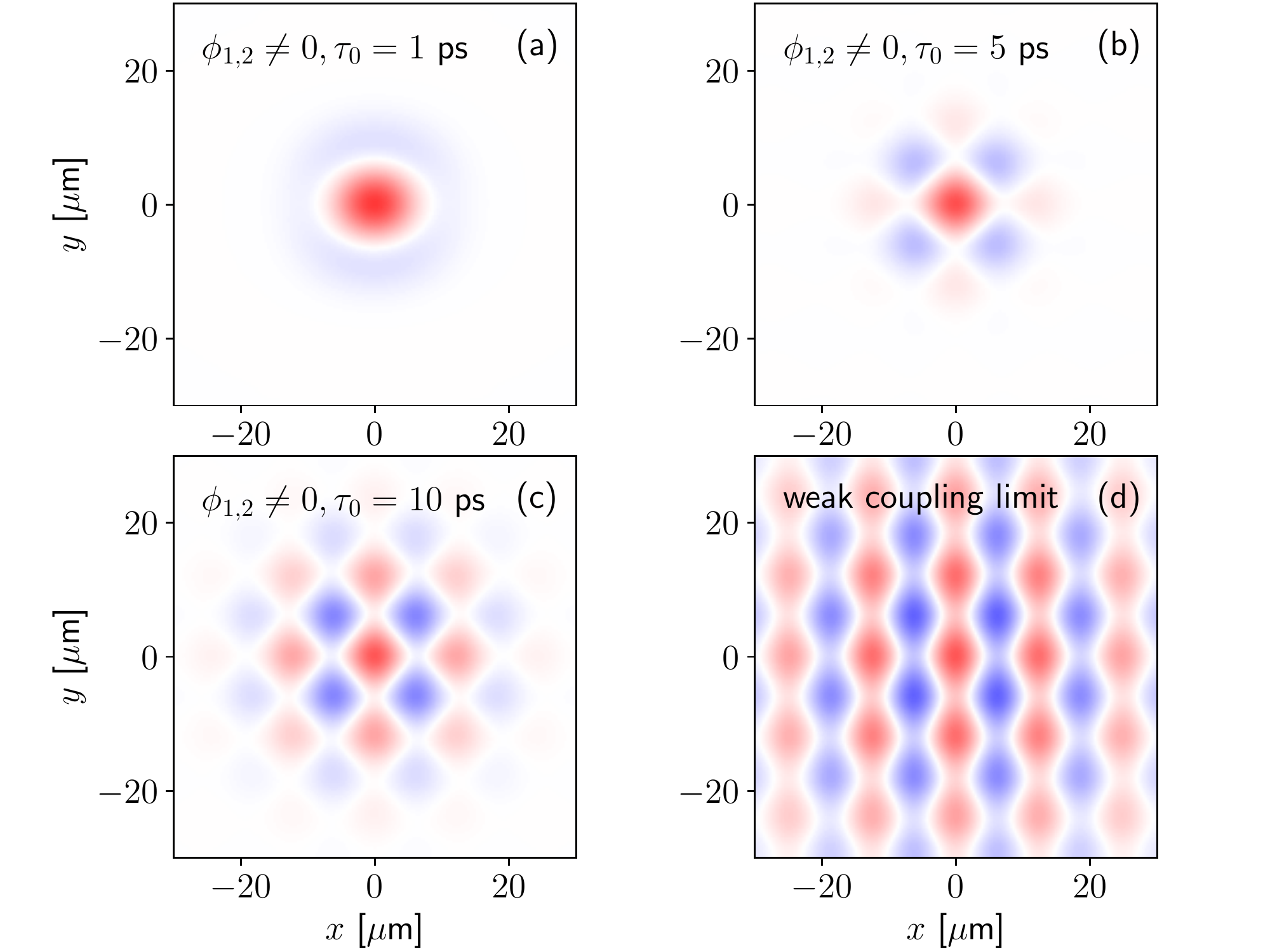}
    \caption{
    Spin patterns $S_z(\bm{r}, t)$ for $F_z = 0.084$~V/$\mu$m calculated at large $t = 2\tau_S$. As $\tau_0$ increases from (a) 1~ps, (b) 5~ps, and (c) 10~ps, the spin dynamics goes from (a) the strong coupling with a nearly circular pattern and approaches (d) the weak-coupling regime, where the checkerboard pattern of the PSL \cite{Fu2015Skyrmion} emerges. For each panel, $\tau_S$ can be extracted from Fig.~\ref{fig:tsegues2}, yielding (a) $\tau_S = 0.377$~ns, (b) 0.115~ns, and (c) 0.137~ns.
    }
    \label{fig:SzmapsEgues}
\end{figure}

The transition from strong- to weak-coupling regimes is also clearly seen in the spin density patterns in Fig.~\ref{fig:SzmapsEgues}. These were calculated near the PSL condition for $F_z \sim 0.084$~V/$\mu$m. First, for $\tau_0 = 1$~ps the system is at the strong-coupling regime with $t_c = 8.8$~ps, and its spin pattern in Fig.~\ref{fig:SzmapsEgues}(a) shows the nearly circular pattern expected due to the subband-averaged dynamics. Then, as we increase $\tau_0$ to 5 and 10~ps in Figs.~\ref{fig:SzmapsEgues}(b) and \ref{fig:SzmapsEgues}(c), the characteristic checkerboard pattern of the PSL regime emerges as the system approaches the weak-coupling regime with $t_c = 44$ and 88~ps, respectively. The extreme limit of the weak-coupling with $\phi_{1,2} = 0$ is shown in Fig.~\ref{fig:SzmapsEgues}(d) for comparison.

This system clearly shows how $t_c$ defines the diffusion regime between weak and strong subband couplings. However, one must also pay attention to the validity of diffusion picture, which requires $\tau_S \gg \tau_0$, \textit{i.e.,} the spin relaxation must be slow in comparison with the momentum scattering time. From Fig.~\ref{fig:tsegues2}, for $F_z \sim 0.084$~V/$\mu$m, we see that this criterion is certainly valid for $\tau_0 = 1$~ps and $\tau_S \approx 377$~ps, while for $\tau_0 = 5$ and $10$~ps we get $\tau_S \approx 115$~ps and $137$~ps, respectively, corresponding to factors of $\tau_S/\tau_0 \sim 23$ and $13.7$, respectively. Therefore, for large $\tau_0$ one might fall towards a (quasi-) ballistic picture and deviations from our model would be expected.

\section{Conclusions}
\label{sec:conclusions}

We have extended the spin drift-diffusion model to the case of two-subbands 2DEGs following the kinetic equation approach from the Keldysh formalism and semi-classical approximations to achieve the quantum Boltzmann equation and its linearization towards the diffusion equation. Considering the intrasubband and intersubband scattering by impurities, we show that the subband dynamics is controlled by the time scale $t_c \approx \tau_0/(2\phi_{1,2})$, which defines a new knob to control the spin lifetime. We find that for small $t_c$ the subbands are strong-coupled and the spin follows a subband-averaged dynamics, while for large $t_c$ the subbands are nearly independent in the weak-coupling regime. Applying our model to the \textit{persistent skyrmion lattice} setup \cite{Fu2015Skyrmion}, we show that its characteristic checkerboard pattern is killed by strong inter-subband coupling and reemerges for large $t_c$ as it approaches the weak-coupling regime. Moreover, for any non-degenerate two-subband system we find that the spin lifetime peaks around the symmetric well configuration due to Matthiessen's rule, which follows the corresponding peak in intersubband scattering coupling.

Our results apply for non degenerate subbands, meaning that we assume that the broadening of the spectral function is small compared to the subband splitting, which justifies the quasi-particle approximation. This leads to the block-diagonal approximations in the Green's functions and self-energy. In the opposite limit, for near-degenerate subbands the block-diagonal approximation fails and a model for this scenario shall remain a challenge for future works. 

\section{Acknowledgements}

G.J.F.~thanks Felix G.~G.~Hernandez for useful discussions. I.R.A.~and G.J.F.~acknowledge financial support from the Brazilian funding agencies CNPq, CAPES, and FAPEMIG.

\appendix

\section{Self-energy and impurity self-average}
\label{app:selfenergy}

Here we derive the self-energy $\Sigma$ within the self-consistent Born approximation, considering the weak SOC limit, and neglecting intersubband SOC. We follow a similar derivation as in Refs.~\cite{rammer2007quantum, bruus2004many}. However, due to the two-subband structure, one must consider a basis set by a discrete subband index $j = \{1,2\}$ and the usual in-plane plane-wave continuum momentum $\bm{k} = (k_x, k_y)$, i.e. $\braket{\bm{r}}{j,\bm{k}} = A^{-1/2}e^{i\bm{k}\cdot\bm{r}} \varphi_j(z)$, where $A$ is the normalization area for the plane-wave and $\varphi_j(z)$ is the quantum well solution of each subband $j$. Since we will be considering only scalar impurities, we neglect the spin quantum number. 

Within the $\ket{j,\bm{k}}$ basis, the Dyson equation reads as
\begin{align}
	G_{b,a} &= \delta_{b,a} G_b^0 + \sum_c G_b^0 V_{b,c} G_{c,a},
	\label{eqa:dyson}
	\\
	\nonumber
	&=\delta_{b,a} G_b^0 + G_b^0 V_{b,a} G_a^0 + \sum_c G_b^0 V_{b,c} G_c^0 V_{c,a} G_a^0 + \cdots
\end{align}
such that the random impurity potential matrix element is $V_{b,c} = V_{j_b, j_c}(\bm{k}_b, \bm{k}_c) = \bra{j_b, \bm{k}_b} V \ket{j_c, \bm{k}_c}$. As previously mentioned in the main text, we have introduced a compact notation for the indices, e.g., $a \rightarrow {j_a, \bm{k}_a, \sigma_a}$.

Here, we consider random scalar short-range impurities set by $V(\bm{r}) = \sum_i v_0 \delta(\bm{r}-\bm{R}_i)$, where $\bm{R}_i = (x_i, y_i, z_i)$ is the position of each impurity $i$, and $v_0$ is the intensity (in units of energy $\times$ volume). The matrix element $V_{b,c}$ for a single set of $N$ random impurities reads as
\begin{align}
	V_{b,c} &= \sum_{i=1}^N \tilde{v}_{j_b, j_c}(z_i)e^{-i(\bm{k}_b-\bm{k}_c)\cdot\bm{r}_i},
	\\
	\tilde{v}_{j_b, j_c}(z_i) &= \dfrac{v_0}{A}\varphi_{j_b}^\dagger(z_i)\varphi_{j_c}(z_i).
\end{align}
To consider an ensemble of random impurities set of identical impurity concentrations $n_{\rm imp} = N/\Omega$ (where $\Omega$ is the volume), one defines the ensemble average of any quantity $\mathcal{Q}$ as \cite{bruus2004many, rammer2007quantum}
\begin{align}
	\mean{\mathcal{Q}} \equiv \int \Big[ \prod_{i=1}^{N} \dfrac{d \bm{r}_i}{\Omega} \Big] \mathcal{Q}(\bm{r}_1, \bm{r}_2,...,\bm{r}_N).
	\label{eqa:ensemble}
\end{align}

To apply the ensemble average on the Dyson equation from Eq.~\eqref{eqa:dyson}, it is sufficient to consider $\mean{V_{b,a}}$ and $\mean{V_{b,c}V_{c,a}}$. The ensemble average on $\mean{V_{b,a}} $ straight-forwardly yields
\begin{align}
	\mean{V_{b,a}} &= \delta_{b,a} n_{\rm imp} v_0 = \includegraphics[height=0.7cm,valign=b]{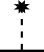},
\end{align}
which is a trivial constant term that can be incorporated into the chemical potential. We shall neglect it from now on. The second-order term, $\mean{V_{b,c}V_{c,a}}$ reads as
\begin{multline}
	\mean{V_{b,c}V_{c,a}} = \sum_{n,p=1}^N \int \Big[\prod_{m=1}^N \dfrac{d^3 r_m}{\Omega}\Big]
	\\
	\times \tilde{v}_{j_b,j_c}(z_n)e^{-i(\bm{k}_b-\bm{k}_c)\cdot\bm{r}_n}
	\tilde{v}_{j_c,j_a}(z_p)e^{-i(\bm{k}_c-\bm{k}_a)\cdot\bm{r}_p}.
	\label{eqa:V2raw}
\end{multline}
Here one might have two scattering events with the same impurity ($n = p$), or with distinct impurities ($n \neq p$). These two cases have to be considered separately. 

For $n\neq p$ the integrals in Eq.~\eqref{eqa:V2raw} simplify into pairs of products of two integrals equivalent to the one on the first-order term, i.e., $\mean{V_{b,c} V_{c,a}}_{n\neq p} = \mean{V_{b,c}}_N \mean{V_{c,a}}_{N-1} \approx \mean{V_{b,c}} \mean{V_{c,a}}$. The indices $N$ and $N-1$ indicate that one of the sums runs up to $N-1$, since we are removing the $n=p$ contribution. Moreover $N \gg 1$, thus, we can neglect this detail and assume the approximate expression above. Since this term corresponds to a reducible sequence of two first-order scattering events, it can be neglected as it vanishes by a chemical potential renormalization.

The $n=p$ contributions from Eq.~\eqref{eqa:V2raw} simplify into $N$ equivalent integrals, yielding
\begin{align}
	\mean{V_{b,c}V_{c,a}}_{n=p} &= \dfrac{n_{\rm imp} v_0^2}{A}
	\Lambda_{b,c,a}
	\delta_{\bm{k}_b, \bm{k}_a},
	\\
	\label{eqa:lambdaoverlap}
	\Lambda_{b,c,a} &=
	\int dz
	\varphi_{j_b}^\dagger(z) 
	|\varphi_{j_c}(z)|^2
	\varphi_{j_a}(z).
\end{align}
Later we'll also use a short notation for $\Lambda_{a,c} \equiv \Lambda_{a,c,a}$. Here, the momentum conservation  implies subband conservation, i.e., $\delta_{\bm{k}_b, \bm{k}_a} \rightarrow \delta_{a,b}$, since we are considering that the subbands Fermi circles do not overlap, as shown in Fig.~\ref{fig:system}.

Back on the Dyson equation, we get
\begin{multline}
	\mean{G_a} = G_a^0 + G_a^0 \Big[
	\dfrac{n_{\rm imp} v_0^2}{A} \sum_c \Lambda_{a,c}
	G_c^0
	\Big] G_a^0 + \cdots,
\end{multline}
where we have already used $\mean{G_{b,a}} \approx \delta_{b,a} \mean{G_a}$. This block-diagonal form is only valid due to the approximation $\delta_{\bm{k}_b, \bm{k}_a} \rightarrow \delta_{a,b}$ above.

The term above between square-brackets $[\cdots]$ is the self-energy in the first Born approximation,
\begin{align}
    \Sigma_a^{\rm 1stBA} = \dfrac{n_{\rm imp} v_0^2}{A} \sum_c \Lambda_{a,c}
	G_c^0 = \includegraphics[height=0.7cm,valign=b]{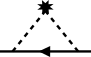}.
\end{align}
Finally, for the self-consistent Born approximation, the usual replacement $G_c^0 \rightarrow \mean{G_c}$ yields
\begin{align}
	\Sigma_a^{\rm SCBA} = \dfrac{n_{\rm imp} v_0^2}{A} \sum_c  \Lambda_{a,c} \mean{G_c} = \includegraphics[height=0.7cm,valign=b]{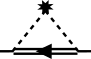}.
	\label{eqa:sigma}
\end{align}

\section{Collision integral}
\label{app:collision}

Since $H$ has the block-diagonal form of Eq.~\eqref{eq:Hclean}, here it is sufficient to analyze each block as $H_j = \varepsilon_j(k)\sigma_0 + H_j^{\rm soc}(\bm{k})$. Consequently, the equilibrium $G_j^{R(A)}$ are
\begin{align}
	G_j^{R(A)}(E,\bm{k}) = \dfrac{1}{E-\varepsilon_j(k)\sigma_0 - H_j^{\rm soc}(\bm{k}) \pm i \eta}.
\end{align}
The spectral function $A_j(E,\bm{k}) = i(G_j^R-G_j^A)$ might take a complicated form due $H_j^{\rm soc}$. However, for small SOC ($\varepsilon_F \gg |H_j^{\rm soc}|$), we can consider a series expansion, yielding
\begin{align}
	A_j(E,\bm{k}) 
	&= 2\pi \exp\Big[-H_j^{\rm soc}(\bm{k}) \dfrac{\partial}{\partial E}\Big]\delta(E-\varepsilon_j(k)), 
	\label{eqa:fullA}
	\\
	\nonumber
	&\approx 2\pi\Big[\delta(E-\varepsilon_j(k)) - H_k^{\rm soc}(\bm{k}) \delta'(E-\varepsilon_j(k)) \Big],
\end{align}
where the second expression considers the approximation up to linear order in $H_j^{\rm soc}$. We shall consider this expression for $A_j(E,\bm{k})$ for the quasi-classical approximation. Namely, we take the Kadanoff-Baym ansatz as \cite{Rammer1986RMP}
\begin{align}
    \label{eqa:KBA}
	G_j^K(E,\bm{k},\bm{R}, T) = \dfrac{-i}{2} \Big\{A_j(E,\bm{k}) , \rho_j(\bm{k},\bm{R}, T)\Big\},
\end{align}
where $\rho_j(\bm{k},\bm{R}, T) = 1-2F_j(\bm{k},\bm{R}, T)$ is the non-equilibrium distribution function. At equilibrium $F_j(\bm{k},\bm{R}, T) \rightarrow f_j(k)$ becomes the Fermi-Dirac distribution $f_j(k) \approx (1+\exp[\beta(\varepsilon_j(k) - \mu)])^{-1}$. Assuming we are not far from equilibrium, it is reasonable to consider a linear response regime, with
\begin{align}
	\rho_j(\bm{k},\bm{R}, T) 
	&\approx \rho^{(0)}_j(\bm{k}) - \dfrac{1}{N_0} f'_j(\bm{k}) \rho^{(1)}_j(\bm{k},\bm{R},T),
	\label{eqa:rholinres}
	\\
	\nonumber
	&\approx \rho^{(0)}_j(\bm{k}) + \dfrac{1}{N_0} \delta(\varepsilon_j(k)-\varepsilon_F)\rho^{(1)}_j(\bm{k},\bm{R},T),
\end{align}
where the second form is taken in the zero temperature limit ($\beta \rightarrow \infty$, $\mu \rightarrow \varepsilon_F$). Here, the 2DEG density of states $N_0 = A m/2\pi\hbar^2$ is introduced to simplify future expressions. The equilibrium contribution $\rho_j^{(0)}(\bm{k})$ vanishes in Eq.~\eqref{eq:boltzmann}, therefore, to focus on the relevant contributions, we may write our non-equilibrium distribution function as
\begin{align}
	\rho_j(\bm{k},\bm{R}, T) &\approx \dfrac{1}{N_0} \delta(\varepsilon_j(k)-\varepsilon_F)\rho'_j(\bm{k},\bm{R},T).
	\label{eqa:rho}
\end{align}

The spectral function $A$ in Eq.~\eqref{eqa:fullA} carries two terms. Therefore, we split the collision integral in Eq.~\eqref{eq:Icoll} into zeroth and first order terms $I^{\rm coll}_j(\bm{k}) = I^{(0)}_j(\bm{k}) + I^{(1)}_j(\bm{k})$. 

The zeroth order term $I^{(0)}_j(\bm{k})$, accounts the contributions from the leading order of $A_j(E,\bm{k}) \approx 2\pi \delta[E-\varepsilon_j(k)]$. As a result, both $A_j$ and $\Gamma_j$ become scalar matrices and the anticommutators simplify, e.g., $\{A_j, \Sigma_j^K\} = 2A_j \Sigma_j^K$. Namely, $\Gamma_j= (n_{\rm imp} / v_0^2A) \sum_d  \Lambda_{j,d} A_d(E,\bm{k}_d).$ Therefore, the collision integral in Eq.\eqref{eq:I0coll} reads as
\begin{multline}
\label{eqa:collisionintegralrho}
	I^{(0)}_j(\bm{k})
	= \dfrac{2\pi n_{\rm imp} v_0^2}{A\hbar} 
	\sum_d \Lambda_{j,d} \delta[\varepsilon_j(k) - \varepsilon_d(k_d)]
	\\
	\times \Big(\rho_d(\bm{k}_d,\bm{R},T)
	- 
	\rho_j(\bm{k},\bm{R},T)
	\Big),
\end{multline}
where we have expressed $G_{j}^{K}$ in terms of $\rho_j(\bm{k},\bm{R},T)$ using  Eq.~\eqref{eqa:KBA}.

The sum in Eq.~\eqref{eqa:collisionintegralrho} can be written as $\sum_d = \sum_{j_d, \bm{k}_d} = \frac{N_0}{2\pi} \sum_{j_d} \int d\theta_d \int d\varepsilon_d$. The energy conservation implied by the Dirac delta above enforces $\varepsilon_j(k) = \varepsilon_d(k_d)$, which constrains the absolute values $k$ and $k_d$. Since $\varepsilon_j(k)$ is monotonic in $k$, we may change variables to cast $\rho_j(\bm{k},\bm{R},T) \rightarrow \rho_j(\varepsilon_j(k),\theta,\bm{R},T)$, leading to 
\begin{align}
\nonumber
	I^{(0)}_j(\bm{k})
	= 
	\sum_{j_d} \dfrac{
	\mean{\rho_{j_d}(\varepsilon_j(k), \bm{R},T)}
	- 
	\rho_j(\varepsilon_j(k), \theta,\bm{R},T)
	}{\tau_{j,j_d}}.
\end{align}

 Finally, $I^{(1)}_j(\bm{k})$  accounts for first order correction in $H_j^{\rm soc}$ from Eq.~\eqref{eqa:fullA}. Following the same procedures applied on $I^{(0)}_j(\bm{k})$, we find
\begin{multline}
    I_j^{(1)}(\bm{k}) = \sum_{j_d} \dfrac{1}{2\tau_{j,j_d}}
    \Bigg[
    \Big\{
    H_j^{\rm soc}(\bm{k}), \dfrac{\partial}{\partial \varepsilon} \mean{\rho_{j_d}(\varepsilon,\theta,\bm{R},T)}
    \Big\}
    \\
    -\dfrac{\partial}{\partial \varepsilon}
    \mean{
    \Big\{
    H_{j_d}^{\rm soc}(\varepsilon,\theta_d), \rho_{j_d}(\varepsilon,\theta_d,\bm{R},T)
    \Big\}
    }_{\theta_d}
    \Bigg]_{\varepsilon=\varepsilon_j(k)},
\end{multline}
where we have applied the change of variables $\bm{k} \rightarrow (\varepsilon_j(k), \theta)$, and on the second term the average is over $\theta_d$. Since $\mean{H^{\rm soc}} = 0$, the second term yields a negligible coupling to the ``p-wave'' part of $\rho_{j_d}(\varepsilon,\theta_d,\bm{R},T)$. Hereafter we neglect this term, yielding
\begin{align}
    I_j^{(1)}(\bm{k}) &\approx \sum_{j_d} \dfrac{1}{2\tau_{j,j_d}}
    \Big\{
    H_j^{\rm soc}(\bm{k}), \dfrac{\partial}{\partial \varepsilon} \mean{\rho_{j_d}(\varepsilon,\theta,\bm{R},T)}
    \Big\}_{\varepsilon=\varepsilon_j(k)}.
    \label{eqa:I1coll}
\end{align}

\begin{widetext}
\section{Boltzmann equation in the matrix form}
\label{app:matrices}

Here we present the matrices in Eq.~\eqref{eq:matrixform}. The $\mathcal{K}_j$ originates by replacing the Hamiltonian in Eq.~\eqref{eq:Hj} on the left hand side of Eq.~\eqref{eq:boltzmann}. Similarly, the $\mathcal{T}_j$ arises by replacing the SOC Hamiltonian in Eq.\eqref{eq:I1coll}. The $ \mathcal{K}_{j}$ and $\mathcal{T}_j$ matrices are given respectively by 
\begin{multline}
    \label{eqa:kmatrix}
    \mathcal{K}_{j} =
    \dfrac{\tau_0}{\hbar}
    \begin{bmatrix}
    	\dfrac{\hbar \Omega}{\tau_0} & i q_y \lambda_{j,+} & i q_x \lambda_{j,-} & 0
    	\\
    	i q_y \lambda_{j,+} & \dfrac{\hbar \Omega}{\tau_0} & 0 & -2 k_x \lambda_{j,-}
    	\\
    	i q_x \lambda_{j,-} & 0 & \dfrac{\hbar \Omega}{\tau_0} & +2 k_y \lambda_{j,+} 
    	\\
        0 & +2 k_x \lambda_{j,-} & -2 k_y \lambda_{j,+} & \dfrac{\hbar \Omega}{\tau_0}
    \end{bmatrix}
    +\\+\dfrac{\tau_0 \gamma_D}{\hbar}
    \begin{bmatrix}
        0 & 2i[2k_xk_yq_x+q_y(k_x^2-3k_y^2)] & 2i[2k_xk_yq_y+q_x(k_y^2-3k_x^2)] & 0
        \\
        2i[2k_xk_yq_x+q_y(k_x^2-3k_y^2)] & 0 & 0 & 4k_x(k_x^2-k_y^2)
        \\
        2i[2k_xk_yq_y+q_x(k_y^2-3k_x^2)] & 0 & 0 & 4k_y(k_x^2-k_y^2)
        \\
        0 & -4k_x(k_x^2-k_y^2) & -4k_y(k_x^2-k_y^2) & 0
    \end{bmatrix},
\end{multline}
\begin{align}
    \mathcal{T}_j &=
    \begin{bmatrix}
        0 & [\lambda_{j,+} +2(k_x^2-k_y^2)\gamma_D]k_y & (\lambda_{j,-} - (k_x^2-k_y^2)\gamma_D)k_x & 0
        \\
        [\lambda_{j,+} +2(k_x^2-k_y^2)\gamma_D]k_y & 0 & 0 & 0
        \\
        [\lambda_{j,-} - (k_x^2-k_y^2)\gamma_D]k_x & 0 & 0 & 0
        \\
        0 & 0 & 0 & 0
    \end{bmatrix} \dfrac{\partial}{\partial \varepsilon_j(k)},
    \label{eq:Tj}
\end{align}
where $\Omega = \tau_0\partial_t + i \frac{\tau_0 \hbar}{m} \bm{k} \cdot \mathbf{q}$, and $\gamma_D$ is the cubic Dresselhaus SOC coefficient that defines $\beta_{3,j} = \gamma_D k_{F,j}^2/4$.
\end{widetext}

\begin{widetext}
\section{Spin-charge coupling and the spin Hall angle}
\label{app:spincharge}

Here we present the theta matrices in Eq.~\eqref{eq:Djmat}. These matrices are responsible for the spin-charge coupling and are given by
\begin{align}
    \Theta_0 &= 
    \begin{pmatrix}
        -iD_j q_y\Big[ \theta_{+-}^j (Q_{j,-} -3Q_{j,3}) + 2\theta_{3,3}^j [Q_{j,+} - 2Q_{j,-}] \Big]
        \\
        -iD_j q_x\Big[ \theta_{+-}^j (Q_{j,+} -3Q_{j,3}) + 2\theta_{3,3}^j [Q_{j,-} - 2Q_{j,+}] \Big]
        \\
        -i\Big[(\theta_{+-}^j-2\theta_{3,+}^j-2\theta_{3,-}^j-6\theta_{3,3}^{j})(\bm{v} \times \bm{q})_z
        -4(\theta_{3,+}^j-\theta_{3,-}^j)(v_x q_y + v_y q_x)\Big]
    \end{pmatrix},
    \\
    \Theta_1 &= 
    \begin{pmatrix}
        -v_y\Big[\theta_{+-}^{j} (Q_{j,-}-6Q_{j,3})
        +6 \theta_{3,3}^j [Q_{j,+} - 2Q_{j,-}] \Big]
        \\
        -v_x\Big[\theta_{+-}^{j} (Q_{j,+} -6Q_{j,3})
        +6 \theta_{3,3}^j [Q_{j,-} - 2Q_{j,+}] \Big]
        \\
        -12i(\theta_{3+}^j - \theta_{3-}^j)(v_x q_y + v_y q_x)
    \end{pmatrix},
\end{align}
where $\theta_{\mu,\nu}^j = (\hbar \tau_j/2m) Q_{j,\mu} Q_{j,\nu}$ is the spin Hall angle \cite{shen2014theory}.
\end{widetext}

\section{Intersubband relaxation rate}
\label{app:weakstrong}

To establish a criterion to distinguish between the strong- and weak-coupling regimes, we look at the subband relaxation rates $\gamma_j$. In Eq.~\eqref{eq:spindiff}, the $\gamma_j$ enter as diagonal blocks that dictate the coupling between the subbands. If we neglect the matrices $\mathcal{D}_j$ for simplicity, the equation splits into $2\times2$ diagonal blocks as
\begin{align}
    \begin{pmatrix}
      1 & \frac{\phi_{1,2}}{\phi_{1,1}}
      \\
      \frac{\phi_{2,1}}{\phi_{2,2}} & 1
    \end{pmatrix}
    \dfrac{\partial \vec{F}}{\partial t}
    =
    \begin{pmatrix}
      -\gamma_1 & +\gamma_1
      \\
      +\gamma_2 & -\gamma_2
    \end{pmatrix}
    \vec{F},
\end{align}
where $\vec{F}$ is any pair of coupled charge or spin components, i.e., $(N_1, N_2)$, or $(\vec{S}_1, \vec{S}_2)$. The general solution is
\begin{align}
    \vec{F} = c_0 \vec{F}_0 + c_1 e^{-t/t_c} \vec{F}_1,
    \label{eqa:solF}
\end{align}
where $\vec{F}_0 = [1, 1]^T$, and $\vec{F}_1 \approx [1, -1]^T$ for $\phi_{2,2} \approx 1$. The relaxation time is
\begin{align}
    t_c &=
    \dfrac{
        \phi_{2,2} - \phi_{1,2}^2
    }{
        2\phi_{1,2}(1+\phi_{1,2})(\phi_{1,2}+\phi_{2,2})
    } \tau_0
    \label{eqa:tc}
\end{align}
For $\phi_{1,2} \rightarrow 0$, $t_c \rightarrow \infty$ and the linear combination of $\vec{F}_0 \pm \vec{F}_1$ in Eq.~\eqref{eqa:solF} indicates that the dynamics of the first and second subbands can be decoupled as in the weak-coupling regime. On the other hand, for $\phi_{1,2} \rightarrow \sqrt{\phi_{2,2}}$ we see that $t_c \rightarrow 0$, which tell us that the dynamics quickly falls into the subband-averaged strong-coupling regime. Therefore, $t_c$ defines how fast the dynamics relaxes towards the strong-coupling regime, as discussed in the main text.

\section{Approximate expressions for the spin relaxation times}
\label{app:spinrelax}

To find analytical solutions for the spin lifetimes, we neglect the small spin-charge coupling and the drift velocity in Eq.~\eqref{eq:Djmat}. This simplifies the spin-diffusion equations for the weak and strong coupling, Eqs.~\eqref{eq:spindiffNST} and \eqref{eq:spindiff0}, which take the $3\times3$ form
\begin{align}
    \dfrac{\partial \vec{\mathcal{S}}}{\partial t} &=
    -\mathcal{D} \vec{\mathcal{S}},
    \label{eqa:gendiff}
\end{align}
where $\mathcal{D} = D q^2 + \mathcal{D}^S$, with
\begin{align}
    \mathcal{D}^S
    &= 
    \begin{pmatrix}
      \dfrac{1}{\tau_-} &
      0 & 
      +2iDQ_-^*q_x
    \\
      0 &
      \dfrac{1}{\tau_+} &
      -2iDQ_+^*q_y
    \\
      -2iDQ_-^*q_x &
      +2iDQ_+^*q_y &
      \dfrac{1}{\tau_-} + \dfrac{1}{\tau_+}
    \end{pmatrix}.
    \label{eqa:genDmat}
\end{align}
Notice that here these are written in a generic form. Namely, for the weak subband regime one must consider the independent subbands by adding the $j$ index to the quantities above: $\vec{\mathcal{S}}_j$, $D_j$, $Q_{j,\pm}^*$, $\tau_{j,\pm}$. On the other hand, for the strong-coupling regime, $\mathcal{D}$ must be taken as the subband-averaged diffusion matrix, yielding 
\begin{align}
    \vec{\mathcal{S}} & \rightarrow \dfrac{\vec{\mathcal{S}}_1 + \vec{\mathcal{S}}_2}{2},
    \\
    D &\rightarrow \dfrac{D_1+D_2}{2},
    \\
    Q_{\pm}^* &\rightarrow \dfrac{D_1 Q_{1,\pm}^* + D_2 Q_{2,\pm}^*}{D_1+D_2},
    \\
    \dfrac{1}{\tau_{\pm}} &\rightarrow \dfrac{1}{2}\left(\dfrac{1}{\tau_{1,\pm}} + \dfrac{1}{\tau_{2,\pm}}\right).
\end{align}
In this section we proceed with the generic form of Eqs.~\eqref{eqa:gendiff} and \eqref{eqa:genDmat}. 

A generic solution of Eq.~\eqref{eqa:gendiff} can be written in terms of its eigenvalues as
\begin{align}
    \vec{\mathcal{S}}(\bm{q},t) &= \sum_n c_n e^{-\omega_n t} \phi_n,
\end{align}
where $\omega_n \equiv \omega_n(\bm{q})$ and $\phi_n \equiv \phi_n(\bm{q})$ are the eigenvalues and eigenvectors of $\mathcal{D}$, and $c_n \equiv c_n(\bm{q})$ are the linear combination coefficients to be set by the initial conditions. For the generic matrix $\mathcal{D}$ in Eq.~\eqref{eqa:genDmat} the eigenvalues and eigenvectors expressions are cumbersome, but simple analytical solutions exist for the cases where: (i) $Q_{-}^* = 0$ or $Q_{+}^* =0$; and (ii) $Q_{-}^* = Q_{+}^* \equiv Q^*$ and $\tau_+ = \tau_- \equiv \tau$. These solutions are well known \cite{Bernevig2006PSH, Stanescu2007}, and below we simply revise and write them in a convenient and generic form.

To obtain the spin lifetimes in each of these cases above, we seek for the the $\bm{q}$ that extremizes $\Re[\omega_n(\bm{q})]$. For instance, let us consider the first case of $Q_{-}^* = 0$ and focus on the decoupled $2\times2$ block of $\mathcal{D}$. Its eigenvalues are
\begin{align}
    \omega_\pm(\bm{q}) &= -\dfrac{1}{2\tau_-} - \dfrac{1}{\tau_+} - Dq^2 \pm \dfrac{\sqrt{1 + 16 (Q_+^* \tau_- D q_y)^2}}{2\tau_-},
\end{align}
which are maximum (least negative) at $\bm{q}^0 = (0, \pm q_y^0)$, with $q_y^0 = \sqrt{16 (Q_+^*)^4(D \tau_-)^2 - 1}/(4 Q_+^* D \tau_-)$.
From these, the spin lifetime is defined as $\tau_S = -1/\omega_+(\bm{q}^0)$. Namely,
\begin{align}
    \dfrac{1}{\tau_S^{(i)}} &\approx
    \dfrac{1}{2 \tau_{\mp}}
    + \dfrac{1}{\tau_{\pm}}
    - \dfrac{1}{(4Q^*_{\pm} \tau_{\mp})^2 D}
    - (Q_{\pm}^*)^2 D,
    \label{eqa:tauSi}
\end{align}
where the top and bottom signs refer to the cases of $Q^*_- = 0$ or $Q^*_+ = 0$, respectively. Similarly, for the case (ii) we find
\begin{align}
    \dfrac{1}{\tau_S^{(ii)}} &\approx
    \dfrac{3}{2 \tau}
    - \dfrac{1}{(4Q^* \tau)^2 D}
    - (Q^*)^2 D.
    \label{eqa:tauSii}
\end{align}
Notice that the expression for $\tau_S^{(i)}$ reduces to $\tau_S^{(ii)}$ in the appropriate limit of $Q_{-}^* = Q_{+}^* \equiv Q^*$ and $\tau_+ = \tau_- \equiv \tau$. Therefore, it is sufficient to consider only Eq.~\eqref{eqa:tauSi} as a generic expression for the spin lifetime that incorporates both cases.

In the single-subband case, or in the weak-coupling regime, case (i) corresponds to the PSH regime and $\tau_S^{\rm (i)}$ simplifies to
\begin{align}
    \dfrac{1}{\tau_S^{{\rm PSH}_j^\pm}} &=
    \dfrac{2m^2 D_j}{\hbar^4} 
    \Big[
        3\beta_{j,3}^2 + (\beta_{j}^* \pm \alpha_j)\red{^2}
    \Big].
    \label{eqa:tauSPSH}
\end{align}

\subsection{Expressions for the spin dynamics}

The full expressions for $S_z(\bm{q},t)$ or its Fourier transform $S_z(\bm{r},t) = \mathcal{F}[S_z(\bm{q},t)]$ are also well known \cite{Bernevig2006PSH, Stanescu2007}. For the case (i) with $Q^*_- = 0$, the solution in $q$ space have two poles at $\bm{q} = \pm (0, q_y^0)$, which leads to the stripped pattern in the $S_z(\bm{r},t)$ as these poles form a cosine function in the Fourier transform, which reads as
\begin{align}
    \label{eqa:SzPSH}
    S_z(\bm{r},t) &\approx \dfrac{e^{-\frac{r^2}{2\Gamma^2}}}{\Gamma^2} e^{-t/\tau_S} \cos(\kappa y),
\end{align}
where the diffusion broadening $\Gamma = \sqrt{2 D t}$, the spin lifetime $\tau_S$ is given by Eq.~\eqref{eqa:tauSi}, and $\kappa \approx Q_+^*$. Similarly, for $Q^*_+ = 0$ the solution follows from the one above replacing the last term by $\cos(\kappa x)$, with $\kappa \approx Q^*_-$. 

For the isotropic limit of case (ii) above, with $Q_{-}^* = Q_{+}^* \equiv Q^*$ and $\tau_+ = \tau_- \equiv \tau$, an approximate solution for $S_z(\bm{r},t)$ in the asymptotic limit of large times $t \gg \tau_S$ is shown in Ref.~\cite{Stanescu2007} [see their Eqs.~(28)-(30)]. It is also possible to obtain an exact solution for $S_z(\bm{r}=0,t)$ by direct integration of their Eq.~(29) with $\bm{r}=0$, which reads as
\begin{align}
    \label{eqa:SzSG}
    S_z(0, t) &= \dfrac{e^{-2t/\tau} + e^{-t/\tau}}{16\pi t/\tau}
    + \zeta(t)
    \dfrac{e^{-t/\tau_{S}}}{\sqrt{\pi t/\tau}},
    \\
    \zeta(t) &=
    \dfrac{(\gamma^2-1)}{32 \gamma}
    \sum_\pm 
    \erf\left(\frac{\left(\gamma^2 \pm 1\right)}{2 \gamma }\sqrt{\dfrac{t}{\tau}}\right),
\end{align}
where $\tau_S \equiv \tau_S^{(ii)}$ from Eq.~\eqref{eqa:tauSii}, and $\gamma = 2 Q^* \sqrt{D \tau}$.

Notice that at $\bm{r}=0$ the denominator of Eq.~\eqref{eqa:SzPSH} is proportional to $t$, while in Eq.~\eqref{eqa:SzSG} it has $t$ and $\sqrt{t}$ contributions at short times, and asymptotically approaches $\sqrt{t}$ as the second term dominates for large $t$. These features are shown and discussed in Fig.~\ref{fig:lifetime} in the main text. Particularly, for the weak-coupling case of Fig.~\ref{fig:lifetime}(c), the solution is the sum of the contributions for each subband, each in the form of Eq.~\eqref{eqa:SzPSH}, which at $\bm{r}=0$ reads as
\begin{align}
    \label{eqa:Sz12}
    S_z(0, t) &= \dfrac{e^{-t/\tau_1}}{2D_1 t} + \dfrac{e^{-t/\tau_2}}{2D_2 t}.
\end{align}

\bibliography{references} 

\begin{thebibliography}{59}%
\makeatletter
\providecommand \@ifxundefined [1]{%
 \@ifx{#1\undefined}
}%
\providecommand \@ifnum [1]{%
 \ifnum #1\expandafter \@firstoftwo
 \else \expandafter \@secondoftwo
 \fi
}%
\providecommand \@ifx [1]{%
 \ifx #1\expandafter \@firstoftwo
 \else \expandafter \@secondoftwo
 \fi
}%
\providecommand \natexlab [1]{#1}%
\providecommand \enquote  [1]{``#1''}%
\providecommand \bibnamefont  [1]{#1}%
\providecommand \bibfnamefont [1]{#1}%
\providecommand \citenamefont [1]{#1}%
\providecommand \href@noop [0]{\@secondoftwo}%
\providecommand \href [0]{\begingroup \@sanitize@url \@href}%
\providecommand \@href[1]{\@@startlink{#1}\@@href}%
\providecommand \@@href[1]{\endgroup#1\@@endlink}%
\providecommand \@sanitize@url [0]{\catcode `\\12\catcode `\$12\catcode
  `\&12\catcode `\#12\catcode `\^12\catcode `\_12\catcode `\%12\relax}%
\providecommand \@@startlink[1]{}%
\providecommand \@@endlink[0]{}%
\providecommand \url  [0]{\begingroup\@sanitize@url \@url }%
\providecommand \@url [1]{\endgroup\@href {#1}{\urlprefix }}%
\providecommand \urlprefix  [0]{URL }%
\providecommand \Eprint [0]{\href }%
\providecommand \doibase [0]{https://doi.org/}%
\providecommand \selectlanguage [0]{\@gobble}%
\providecommand \bibinfo  [0]{\@secondoftwo}%
\providecommand \bibfield  [0]{\@secondoftwo}%
\providecommand \translation [1]{[#1]}%
\providecommand \BibitemOpen [0]{}%
\providecommand \bibitemStop [0]{}%
\providecommand \bibitemNoStop [0]{.\EOS\space}%
\providecommand \EOS [0]{\spacefactor3000\relax}%
\providecommand \BibitemShut  [1]{\csname bibitem#1\endcsname}%
\let\auto@bib@innerbib\@empty
\bibitem [{\citenamefont {Datta}\ and\ \citenamefont {Das}(1990)}]{DattaDas}%
  \BibitemOpen
  \bibfield  {author} {\bibinfo {author} {\bibfnamefont {S.}~\bibnamefont
  {Datta}}\ and\ \bibinfo {author} {\bibfnamefont {B.}~\bibnamefont {Das}},\
  }\bibfield  {title} {\bibinfo {title} {Electronic analog of the
  electro‐optic modulator},\ }\href {https://doi.org/10.1063/1.102730}
  {\bibfield  {journal} {\bibinfo  {journal} {Appl. Phys. Lett.}\ }\textbf
  {\bibinfo {volume} {56}},\ \bibinfo {pages} {665} (\bibinfo {year}
  {1990})}\BibitemShut {NoStop}%
\bibitem [{\citenamefont {Chuang}\ \emph {et~al.}(2014)\citenamefont {Chuang},
  \citenamefont {Ho}, \citenamefont {Smith}, \citenamefont {Sfigakis},
  \citenamefont {Pepper}, \citenamefont {Chen}, \citenamefont {Fan},
  \citenamefont {Griffiths}, \citenamefont {Farrer}, \citenamefont {Beere},
  \citenamefont {Jones}, \citenamefont {Ritchie},\ and\ \citenamefont
  {Chen}}]{Chuang2014SFET}%
  \BibitemOpen
  \bibfield  {author} {\bibinfo {author} {\bibfnamefont {P.}~\bibnamefont
  {Chuang}}, \bibinfo {author} {\bibfnamefont {S.-C.}\ \bibnamefont {Ho}},
  \bibinfo {author} {\bibfnamefont {L.~W.}\ \bibnamefont {Smith}}, \bibinfo
  {author} {\bibfnamefont {F.}~\bibnamefont {Sfigakis}}, \bibinfo {author}
  {\bibfnamefont {M.}~\bibnamefont {Pepper}}, \bibinfo {author} {\bibfnamefont
  {C.-H.}\ \bibnamefont {Chen}}, \bibinfo {author} {\bibfnamefont {J.-C.}\
  \bibnamefont {Fan}}, \bibinfo {author} {\bibfnamefont {J.~P.}\ \bibnamefont
  {Griffiths}}, \bibinfo {author} {\bibfnamefont {I.}~\bibnamefont {Farrer}},
  \bibinfo {author} {\bibfnamefont {H.~E.}\ \bibnamefont {Beere}}, \bibinfo
  {author} {\bibfnamefont {G.~A.~C.}\ \bibnamefont {Jones}}, \bibinfo {author}
  {\bibfnamefont {D.~A.}\ \bibnamefont {Ritchie}},\ and\ \bibinfo {author}
  {\bibfnamefont {T.-M.}\ \bibnamefont {Chen}},\ }\bibfield  {title} {\bibinfo
  {title} {All-electric all-semiconductor spin field-effect transistors},\
  }\href {https://doi.org/10.1038/nnano.2014.296} {\bibfield  {journal}
  {\bibinfo  {journal} {Nat. Nanotechnol.}\ }\textbf {\bibinfo {volume} {10}},\
  \bibinfo {pages} {35} (\bibinfo {year} {2014})}\BibitemShut {NoStop}%
\bibitem [{\citenamefont {Kikkawa}\ and\ \citenamefont
  {Awschalom}(1999)}]{Kikkawa1999SpinCoherence}%
  \BibitemOpen
  \bibfield  {author} {\bibinfo {author} {\bibfnamefont {J.~M.}\ \bibnamefont
  {Kikkawa}}\ and\ \bibinfo {author} {\bibfnamefont {D.~D.}\ \bibnamefont
  {Awschalom}},\ }\bibfield  {title} {\bibinfo {title} {Lateral drag of spin
  coherence in gallium arsenide},\ }\href {https://doi.org/10.1038/16420}
  {\bibfield  {journal} {\bibinfo  {journal} {Nature (London)}\ }\textbf
  {\bibinfo {volume} {397}},\ \bibinfo {pages} {139} (\bibinfo {year}
  {1999})}\BibitemShut {NoStop}%
\bibitem [{\citenamefont {Wolf}(2001)}]{Wolf2001Spintronics}%
  \BibitemOpen
  \bibfield  {author} {\bibinfo {author} {\bibfnamefont {S.~A.}\ \bibnamefont
  {Wolf}},\ }\bibfield  {title} {\bibinfo {title} {Spintronics: A spin-based
  electronics vision for the future},\ }\href
  {https://doi.org/10.1126/science.1065389} {\bibfield  {journal} {\bibinfo
  {journal} {Science}\ }\textbf {\bibinfo {volume} {294}},\ \bibinfo {pages}
  {1488} (\bibinfo {year} {2001})}\BibitemShut {NoStop}%
\bibitem [{\citenamefont {{\ifmmode \check{Z}\else {\v Z}\fi{}uti\ifmmode
  \acute{c}\else {\'c}\fi{}}}\ \emph {et~al.}(2004)\citenamefont {{\ifmmode
  \check{Z}\else {\v Z}\fi{}uti\ifmmode \acute{c}\else {\'c}\fi{}}},
  \citenamefont {Fabian},\ and\ \citenamefont {{Das
  Sarma}}}]{Zutic2004Spintronics}%
  \BibitemOpen
  \bibfield  {author} {\bibinfo {author} {\bibfnamefont {I.}~\bibnamefont
  {{\ifmmode \check{Z}\else {\v Z}\fi{}uti\ifmmode \acute{c}\else
  {\'c}\fi{}}}}, \bibinfo {author} {\bibfnamefont {J.}~\bibnamefont {Fabian}},\
  and\ \bibinfo {author} {\bibfnamefont {S.}~\bibnamefont {{Das Sarma}}},\
  }\bibfield  {title} {\bibinfo {title} {{Spintronics: Fundamentals and
  applications}},\ }\href {https://doi.org/10.1103/RevModPhys.76.323}
  {\bibfield  {journal} {\bibinfo  {journal} {Rev. Mod. Phys.}\ }\textbf
  {\bibinfo {volume} {76}},\ \bibinfo {pages} {323} (\bibinfo {year}
  {2004})}\BibitemShut {NoStop}%
\bibitem [{\citenamefont {Fabian}\ \emph {et~al.}(2007)\citenamefont {Fabian},
  \citenamefont {Matos-Abiague}, \citenamefont {Ertler}, \citenamefont
  {Stano},\ and\ \citenamefont {{\v Z}uti{\'c}}}]{Fabian2007Review}%
  \BibitemOpen
  \bibfield  {author} {\bibinfo {author} {\bibfnamefont {J.}~\bibnamefont
  {Fabian}}, \bibinfo {author} {\bibfnamefont {A.}~\bibnamefont
  {Matos-Abiague}}, \bibinfo {author} {\bibfnamefont {C.}~\bibnamefont
  {Ertler}}, \bibinfo {author} {\bibfnamefont {P.}~\bibnamefont {Stano}},\ and\
  \bibinfo {author} {\bibfnamefont {I.}~\bibnamefont {{\v Z}uti{\'c}}},\
  }\bibfield  {title} {\bibinfo {title} {Semiconductor spintronics},\ }\href
  {https://doi.org/10.2478/v10155-010-0086-8} {\bibfield  {journal} {\bibinfo
  {journal} {Acta Phys. Slovaca}\ }\textbf {\bibinfo {volume} {57}},\ \bibinfo
  {pages} {565} (\bibinfo {year} {2007})}\BibitemShut {NoStop}%
\bibitem [{\citenamefont {Awschalom}\ and\ \citenamefont
  {Flatt{\'e}}(2007)}]{Awschalom2007Spintronics}%
  \BibitemOpen
  \bibfield  {author} {\bibinfo {author} {\bibfnamefont {D.~D.}\ \bibnamefont
  {Awschalom}}\ and\ \bibinfo {author} {\bibfnamefont {M.~E.}\ \bibnamefont
  {Flatt{\'e}}},\ }\bibfield  {title} {\bibinfo {title} {Challenges for
  semiconductor spintronics},\ }\href {https://doi.org/10.1038/nphys551}
  {\bibfield  {journal} {\bibinfo  {journal} {Nat. Phys.}\ }\textbf {\bibinfo
  {volume} {3}},\ \bibinfo {pages} {153} (\bibinfo {year} {2007})}\BibitemShut
  {NoStop}%
\bibitem [{\citenamefont {Wu}\ \emph {et~al.}(2010)\citenamefont {Wu},
  \citenamefont {Jiang},\ and\ \citenamefont {Weng}}]{Wu2010ReviewSpin}%
  \BibitemOpen
  \bibfield  {author} {\bibinfo {author} {\bibfnamefont {M.}~\bibnamefont
  {Wu}}, \bibinfo {author} {\bibfnamefont {J.}~\bibnamefont {Jiang}},\ and\
  \bibinfo {author} {\bibfnamefont {M.}~\bibnamefont {Weng}},\ }\bibfield
  {title} {\bibinfo {title} {Spin dynamics in semiconductors},\ }\href
  {https://doi.org/10.1016/j.physrep.2010.04.002} {\bibfield  {journal}
  {\bibinfo  {journal} {Phys. Rep.}\ }\textbf {\bibinfo {volume} {493}},\
  \bibinfo {pages} {61} (\bibinfo {year} {2010})}\BibitemShut {NoStop}%
\bibitem [{\citenamefont {Manchon}\ \emph {et~al.}(2015)\citenamefont
  {Manchon}, \citenamefont {Koo}, \citenamefont {Nitta}, \citenamefont
  {Frolov},\ and\ \citenamefont {Duine}}]{Manchon2015PerspectRashba}%
  \BibitemOpen
  \bibfield  {author} {\bibinfo {author} {\bibfnamefont {A.}~\bibnamefont
  {Manchon}}, \bibinfo {author} {\bibfnamefont {H.~C.}\ \bibnamefont {Koo}},
  \bibinfo {author} {\bibfnamefont {J.}~\bibnamefont {Nitta}}, \bibinfo
  {author} {\bibfnamefont {S.~M.}\ \bibnamefont {Frolov}},\ and\ \bibinfo
  {author} {\bibfnamefont {R.~A.}\ \bibnamefont {Duine}},\ }\bibfield  {title}
  {\bibinfo {title} {{New perspectives for Rashba spin--orbit coupling}},\
  }\href {https://doi.org/10.1038/nmat4360} {\bibfield  {journal} {\bibinfo
  {journal} {Nat. Mater.}\ }\textbf {\bibinfo {volume} {14}},\ \bibinfo {pages}
  {871} (\bibinfo {year} {2015})}\BibitemShut {NoStop}%
\bibitem [{\citenamefont {Winkler}(2003)}]{Winkler2003book}%
  \BibitemOpen
  \bibfield  {author} {\bibinfo {author} {\bibfnamefont {R.}~\bibnamefont
  {Winkler}},\ }\href {https://doi.org/10.1007/b13586} {\emph {\bibinfo {title}
  {Spin--Orbit Coupling Effects in Two-Dimensional Electron and Hole
  Systems}}}\ (\bibinfo  {publisher} {Springer Berlin Heidelberg},\ \bibinfo
  {year} {2003})\BibitemShut {NoStop}%
\bibitem [{\citenamefont {Dresselhaus}(1955)}]{Dresselhaus55}%
  \BibitemOpen
  \bibfield  {author} {\bibinfo {author} {\bibfnamefont {G.}~\bibnamefont
  {Dresselhaus}},\ }\bibfield  {title} {\bibinfo {title} {{Spin-Orbit Coupling
  Effects in Zinc Blende Structures}},\ }\href
  {https://doi.org/10.1103/PhysRev.100.580} {\bibfield  {journal} {\bibinfo
  {journal} {Phys. Rev.}\ }\textbf {\bibinfo {volume} {100}},\ \bibinfo {pages}
  {580} (\bibinfo {year} {1955})}\BibitemShut {NoStop}%
\bibitem [{\citenamefont {Bychkov}\ and\ \citenamefont
  {Rashba}(1984)}]{Rashba84}%
  \BibitemOpen
  \bibfield  {author} {\bibinfo {author} {\bibfnamefont {Y.~A.}\ \bibnamefont
  {Bychkov}}\ and\ \bibinfo {author} {\bibfnamefont {E.~I.}\ \bibnamefont
  {Rashba}},\ }\bibfield  {title} {\bibinfo {title} {Oscillatory effects and
  the magnetic susceptibility of carriers in inversion layers},\ }\href
  {https://doi.org/10.1088/0022-3719/17/33/015} {\bibfield  {journal} {\bibinfo
   {journal} {J. Phys. C: Solid State Phys.}\ }\textbf {\bibinfo {volume}
  {17}},\ \bibinfo {pages} {6039} (\bibinfo {year} {1984})}\BibitemShut
  {NoStop}%
\bibitem [{DP7()}]{DP71a}%
  \BibitemOpen
  \href@noop {} {}\bibinfo {note} {M. I. D’yakonov and V. I. Perel’,
  Possibility of orienting electron spins with current,
  \href{http://jetpletters.ru/ps/720/article_11168.shtml}{ZhETF Pis. Red.
  \textbf{13}, 657 (1971)} [English version:
  \href{http://jetpletters.ru/ps/1587/article_24366.shtml}{Sov. Phys. JETP
  Lett. \textbf{13}, 467 (1971)}]; Current-induced spin orientation of
  electrons in semiconductors,
  \href{https://doi.org/10.1016/0375-9601(71)90196-4}{Phys. Lett. A
  \textbf{35}, 459 (1971)}.}\BibitemShut {Stop}%
\bibitem [{\citenamefont {Elliott}(1954)}]{Elliott1954EY}%
  \BibitemOpen
  \bibfield  {author} {\bibinfo {author} {\bibfnamefont {R.~J.}\ \bibnamefont
  {Elliott}},\ }\bibfield  {title} {\bibinfo {title} {Theory of the effect of
  spin-orbit coupling on magnetic resonance in some semiconductors},\ }\href
  {https://doi.org/10.1103/PhysRev.96.266} {\bibfield  {journal} {\bibinfo
  {journal} {Phys. Rev.}\ }\textbf {\bibinfo {volume} {96}},\ \bibinfo {pages}
  {266} (\bibinfo {year} {1954})}\BibitemShut {NoStop}%
\bibitem [{\citenamefont {Yafet}(1963)}]{Yafet1963EY}%
  \BibitemOpen
  \bibfield  {author} {\bibinfo {author} {\bibfnamefont {Y.}~\bibnamefont
  {Yafet}},\ }\bibfield  {title} {\bibinfo {title} {g factors and spin-lattice
  relaxation of conduction electrons},\ }in\ \href
  {https://doi.org/10.1016/s0081-1947(08)60259-3} {\emph {\bibinfo {booktitle}
  {Solid State Physics}}}\ (\bibinfo  {publisher} {Elsevier, Amsterdam},\
  \bibinfo {year} {1963})\ pp.\ \bibinfo {pages} {1--98}\BibitemShut {NoStop}%
\bibitem [{\citenamefont {Schliemann}\ \emph {et~al.}(2003)\citenamefont
  {Schliemann}, \citenamefont {Egues},\ and\ \citenamefont
  {Loss}}]{Schliemann2003}%
  \BibitemOpen
  \bibfield  {author} {\bibinfo {author} {\bibfnamefont {J.}~\bibnamefont
  {Schliemann}}, \bibinfo {author} {\bibfnamefont {J.~C.}\ \bibnamefont
  {Egues}},\ and\ \bibinfo {author} {\bibfnamefont {D.}~\bibnamefont {Loss}},\
  }\bibfield  {title} {\bibinfo {title} {Nonballistic spin-field-effect
  transistor},\ }\href {https://doi.org/10.1103/PhysRevLett.90.146801}
  {\bibfield  {journal} {\bibinfo  {journal} {Phys. Rev. Lett.}\ }\textbf
  {\bibinfo {volume} {90}},\ \bibinfo {pages} {146801} (\bibinfo {year}
  {2003})}\BibitemShut {NoStop}%
\bibitem [{\citenamefont {Bernevig}\ \emph {et~al.}(2006)\citenamefont
  {Bernevig}, \citenamefont {Orenstein},\ and\ \citenamefont
  {Zhang}}]{Bernevig2006PSH}%
  \BibitemOpen
  \bibfield  {author} {\bibinfo {author} {\bibfnamefont {B.~A.}\ \bibnamefont
  {Bernevig}}, \bibinfo {author} {\bibfnamefont {J.}~\bibnamefont
  {Orenstein}},\ and\ \bibinfo {author} {\bibfnamefont {S.-C.}\ \bibnamefont
  {Zhang}},\ }\bibfield  {title} {\bibinfo {title} {{Exact SU(2) Symmetry and
  Persistent Spin Helix in a Spin-Orbit Coupled System}},\ }\href
  {https://doi.org/10.1103/PhysRevLett.97.236601} {\bibfield  {journal}
  {\bibinfo  {journal} {Phys. Rev. Lett.}\ }\textbf {\bibinfo {volume} {97}},\
  \bibinfo {pages} {236601} (\bibinfo {year} {2006})}\BibitemShut {NoStop}%
\bibitem [{\citenamefont {Koralek}\ \emph {et~al.}(2009)\citenamefont
  {Koralek}, \citenamefont {Weber}, \citenamefont {Orenstein}, \citenamefont
  {Bernevig}, \citenamefont {Zhang}, \citenamefont {Mack},\ and\ \citenamefont
  {Awschalom}}]{koralek2009emergence}%
  \BibitemOpen
  \bibfield  {author} {\bibinfo {author} {\bibfnamefont {J.~D.}\ \bibnamefont
  {Koralek}}, \bibinfo {author} {\bibfnamefont {C.~P.}\ \bibnamefont {Weber}},
  \bibinfo {author} {\bibfnamefont {J.}~\bibnamefont {Orenstein}}, \bibinfo
  {author} {\bibfnamefont {B.~A.}\ \bibnamefont {Bernevig}}, \bibinfo {author}
  {\bibfnamefont {S.-C.}\ \bibnamefont {Zhang}}, \bibinfo {author}
  {\bibfnamefont {S.}~\bibnamefont {Mack}},\ and\ \bibinfo {author}
  {\bibfnamefont {D.}~\bibnamefont {Awschalom}},\ }\bibfield  {title} {\bibinfo
  {title} {Emergence of the persistent spin helix in semiconductor quantum
  wells},\ }\href {https://doi.org/10.1038/nature07871} {\bibfield  {journal}
  {\bibinfo  {journal} {Nature (London)}\ }\textbf {\bibinfo {volume} {458}},\
  \bibinfo {pages} {610} (\bibinfo {year} {2009})}\BibitemShut {NoStop}%
\bibitem [{\citenamefont {Weber}\ \emph {et~al.}(2007)\citenamefont {Weber},
  \citenamefont {Orenstein}, \citenamefont {Bernevig}, \citenamefont {Zhang},
  \citenamefont {Stephens},\ and\ \citenamefont
  {Awschalom}}]{weber2007nondiffusive}%
  \BibitemOpen
  \bibfield  {author} {\bibinfo {author} {\bibfnamefont {C.~P.}\ \bibnamefont
  {Weber}}, \bibinfo {author} {\bibfnamefont {J.}~\bibnamefont {Orenstein}},
  \bibinfo {author} {\bibfnamefont {B.~A.}\ \bibnamefont {Bernevig}}, \bibinfo
  {author} {\bibfnamefont {S.-C.}\ \bibnamefont {Zhang}}, \bibinfo {author}
  {\bibfnamefont {J.}~\bibnamefont {Stephens}},\ and\ \bibinfo {author}
  {\bibfnamefont {D.~D.}\ \bibnamefont {Awschalom}},\ }\bibfield  {title}
  {\bibinfo {title} {Nondiffusive spin dynamics in a two-dimensional electron
  gas},\ }\href {https://doi.org/10.1103/PhysRevLett.98.076604} {\bibfield
  {journal} {\bibinfo  {journal} {Phys. Rev. Lett.}\ }\textbf {\bibinfo
  {volume} {98}},\ \bibinfo {pages} {076604} (\bibinfo {year}
  {2007})}\BibitemShut {NoStop}%
\bibitem [{\citenamefont {Walser}\ \emph {et~al.}(2012)\citenamefont {Walser},
  \citenamefont {Reichl}, \citenamefont {Wegscheider},\ and\ \citenamefont
  {Salis}}]{walser2012direct}%
  \BibitemOpen
  \bibfield  {author} {\bibinfo {author} {\bibfnamefont {M.}~\bibnamefont
  {Walser}}, \bibinfo {author} {\bibfnamefont {C.}~\bibnamefont {Reichl}},
  \bibinfo {author} {\bibfnamefont {W.}~\bibnamefont {Wegscheider}},\ and\
  \bibinfo {author} {\bibfnamefont {G.}~\bibnamefont {Salis}},\ }\bibfield
  {title} {\bibinfo {title} {Direct mapping of the formation of a persistent
  spin helix},\ }\href {https://doi.org/10.1038/nphys2383} {\bibfield
  {journal} {\bibinfo  {journal} {Nat. Phys.}\ }\textbf {\bibinfo {volume}
  {8}},\ \bibinfo {pages} {757} (\bibinfo {year} {2012})}\BibitemShut {NoStop}%
\bibitem [{\citenamefont {Ishihara}\ \emph {et~al.}(2013)\citenamefont
  {Ishihara}, \citenamefont {Ohno},\ and\ \citenamefont
  {Ohno}}]{Ishihara2013DirectImage}%
  \BibitemOpen
  \bibfield  {author} {\bibinfo {author} {\bibfnamefont {J.}~\bibnamefont
  {Ishihara}}, \bibinfo {author} {\bibfnamefont {Y.}~\bibnamefont {Ohno}},\
  and\ \bibinfo {author} {\bibfnamefont {H.}~\bibnamefont {Ohno}},\ }\bibfield
  {title} {\bibinfo {title} {Direct imaging of gate-controlled persistent spin
  helix state in a modulation-doped {GaAs}/{AlGaAs} quantum well},\ }\href
  {https://doi.org/10.7567/apex.7.013001} {\bibfield  {journal} {\bibinfo
  {journal} {Appl. Phys. Express}\ }\textbf {\bibinfo {volume} {7}},\ \bibinfo
  {pages} {013001} (\bibinfo {year} {2013})}\BibitemShut {NoStop}%
\bibitem [{\citenamefont {Mishchenko}\ and\ \citenamefont
  {Halperin}(2003)}]{Mishchenko2003}%
  \BibitemOpen
  \bibfield  {author} {\bibinfo {author} {\bibfnamefont {E.~G.}\ \bibnamefont
  {Mishchenko}}\ and\ \bibinfo {author} {\bibfnamefont {B.~I.}\ \bibnamefont
  {Halperin}},\ }\bibfield  {title} {\bibinfo {title} {Transport equations for
  a two-dimensional electron gas with spin-orbit interaction},\ }\href
  {https://doi.org/10.1103/PhysRevB.68.045317} {\bibfield  {journal} {\bibinfo
  {journal} {Phys. Rev. B}\ }\textbf {\bibinfo {volume} {68}},\ \bibinfo
  {pages} {045317} (\bibinfo {year} {2003})}\BibitemShut {NoStop}%
\bibitem [{\citenamefont {Mishchenko}\ \emph {et~al.}(2004)\citenamefont
  {Mishchenko}, \citenamefont {Shytov},\ and\ \citenamefont
  {Halperin}}]{Mishchenko2004Kinetic}%
  \BibitemOpen
  \bibfield  {author} {\bibinfo {author} {\bibfnamefont {E.~G.}\ \bibnamefont
  {Mishchenko}}, \bibinfo {author} {\bibfnamefont {A.~V.}\ \bibnamefont
  {Shytov}},\ and\ \bibinfo {author} {\bibfnamefont {B.~I.}\ \bibnamefont
  {Halperin}},\ }\bibfield  {title} {\bibinfo {title} {Spin current and
  polarization in impure two-dimensional electron systems with spin-orbit
  coupling},\ }\href {https://doi.org/10.1103/PhysRevLett.93.226602} {\bibfield
   {journal} {\bibinfo  {journal} {Phys. Rev. Lett.}\ }\textbf {\bibinfo
  {volume} {93}},\ \bibinfo {pages} {226602} (\bibinfo {year}
  {2004})}\BibitemShut {NoStop}%
\bibitem [{\citenamefont {Saikin}(2004)}]{Saikin2004}%
  \BibitemOpen
  \bibfield  {author} {\bibinfo {author} {\bibfnamefont {S.}~\bibnamefont
  {Saikin}},\ }\bibfield  {title} {\bibinfo {title} {A drift-diffusion model
  for spin-polarized transport in a two-dimensional non-degenerate electron gas
  controlled by spin{\textendash}orbit interaction},\ }\href
  {https://doi.org/10.1088/0953-8984/16/28/025} {\bibfield  {journal} {\bibinfo
   {journal} {J. Phys.: Condens. Matter}\ }\textbf {\bibinfo {volume} {16}},\
  \bibinfo {pages} {5071} (\bibinfo {year} {2004})}\BibitemShut {NoStop}%
\bibitem [{\citenamefont {Rammer}\ and\ \citenamefont
  {Smith}(1986)}]{Rammer1986RMP}%
  \BibitemOpen
  \bibfield  {author} {\bibinfo {author} {\bibfnamefont {J.}~\bibnamefont
  {Rammer}}\ and\ \bibinfo {author} {\bibfnamefont {H.}~\bibnamefont {Smith}},\
  }\bibfield  {title} {\bibinfo {title} {Quantum field-theoretical methods in
  transport theory of metals},\ }\href
  {https://doi.org/10.1103/RevModPhys.58.323} {\bibfield  {journal} {\bibinfo
  {journal} {Rev. Mod. Phys.}\ }\textbf {\bibinfo {volume} {58}},\ \bibinfo
  {pages} {323} (\bibinfo {year} {1986})}\BibitemShut {NoStop}%
\bibitem [{\citenamefont {Rammer}(2007)}]{rammer2007quantum}%
  \BibitemOpen
  \bibfield  {author} {\bibinfo {author} {\bibfnamefont {J.}~\bibnamefont
  {Rammer}},\ }\href {https://doi.org/10.1017/CBO9780511618956} {\emph
  {\bibinfo {title} {Quantum Field Theory of Non-equilibrium States}}}\
  (\bibinfo  {publisher} {Cambridge University Press, Cambridge},\ \bibinfo
  {year} {2007})\BibitemShut {NoStop}%
\bibitem [{\citenamefont {Haug}\ and\ \citenamefont
  {Jauho}(2008)}]{HaugJauhoBook}%
  \BibitemOpen
  \bibfield  {author} {\bibinfo {author} {\bibfnamefont {H.}~\bibnamefont
  {Haug}}\ and\ \bibinfo {author} {\bibfnamefont {A.-P.}\ \bibnamefont
  {Jauho}},\ }\href {https://doi.org/10.1007/978-3-540-73564-9} {\emph
  {\bibinfo {title} {Quantum Kinetics in Transport and Optics of
  Semiconductors}}},\ \bibinfo {series} {Springer Series in Solid-State
  Sciences}, Vol.\ \bibinfo {volume} {123}\ (\bibinfo  {publisher} {Springer,
  Berlin},\ \bibinfo {year} {2008})\BibitemShut {NoStop}%
\bibitem [{\citenamefont {Yang}\ \emph {et~al.}(2010)\citenamefont {Yang},
  \citenamefont {Orenstein},\ and\ \citenamefont {Lee}}]{Yang2010RandomWalk}%
  \BibitemOpen
  \bibfield  {author} {\bibinfo {author} {\bibfnamefont {L.}~\bibnamefont
  {Yang}}, \bibinfo {author} {\bibfnamefont {J.}~\bibnamefont {Orenstein}},\
  and\ \bibinfo {author} {\bibfnamefont {D.-H.}\ \bibnamefont {Lee}},\
  }\bibfield  {title} {\bibinfo {title} {Random walk approach to spin dynamics
  in a two-dimensional electron gas with spin-orbit coupling},\ }\href
  {https://doi.org/10.1103/PhysRevB.82.155324} {\bibfield  {journal} {\bibinfo
  {journal} {Phys. Rev. B}\ }\textbf {\bibinfo {volume} {82}},\ \bibinfo
  {pages} {155324} (\bibinfo {year} {2010})}\BibitemShut {NoStop}%
\bibitem [{\citenamefont {Ferreira}\ \emph {et~al.}(2017)\citenamefont
  {Ferreira}, \citenamefont {Hernandez}, \citenamefont {Altmann},\ and\
  \citenamefont {Salis}}]{Ferreira2017SpinDiff2}%
  \BibitemOpen
  \bibfield  {author} {\bibinfo {author} {\bibfnamefont {G.~J.}\ \bibnamefont
  {Ferreira}}, \bibinfo {author} {\bibfnamefont {F.~G.~G.}\ \bibnamefont
  {Hernandez}}, \bibinfo {author} {\bibfnamefont {P.}~\bibnamefont {Altmann}},\
  and\ \bibinfo {author} {\bibfnamefont {G.}~\bibnamefont {Salis}},\ }\bibfield
   {title} {\bibinfo {title} {Spin drift and diffusion in one- and two-subband
  helical systems},\ }\href {https://doi.org/10.1103/PhysRevB.95.125119}
  {\bibfield  {journal} {\bibinfo  {journal} {Phys. Rev. B}\ }\textbf {\bibinfo
  {volume} {95}},\ \bibinfo {pages} {125119} (\bibinfo {year}
  {2017})}\BibitemShut {NoStop}%
\bibitem [{\citenamefont {Froltsov}(2001)}]{Froltsov2001}%
  \BibitemOpen
  \bibfield  {author} {\bibinfo {author} {\bibfnamefont {V.~A.}\ \bibnamefont
  {Froltsov}},\ }\bibfield  {title} {\bibinfo {title} {{Diffusion of
  inhomogeneous spin distribution in a magnetic field parallel to interfaces of
  a III-V semiconductor quantum well}},\ }\href
  {https://doi.org/10.1103/PhysRevB.64.045311} {\bibfield  {journal} {\bibinfo
  {journal} {Phys. Rev. B}\ }\textbf {\bibinfo {volume} {64}},\ \bibinfo
  {pages} {045311} (\bibinfo {year} {2001})}\BibitemShut {NoStop}%
\bibitem [{\citenamefont {Pershin}(2005)}]{Pershin2005}%
  \BibitemOpen
  \bibfield  {author} {\bibinfo {author} {\bibfnamefont {Y.~V.}\ \bibnamefont
  {Pershin}},\ }\bibfield  {title} {\bibinfo {title} {Long-lived spin coherence
  states in semiconductor heterostructures},\ }\href
  {https://doi.org/10.1103/PhysRevB.71.155317} {\bibfield  {journal} {\bibinfo
  {journal} {Phys. Rev. B}\ }\textbf {\bibinfo {volume} {71}},\ \bibinfo
  {pages} {155317} (\bibinfo {year} {2005})}\BibitemShut {NoStop}%
\bibitem [{\citenamefont {Schwab}\ \emph {et~al.}(2006)\citenamefont {Schwab},
  \citenamefont {Dzierzawa}, \citenamefont {Gorini},\ and\ \citenamefont
  {Raimondi}}]{Schwab2006}%
  \BibitemOpen
  \bibfield  {author} {\bibinfo {author} {\bibfnamefont {P.}~\bibnamefont
  {Schwab}}, \bibinfo {author} {\bibfnamefont {M.}~\bibnamefont {Dzierzawa}},
  \bibinfo {author} {\bibfnamefont {C.}~\bibnamefont {Gorini}},\ and\ \bibinfo
  {author} {\bibfnamefont {R.}~\bibnamefont {Raimondi}},\ }\bibfield  {title}
  {\bibinfo {title} {Spin relaxation in narrow wires of a two-dimensional
  electron gas},\ }\href {https://doi.org/10.1103/PhysRevB.74.155316}
  {\bibfield  {journal} {\bibinfo  {journal} {Phys. Rev. B}\ }\textbf {\bibinfo
  {volume} {74}},\ \bibinfo {pages} {155316} (\bibinfo {year}
  {2006})}\BibitemShut {NoStop}%
\bibitem [{\citenamefont {Stanescu}\ and\ \citenamefont
  {Galitski}(2007)}]{Stanescu2007}%
  \BibitemOpen
  \bibfield  {author} {\bibinfo {author} {\bibfnamefont {T.~D.}\ \bibnamefont
  {Stanescu}}\ and\ \bibinfo {author} {\bibfnamefont {V.}~\bibnamefont
  {Galitski}},\ }\bibfield  {title} {\bibinfo {title} {Spin relaxation in a
  generic two-dimensional spin-orbit coupled system},\ }\href
  {https://doi.org/10.1103/PhysRevB.75.125307} {\bibfield  {journal} {\bibinfo
  {journal} {Phys. Rev. B}\ }\textbf {\bibinfo {volume} {75}},\ \bibinfo
  {pages} {125307} (\bibinfo {year} {2007})}\BibitemShut {NoStop}%
\bibitem [{\citenamefont {Tokatly}(2008)}]{Tokatly2008}%
  \BibitemOpen
  \bibfield  {author} {\bibinfo {author} {\bibfnamefont {I.~V.}\ \bibnamefont
  {Tokatly}},\ }\bibfield  {title} {\bibinfo {title} {Equilibrium spin
  currents: Non-abelian gauge invariance and color diamagnetism in condensed
  matter},\ }\href {https://doi.org/10.1103/PhysRevLett.101.106601} {\bibfield
  {journal} {\bibinfo  {journal} {Phys. Rev. Lett.}\ }\textbf {\bibinfo
  {volume} {101}},\ \bibinfo {pages} {106601} (\bibinfo {year}
  {2008})}\BibitemShut {NoStop}%
\bibitem [{\citenamefont {Tokatly}\ and\ \citenamefont
  {Sherman}(2010)}]{Tokatly2010}%
  \BibitemOpen
  \bibfield  {author} {\bibinfo {author} {\bibfnamefont {I.}~\bibnamefont
  {Tokatly}}\ and\ \bibinfo {author} {\bibfnamefont {E.}~\bibnamefont
  {Sherman}},\ }\bibfield  {title} {\bibinfo {title} {Gauge theory approach for
  diffusive and precessional spin dynamics in a two-dimensional electron gas},\
  }\href {https://doi.org/10.1016/j.aop.2010.01.007} {\bibfield  {journal}
  {\bibinfo  {journal} {Ann. Phys. (NY)}\ }\textbf {\bibinfo {volume} {325}},\
  \bibinfo {pages} {1104} (\bibinfo {year} {2010})}\BibitemShut {NoStop}%
\bibitem [{\citenamefont {Liu}\ and\ \citenamefont
  {Sinova}(2012)}]{Sinova2012UTheorySpin}%
  \BibitemOpen
  \bibfield  {author} {\bibinfo {author} {\bibfnamefont {X.}~\bibnamefont
  {Liu}}\ and\ \bibinfo {author} {\bibfnamefont {J.}~\bibnamefont {Sinova}},\
  }\bibfield  {title} {\bibinfo {title} {Unified theory of spin dynamics in a
  two-dimensional electron gas with arbitrary spin-orbit coupling strength at
  finite temperature},\ }\href {https://doi.org/10.1103/PhysRevB.86.174301}
  {\bibfield  {journal} {\bibinfo  {journal} {Phys. Rev. B}\ }\textbf {\bibinfo
  {volume} {86}},\ \bibinfo {pages} {174301} (\bibinfo {year}
  {2012})}\BibitemShut {NoStop}%
\bibitem [{\citenamefont {Salis}\ \emph {et~al.}(2014)\citenamefont {Salis},
  \citenamefont {Walser}, \citenamefont {Altmann}, \citenamefont {Reichl},\
  and\ \citenamefont {Wegscheider}}]{Salis2014PSH}%
  \BibitemOpen
  \bibfield  {author} {\bibinfo {author} {\bibfnamefont {G.}~\bibnamefont
  {Salis}}, \bibinfo {author} {\bibfnamefont {M.~P.}\ \bibnamefont {Walser}},
  \bibinfo {author} {\bibfnamefont {P.}~\bibnamefont {Altmann}}, \bibinfo
  {author} {\bibfnamefont {C.}~\bibnamefont {Reichl}},\ and\ \bibinfo {author}
  {\bibfnamefont {W.}~\bibnamefont {Wegscheider}},\ }\bibfield  {title}
  {\bibinfo {title} {Dynamics of a localized spin excitation close to the
  spin-helix regime},\ }\href {https://doi.org/10.1103/PhysRevB.89.045304}
  {\bibfield  {journal} {\bibinfo  {journal} {Phys. Rev. B}\ }\textbf {\bibinfo
  {volume} {89}},\ \bibinfo {pages} {045304} (\bibinfo {year}
  {2014})}\BibitemShut {NoStop}%
\bibitem [{\citenamefont {Shen}\ \emph {et~al.}(2014)\citenamefont {Shen},
  \citenamefont {Raimondi},\ and\ \citenamefont {Vignale}}]{shen2014theory}%
  \BibitemOpen
  \bibfield  {author} {\bibinfo {author} {\bibfnamefont {K.}~\bibnamefont
  {Shen}}, \bibinfo {author} {\bibfnamefont {R.}~\bibnamefont {Raimondi}},\
  and\ \bibinfo {author} {\bibfnamefont {G.}~\bibnamefont {Vignale}},\
  }\bibfield  {title} {\bibinfo {title} {Theory of coupled spin-charge
  transport due to spin-orbit interaction in inhomogeneous two-dimensional
  electron liquids},\ }\href {https://doi.org/10.1103/PhysRevB.90.245302}
  {\bibfield  {journal} {\bibinfo  {journal} {Phys. Rev. B}\ }\textbf {\bibinfo
  {volume} {90}},\ \bibinfo {pages} {245302} (\bibinfo {year}
  {2014})}\BibitemShut {NoStop}%
\bibitem [{\citenamefont {de~Assis}(2019)}]{mastersIsmael}%
  \BibitemOpen
  \bibfield  {author} {\bibinfo {author} {\bibfnamefont {I.~R.}\ \bibnamefont
  {de~Assis}},\ }\emph {\bibinfo {title} {From the Keldysh formalism to the
  Boltzmann equation for spin drift and diffusion}},\ \href
  {https://repositorio.ufu.br/handle/123456789/29403} {Master's thesis},\
  \bibinfo  {school} {Instituto de Física, Universidade Federal de
  Uberlândia} (\bibinfo {year} {2019})\BibitemShut {NoStop}%
\bibitem [{\citenamefont {Altmann}\ \emph {et~al.}(2016)\citenamefont
  {Altmann}, \citenamefont {Hernandez}, \citenamefont {Ferreira}, \citenamefont
  {Kohda}, \citenamefont {Reichl}, \citenamefont {Wegscheider},\ and\
  \citenamefont {Salis}}]{Altmann2016CubicSOCPSH}%
  \BibitemOpen
  \bibfield  {author} {\bibinfo {author} {\bibfnamefont {P.}~\bibnamefont
  {Altmann}}, \bibinfo {author} {\bibfnamefont {F.~G.~G.}\ \bibnamefont
  {Hernandez}}, \bibinfo {author} {\bibfnamefont {G.~J.}\ \bibnamefont
  {Ferreira}}, \bibinfo {author} {\bibfnamefont {M.}~\bibnamefont {Kohda}},
  \bibinfo {author} {\bibfnamefont {C.}~\bibnamefont {Reichl}}, \bibinfo
  {author} {\bibfnamefont {W.}~\bibnamefont {Wegscheider}},\ and\ \bibinfo
  {author} {\bibfnamefont {G.}~\bibnamefont {Salis}},\ }\bibfield  {title}
  {\bibinfo {title} {Current-controlled spin precession of quasistationary
  electrons in a cubic spin-orbit field},\ }\href
  {https://doi.org/10.1103/PhysRevLett.116.196802} {\bibfield  {journal}
  {\bibinfo  {journal} {Phys. Rev. Lett.}\ }\textbf {\bibinfo {volume} {116}},\
  \bibinfo {pages} {196802} (\bibinfo {year} {2016})}\BibitemShut {NoStop}%
\bibitem [{\citenamefont {Kunihashi}\ \emph {et~al.}(2016)\citenamefont
  {Kunihashi}, \citenamefont {Sanada}, \citenamefont {Gotoh}, \citenamefont
  {Onomitsu}, \citenamefont {Kohda}, \citenamefont {Nitta},\ and\ \citenamefont
  {Sogawa}}]{Kunihashi2016Drift}%
  \BibitemOpen
  \bibfield  {author} {\bibinfo {author} {\bibfnamefont {Y.}~\bibnamefont
  {Kunihashi}}, \bibinfo {author} {\bibfnamefont {H.}~\bibnamefont {Sanada}},
  \bibinfo {author} {\bibfnamefont {H.}~\bibnamefont {Gotoh}}, \bibinfo
  {author} {\bibfnamefont {K.}~\bibnamefont {Onomitsu}}, \bibinfo {author}
  {\bibfnamefont {M.}~\bibnamefont {Kohda}}, \bibinfo {author} {\bibfnamefont
  {J.}~\bibnamefont {Nitta}},\ and\ \bibinfo {author} {\bibfnamefont
  {T.}~\bibnamefont {Sogawa}},\ }\bibfield  {title} {\bibinfo {title} {Drift
  transport of helical spin coherence with tailored spin--orbit interactions},\
  }\href {https://doi.org/10.1038/ncomms10722} {\bibfield  {journal} {\bibinfo
  {journal} {Nat. Commun.}\ }\textbf {\bibinfo {volume} {7}},\ \bibinfo {pages}
  {10722} (\bibinfo {year} {2016})}\BibitemShut {NoStop}%
\bibitem [{\citenamefont {Dettwiler}\ \emph {et~al.}(2017)\citenamefont
  {Dettwiler}, \citenamefont {Fu}, \citenamefont {Mack}, \citenamefont
  {Weigele}, \citenamefont {Egues}, \citenamefont {Awschalom},\ and\
  \citenamefont {Zumb{\"u}hl}}]{dettwiler2017stretchable}%
  \BibitemOpen
  \bibfield  {author} {\bibinfo {author} {\bibfnamefont {F.}~\bibnamefont
  {Dettwiler}}, \bibinfo {author} {\bibfnamefont {J.}~\bibnamefont {Fu}},
  \bibinfo {author} {\bibfnamefont {S.}~\bibnamefont {Mack}}, \bibinfo {author}
  {\bibfnamefont {P.~J.}\ \bibnamefont {Weigele}}, \bibinfo {author}
  {\bibfnamefont {J.~C.}\ \bibnamefont {Egues}}, \bibinfo {author}
  {\bibfnamefont {D.~D.}\ \bibnamefont {Awschalom}},\ and\ \bibinfo {author}
  {\bibfnamefont {D.~M.}\ \bibnamefont {Zumb{\"u}hl}},\ }\bibfield  {title}
  {\bibinfo {title} {{Stretchable persistent spin helices in GaAs quantum
  wells}},\ }\href {https://doi.org/10.1103/PhysRevX.7.031010} {\bibfield
  {journal} {\bibinfo  {journal} {Phys. Rev. X}\ }\textbf {\bibinfo {volume}
  {7}},\ \bibinfo {pages} {031010} (\bibinfo {year} {2017})}\BibitemShut
  {NoStop}%
\bibitem [{\citenamefont {Weigele}\ \emph {et~al.}(2020)\citenamefont
  {Weigele}, \citenamefont {Marinescu}, \citenamefont {Dettwiler},
  \citenamefont {Fu}, \citenamefont {Mack}, \citenamefont {Egues},
  \citenamefont {Awschalom},\ and\ \citenamefont
  {Zumb{\"u}hl}}]{Weigele2020WeakLocal}%
  \BibitemOpen
  \bibfield  {author} {\bibinfo {author} {\bibfnamefont {P.~J.}\ \bibnamefont
  {Weigele}}, \bibinfo {author} {\bibfnamefont {D.~C.}\ \bibnamefont
  {Marinescu}}, \bibinfo {author} {\bibfnamefont {F.}~\bibnamefont
  {Dettwiler}}, \bibinfo {author} {\bibfnamefont {J.}~\bibnamefont {Fu}},
  \bibinfo {author} {\bibfnamefont {S.}~\bibnamefont {Mack}}, \bibinfo {author}
  {\bibfnamefont {J.~C.}\ \bibnamefont {Egues}}, \bibinfo {author}
  {\bibfnamefont {D.~D.}\ \bibnamefont {Awschalom}},\ and\ \bibinfo {author}
  {\bibfnamefont {D.~M.}\ \bibnamefont {Zumb{\"u}hl}},\ }\bibfield  {title}
  {\bibinfo {title} {Symmetry breaking of the persistent spin helix in quantum
  transport},\ }\href {https://doi.org/10.1103/PhysRevB.101.035414} {\bibfield
  {journal} {\bibinfo  {journal} {Phys. Rev. B}\ }\textbf {\bibinfo {volume}
  {101}},\ \bibinfo {pages} {035414} (\bibinfo {year} {2020})}\BibitemShut
  {NoStop}%
\bibitem [{\citenamefont {Iizasa}\ \emph {et~al.}(2021)\citenamefont {Iizasa},
  \citenamefont {Aoki}, \citenamefont {Saito}, \citenamefont {Nitta},
  \citenamefont {Salis},\ and\ \citenamefont {Kohda}}]{Iizasa2021Anisotropy}%
  \BibitemOpen
  \bibfield  {author} {\bibinfo {author} {\bibfnamefont {D.}~\bibnamefont
  {Iizasa}}, \bibinfo {author} {\bibfnamefont {A.}~\bibnamefont {Aoki}},
  \bibinfo {author} {\bibfnamefont {T.}~\bibnamefont {Saito}}, \bibinfo
  {author} {\bibfnamefont {J.}~\bibnamefont {Nitta}}, \bibinfo {author}
  {\bibfnamefont {G.}~\bibnamefont {Salis}},\ and\ \bibinfo {author}
  {\bibfnamefont {M.}~\bibnamefont {Kohda}},\ }\bibfield  {title} {\bibinfo
  {title} {Control of spin relaxation anisotropy by spin-orbit-coupled
  diffusive spin motion},\ }\href {https://doi.org/10.1103/PhysRevB.103.024427}
  {\bibfield  {journal} {\bibinfo  {journal} {Phys. Rev. B}\ }\textbf {\bibinfo
  {volume} {103}},\ \bibinfo {pages} {024427} (\bibinfo {year}
  {2021})}\BibitemShut {NoStop}%
\bibitem [{\citenamefont {Anghel}\ \emph {et~al.}(2021)\citenamefont {Anghel},
  \citenamefont {Poshakinskiy}, \citenamefont {Schiller}, \citenamefont
  {Passmann}, \citenamefont {Ruppert}, \citenamefont {Tarasenko}, \citenamefont
  {Yusa}, \citenamefont {Mano}, \citenamefont {Noda},\ and\ \citenamefont
  {Betz}}]{Anghel2021EHoles}%
  \BibitemOpen
  \bibfield  {author} {\bibinfo {author} {\bibfnamefont {S.}~\bibnamefont
  {Anghel}}, \bibinfo {author} {\bibfnamefont {A.~V.}\ \bibnamefont
  {Poshakinskiy}}, \bibinfo {author} {\bibfnamefont {K.}~\bibnamefont
  {Schiller}}, \bibinfo {author} {\bibfnamefont {F.}~\bibnamefont {Passmann}},
  \bibinfo {author} {\bibfnamefont {C.}~\bibnamefont {Ruppert}}, \bibinfo
  {author} {\bibfnamefont {S.~A.}\ \bibnamefont {Tarasenko}}, \bibinfo {author}
  {\bibfnamefont {G.}~\bibnamefont {Yusa}}, \bibinfo {author} {\bibfnamefont
  {T.}~\bibnamefont {Mano}}, \bibinfo {author} {\bibfnamefont {T.}~\bibnamefont
  {Noda}},\ and\ \bibinfo {author} {\bibfnamefont {M.}~\bibnamefont {Betz}},\
  }\bibfield  {title} {\bibinfo {title} {{Anisotropic expansion of drifting
  spin helices in GaAs quantum wells}},\ }\href
  {https://doi.org/10.1103/PhysRevB.103.035429} {\bibfield  {journal} {\bibinfo
   {journal} {Phys. Rev. B}\ }\textbf {\bibinfo {volume} {103}},\ \bibinfo
  {pages} {035429} (\bibinfo {year} {2021})}\BibitemShut {NoStop}%
\bibitem [{\citenamefont {Glazov}\ \emph
  {et~al.}(2010{\natexlab{a}})\citenamefont {Glazov}, \citenamefont {Semina},\
  and\ \citenamefont {Sherman}}]{Glazov2010PRB}%
  \BibitemOpen
  \bibfield  {author} {\bibinfo {author} {\bibfnamefont {M.~M.}\ \bibnamefont
  {Glazov}}, \bibinfo {author} {\bibfnamefont {M.~A.}\ \bibnamefont {Semina}},\
  and\ \bibinfo {author} {\bibfnamefont {E.~Y.}\ \bibnamefont {Sherman}},\
  }\bibfield  {title} {\bibinfo {title} {Spin relaxation in multiple (110)
  quantum wells},\ }\href {https://doi.org/10.1103/physrevb.81.115332}
  {\bibfield  {journal} {\bibinfo  {journal} {Phys. Rev. B}\ }\textbf {\bibinfo
  {volume} {81}},\ \bibinfo {pages} {115332} (\bibinfo {year}
  {2010}{\natexlab{a}})}\BibitemShut {NoStop}%
\bibitem [{\citenamefont {Glazov}\ \emph
  {et~al.}(2010{\natexlab{b}})\citenamefont {Glazov}, \citenamefont {Sherman},\
  and\ \citenamefont {Dugaev}}]{Glazov2010PhysE}%
  \BibitemOpen
  \bibfield  {author} {\bibinfo {author} {\bibfnamefont {M.}~\bibnamefont
  {Glazov}}, \bibinfo {author} {\bibfnamefont {E.}~\bibnamefont {Sherman}},\
  and\ \bibinfo {author} {\bibfnamefont {V.}~\bibnamefont {Dugaev}},\
  }\bibfield  {title} {\bibinfo {title} {Two-dimensional electron gas with
  spin{\textendash}orbit coupling disorder},\ }\href
  {https://doi.org/10.1016/j.physe.2010.04.021} {\bibfield  {journal} {\bibinfo
   {journal} {Phys. E (Amsterdam)}\ }\textbf {\bibinfo {volume} {42}},\
  \bibinfo {pages} {2157} (\bibinfo {year} {2010}{\natexlab{b}})}\BibitemShut
  {NoStop}%
\bibitem [{\citenamefont {Bindel}\ \emph {et~al.}(2016)\citenamefont {Bindel},
  \citenamefont {Pezzotta}, \citenamefont {Ulrich}, \citenamefont {Liebmann},
  \citenamefont {Sherman},\ and\ \citenamefont
  {Morgenstern}}]{Bindel2016NatPhys}%
  \BibitemOpen
  \bibfield  {author} {\bibinfo {author} {\bibfnamefont {J.~R.}\ \bibnamefont
  {Bindel}}, \bibinfo {author} {\bibfnamefont {M.}~\bibnamefont {Pezzotta}},
  \bibinfo {author} {\bibfnamefont {J.}~\bibnamefont {Ulrich}}, \bibinfo
  {author} {\bibfnamefont {M.}~\bibnamefont {Liebmann}}, \bibinfo {author}
  {\bibfnamefont {E.~Y.}\ \bibnamefont {Sherman}},\ and\ \bibinfo {author}
  {\bibfnamefont {M.}~\bibnamefont {Morgenstern}},\ }\bibfield  {title}
  {\bibinfo {title} {Probing variations of the rashba spin{\textendash}orbit
  coupling at the nanometre scale},\ }\href {https://doi.org/10.1038/nphys3774}
  {\bibfield  {journal} {\bibinfo  {journal} {Nat. Phys.}\ }\textbf {\bibinfo
  {volume} {12}},\ \bibinfo {pages} {920} (\bibinfo {year} {2016})}\BibitemShut
  {NoStop}%
\bibitem [{\citenamefont {Schliemann}(2017)}]{schliemann2017colloquium}%
  \BibitemOpen
  \bibfield  {author} {\bibinfo {author} {\bibfnamefont {J.}~\bibnamefont
  {Schliemann}},\ }\bibfield  {title} {\bibinfo {title} {Colloquium: Persistent
  spin textures in semiconductor nanostructures},\ }\href
  {https://doi.org/10.1103/RevModPhys.89.011001} {\bibfield  {journal}
  {\bibinfo  {journal} {Rev. Mod. Phys.}\ }\textbf {\bibinfo {volume} {89}},\
  \bibinfo {pages} {011001} (\bibinfo {year} {2017})}\BibitemShut {NoStop}%
\bibitem [{\citenamefont {Bernardes}\ \emph {et~al.}(2006)\citenamefont
  {Bernardes}, \citenamefont {Schliemann}, \citenamefont {Egues},\ and\
  \citenamefont {Loss}}]{bernardes2006spin}%
  \BibitemOpen
  \bibfield  {author} {\bibinfo {author} {\bibfnamefont {E.}~\bibnamefont
  {Bernardes}}, \bibinfo {author} {\bibfnamefont {J.}~\bibnamefont
  {Schliemann}}, \bibinfo {author} {\bibfnamefont {J.~C.}\ \bibnamefont
  {Egues}},\ and\ \bibinfo {author} {\bibfnamefont {D.}~\bibnamefont {Loss}},\
  }\bibfield  {title} {\bibinfo {title} {Spin orbit interaction and
  zitterbewegung in symmetric wells},\ }\href
  {https://doi.org/10.1002/pssc.200672855} {\bibfield  {journal} {\bibinfo
  {journal} {Phys. Status Solidi C}\ }\textbf {\bibinfo {volume} {3}},\
  \bibinfo {pages} {4330} (\bibinfo {year} {2006})}\BibitemShut {NoStop}%
\bibitem [{\citenamefont {Bernardes}\ \emph {et~al.}(2007)\citenamefont
  {Bernardes}, \citenamefont {Schliemann}, \citenamefont {Lee}, \citenamefont
  {Egues},\ and\ \citenamefont {Loss}}]{EsmerindoPRL2007}%
  \BibitemOpen
  \bibfield  {author} {\bibinfo {author} {\bibfnamefont {E.}~\bibnamefont
  {Bernardes}}, \bibinfo {author} {\bibfnamefont {J.}~\bibnamefont
  {Schliemann}}, \bibinfo {author} {\bibfnamefont {M.}~\bibnamefont {Lee}},
  \bibinfo {author} {\bibfnamefont {J.~C.}\ \bibnamefont {Egues}},\ and\
  \bibinfo {author} {\bibfnamefont {D.}~\bibnamefont {Loss}},\ }\bibfield
  {title} {\bibinfo {title} {Spin-orbit interaction in symmetric wells with two
  subbands},\ }\href {https://doi.org/10.1103/PhysRevLett.99.076603} {\bibfield
   {journal} {\bibinfo  {journal} {Phys. Rev. Lett.}\ }\textbf {\bibinfo
  {volume} {99}},\ \bibinfo {pages} {076603} (\bibinfo {year}
  {2007})}\BibitemShut {NoStop}%
\bibitem [{\citenamefont {Calsaverini}\ \emph {et~al.}(2008)\citenamefont
  {Calsaverini}, \citenamefont {Bernardes}, \citenamefont {Egues},\ and\
  \citenamefont {Loss}}]{calsaverini2008intersubband}%
  \BibitemOpen
  \bibfield  {author} {\bibinfo {author} {\bibfnamefont {R.~S.}\ \bibnamefont
  {Calsaverini}}, \bibinfo {author} {\bibfnamefont {E.}~\bibnamefont
  {Bernardes}}, \bibinfo {author} {\bibfnamefont {J.~C.}\ \bibnamefont
  {Egues}},\ and\ \bibinfo {author} {\bibfnamefont {D.}~\bibnamefont {Loss}},\
  }\bibfield  {title} {\bibinfo {title} {Intersubband-induced spin-orbit
  interaction in quantum wells},\ }\href
  {https://doi.org/10.1103/PhysRevB.78.155313} {\bibfield  {journal} {\bibinfo
  {journal} {Phys. Rev. B}\ }\textbf {\bibinfo {volume} {78}},\ \bibinfo
  {pages} {155313} (\bibinfo {year} {2008})}\BibitemShut {NoStop}%
\bibitem [{\citenamefont {Fu}\ and\ \citenamefont
  {Egues}(2015)}]{Fu2015TwoSubbands}%
  \BibitemOpen
  \bibfield  {author} {\bibinfo {author} {\bibfnamefont {J.}~\bibnamefont
  {Fu}}\ and\ \bibinfo {author} {\bibfnamefont {J.~C.}\ \bibnamefont {Egues}},\
  }\bibfield  {title} {\bibinfo {title} {{Spin-orbit interaction in GaAs wells:
  From one to two subbands}},\ }\href
  {https://doi.org/10.1103/PhysRevB.91.075408} {\bibfield  {journal} {\bibinfo
  {journal} {Phys. Rev. B}\ }\textbf {\bibinfo {volume} {91}},\ \bibinfo
  {pages} {075408} (\bibinfo {year} {2015})}\BibitemShut {NoStop}%
\bibitem [{\citenamefont {Fu}\ \emph {et~al.}(2016)\citenamefont {Fu},
  \citenamefont {Penteado}, \citenamefont {Hachiya}, \citenamefont {Loss},\
  and\ \citenamefont {Egues}}]{Fu2015Skyrmion}%
  \BibitemOpen
  \bibfield  {author} {\bibinfo {author} {\bibfnamefont {J.}~\bibnamefont
  {Fu}}, \bibinfo {author} {\bibfnamefont {P.~H.}\ \bibnamefont {Penteado}},
  \bibinfo {author} {\bibfnamefont {M.~O.}\ \bibnamefont {Hachiya}}, \bibinfo
  {author} {\bibfnamefont {D.}~\bibnamefont {Loss}},\ and\ \bibinfo {author}
  {\bibfnamefont {J.~C.}\ \bibnamefont {Egues}},\ }\bibfield  {title} {\bibinfo
  {title} {Persistent skyrmion lattice of noninteracting electrons with
  spin-orbit coupling},\ }\href
  {https://doi.org/10.1103/PhysRevLett.117.226401} {\bibfield  {journal}
  {\bibinfo  {journal} {Phys. Rev. Lett.}\ }\textbf {\bibinfo {volume} {117}},\
  \bibinfo {pages} {226401} (\bibinfo {year} {2016})}\BibitemShut {NoStop}%
\bibitem [{\citenamefont {Hernandez}\ \emph {et~al.}(2016)\citenamefont
  {Hernandez}, \citenamefont {Ullah}, \citenamefont {Ferreira}, \citenamefont
  {Kawahala}, \citenamefont {Gusev},\ and\ \citenamefont
  {Bakarov}}]{Hernandez2016CISP}%
  \BibitemOpen
  \bibfield  {author} {\bibinfo {author} {\bibfnamefont {F.~G.~G.}\
  \bibnamefont {Hernandez}}, \bibinfo {author} {\bibfnamefont {S.}~\bibnamefont
  {Ullah}}, \bibinfo {author} {\bibfnamefont {G.~J.}\ \bibnamefont {Ferreira}},
  \bibinfo {author} {\bibfnamefont {N.~M.}\ \bibnamefont {Kawahala}}, \bibinfo
  {author} {\bibfnamefont {G.~M.}\ \bibnamefont {Gusev}},\ and\ \bibinfo
  {author} {\bibfnamefont {A.~K.}\ \bibnamefont {Bakarov}},\ }\bibfield
  {title} {\bibinfo {title} {Macroscopic transverse drift of long
  current-induced spin coherence in two-dimensional electron gases},\ }\href
  {https://doi.org/10.1103/PhysRevB.94.045305} {\bibfield  {journal} {\bibinfo
  {journal} {Phys. Rev. B}\ }\textbf {\bibinfo {volume} {94}},\ \bibinfo
  {pages} {045305} (\bibinfo {year} {2016})}\BibitemShut {NoStop}%
\bibitem [{\citenamefont {Luengo-Kovac}\ \emph {et~al.}(2017)\citenamefont
  {Luengo-Kovac}, \citenamefont {Moraes}, \citenamefont {Ferreira},
  \citenamefont {Ribeiro}, \citenamefont {Gusev}, \citenamefont {Bakarov},
  \citenamefont {Sih},\ and\ \citenamefont {Hernandez}}]{Luengo2017Gate}%
  \BibitemOpen
  \bibfield  {author} {\bibinfo {author} {\bibfnamefont {M.}~\bibnamefont
  {Luengo-Kovac}}, \bibinfo {author} {\bibfnamefont {F.~C.~D.}\ \bibnamefont
  {Moraes}}, \bibinfo {author} {\bibfnamefont {G.~J.}\ \bibnamefont
  {Ferreira}}, \bibinfo {author} {\bibfnamefont {A.~S.~L.}\ \bibnamefont
  {Ribeiro}}, \bibinfo {author} {\bibfnamefont {G.~M.}\ \bibnamefont {Gusev}},
  \bibinfo {author} {\bibfnamefont {A.~K.}\ \bibnamefont {Bakarov}}, \bibinfo
  {author} {\bibfnamefont {V.}~\bibnamefont {Sih}},\ and\ \bibinfo {author}
  {\bibfnamefont {F.~G.~G.}\ \bibnamefont {Hernandez}},\ }\bibfield  {title}
  {\bibinfo {title} {Gate control of the spin mobility through the modification
  of the spin-orbit interaction in two-dimensional systems},\ }\href
  {https://doi.org/10.1103/PhysRevB.95.245315} {\bibfield  {journal} {\bibinfo
  {journal} {Phys. Rev. B}\ }\textbf {\bibinfo {volume} {95}},\ \bibinfo
  {pages} {245315} (\bibinfo {year} {2017})}\BibitemShut {NoStop}%
\bibitem [{\citenamefont {Hernandez}\ \emph {et~al.}(2020)\citenamefont
  {Hernandez}, \citenamefont {Ferreira}, \citenamefont {Luengo-Kovac},
  \citenamefont {Sih}, \citenamefont {Kawahala}, \citenamefont {Gusev},\ and\
  \citenamefont {Bakarov}}]{Hernandez2020Anisotropy}%
  \BibitemOpen
  \bibfield  {author} {\bibinfo {author} {\bibfnamefont {F.~G.~G.}\
  \bibnamefont {Hernandez}}, \bibinfo {author} {\bibfnamefont {G.~J.}\
  \bibnamefont {Ferreira}}, \bibinfo {author} {\bibfnamefont {M.}~\bibnamefont
  {Luengo-Kovac}}, \bibinfo {author} {\bibfnamefont {V.}~\bibnamefont {Sih}},
  \bibinfo {author} {\bibfnamefont {N.~M.}\ \bibnamefont {Kawahala}}, \bibinfo
  {author} {\bibfnamefont {G.~M.}\ \bibnamefont {Gusev}},\ and\ \bibinfo
  {author} {\bibfnamefont {A.~K.}\ \bibnamefont {Bakarov}},\ }\bibfield
  {title} {\bibinfo {title} {Electrical control of spin relaxation anisotropy
  during drift transport in a two-dimensional electron gas},\ }\href
  {https://doi.org/10.1103/PhysRevB.102.125305} {\bibfield  {journal} {\bibinfo
   {journal} {Phys. Rev. B}\ }\textbf {\bibinfo {volume} {102}},\ \bibinfo
  {pages} {125305} (\bibinfo {year} {2020})}\BibitemShut {NoStop}%
\bibitem [{\citenamefont {Zaremba}(1992)}]{Zaremba1992Boltzmann2subbands}%
  \BibitemOpen
  \bibfield  {author} {\bibinfo {author} {\bibfnamefont {E.}~\bibnamefont
  {Zaremba}},\ }\bibfield  {title} {\bibinfo {title} {Transverse
  magnetoresistance in quantum wells with multiple subband occupancy},\ }\href
  {https://doi.org/10.1103/PhysRevB.45.14143} {\bibfield  {journal} {\bibinfo
  {journal} {Phys. Rev. B}\ }\textbf {\bibinfo {volume} {45}},\ \bibinfo
  {pages} {14143} (\bibinfo {year} {1992})}\BibitemShut {NoStop}%
\bibitem [{\citenamefont {Bruus}\ and\ \citenamefont
  {Flensberg}(2004)}]{bruus2004many}%
  \BibitemOpen
  \bibfield  {author} {\bibinfo {author} {\bibfnamefont {H.}~\bibnamefont
  {Bruus}}\ and\ \bibinfo {author} {\bibfnamefont {K.}~\bibnamefont
  {Flensberg}},\ }\href@noop {} {\emph {\bibinfo {title} {Many-body Quantum
  Theory in Condensed Matter Physics: An Introduction}}}\ (\bibinfo
  {publisher} {Oxford University Press, Oxford},\ \bibinfo {year}
  {2004})\BibitemShut {NoStop}%
\end{thebibliography}%


%

\end{document}